\pdfoutput=1
\documentclass[aps,prx,floatfix,superscriptaddress,twocolumn,reprint]{revtex4-2}
\usepackage{eurosym}
\usepackage{amsbsy}
\usepackage{latexsym,epsfig,graphicx}
\usepackage{dcolumn}
\usepackage{graphicx}
\usepackage{subfigure}
\usepackage{comment}
\usepackage{color}
\usepackage{bm}
\usepackage{mathrsfs} 
\usepackage{amssymb}
\usepackage{amsthm}
\usepackage{amsfonts}
\usepackage{amsmath}
\usepackage{xspace}
\usepackage{epstopdf}  
\usepackage{tabularx}
\usepackage{longtable}
\usepackage[colorlinks=true, letterpaper=true, pdfstartview=FitV, linkcolor=red, citecolor=blue, urlcolor=blue]{hyperref}
\usepackage[normalem]{ulem}
\usepackage[version=4]{mhchem}
\usepackage{longtable}
\usepackage{ulem}
\usepackage{color}
\newtheorem{lemma}{Lemma}
\newtheorem{lemma*}{Lemma*}
\newtheorem{theorem}{Theorem}
\newcommand{\Z}{$\mathbb{Z}$}
\newcommand{\Zt}{$\mathbb{Z}_{2}$}
\newcommand{\Cze}{ \Z & $0$ & \Z & $0$ & \Z & $0$ & \Z & $0$}
\newcommand{\Czeze}{ \Z\,$\oplus$\,\Z & $0$ & \Z\,$\oplus$\,\Z & $0$ & \Z\,$\oplus$\,\Z & $0$ & \Z\,$\oplus$\,\Z & $0$}
\newcommand{\Con}{ $0$ & \Z & $0$ & \Z & $0$ & \Z & $0$ & \Z}
\newcommand{\Conon}{ $0$ & \Z\,$\oplus$\,\Z & $0$ & \Z\,$\oplus$\,\Z & $0$ & \Z\,$\oplus$\,\Z & $0$ & \Z\,$\oplus$\,\Z}
\newcommand{\Rze}{ \Z & $0$ & $0$ & $0$ & $2$\Z & $0$ & \Zt & \Zt}
\newcommand{\Rzeze}{ \Z\,$\oplus$\,\Z & $0$ & $0$ & $0$ & $2$\Z\,$\oplus$\,$2$\Z & $0$ & \Zt\,$\oplus$\,\Zt & \Zt\,$\oplus$\,\Zt}
\newcommand{\Ron}{ \Zt & \Z & $0$ & $0$ & $0$ & $2$\Z & $0$ & \Zt}
\newcommand{\Ronon}{ \Zt\,$\oplus$\,\Zt & \Z\,$\oplus$\,\Z & $0$ & $0$ & $0$ & $2$\Z\,$\oplus$\,$2$\Z & $0$ & \Zt\,$\oplus$\,\Zt}
\newcommand{\Rtw}{ \Zt & \Zt & \Z & $0$ & $0$ & $0$ & $2$\Z & $0$}
\newcommand{\Rtwtw}{ \Zt\,$\oplus$\,\Zt & \Zt\,$\oplus$\,\Zt & \Z\,$\oplus$\,\Z & $0$ & $0$ & $0$ & $2$\Z\,$\oplus$\,$2$\Z & $0$}
\newcommand{\Rth}{ $0$ & \Zt & \Zt & \Z & $0$ & $0$ & $0$ & $2$\Z}
\newcommand{\Rthth}{ $0$ & \Zt\,$\oplus$\,\Zt & \Zt\,$\oplus$\,\Zt & \Z\,$\oplus$\,\Z & $0$ & $0$ & $0$ & $2$\Z\,$\oplus$\,$2$\Z}
\newcommand{\Rfo}{ $2$\Z & $0$ & \Zt & \Zt & \Z & $0$ & $0$ & $0$}
\newcommand{\Rfofo}{$2$\Z\,$\oplus$\,$2$\Z & $0$ & \Zt\,$\oplus$\,\Zt & \Zt\,$\oplus$\,\Zt & \Z\,$\oplus$\,\Z & $0$ & $0$ & $0$}
\newcommand{\Rfi}{ $0$ & $2$\Z & $0$ & \Zt & \Zt & \Z & $0$ & $0$}
\newcommand{\Rfifi}{ $0$ & $2$\Z\,$\oplus$\,$2$\Z & $0$ & \Zt\,$\oplus$\,\Zt & \Zt\,$\oplus$\,\Zt & \Z\,$\oplus$\,\Z & $0$ & $0$}
\newcommand{\Rsi}{ $0$ & $0$ & $2$\Z & $0$ & \Zt & \Zt & \Z & $0$}
\newcommand{\Rsisi}{$0$ & $0$ & $2$\Z\,$\oplus$\,$2$\Z & $0$ & \Zt\,$\oplus$\,\Zt & \Zt\,$\oplus$\,\Zt & \Z\,$\oplus$\,\Z & $0$}
\newcommand{\Rse}{ $0$ & $0$ & $0$ & $2$\Z & $0$ & \Zt & \Zt & \Z}
\newcommand{\Rsese}{ $0$ & $0$ & $0$ & $2$\Z\,$\oplus$\,$2$\Z & $0$ & \Zt\,$\oplus$\,\Zt & \Zt\,$\oplus$\,\Zt & \Z\,$\oplus$\,\Z}

\begin{document}
\title{Symmetry and topological classification of Floquet non-Hermitian systems}
\author{Chun-Hui Liu}
\email{liuchunhui@iphy.ac.cn}
\affiliation{Beijing National Laboratory for Condensed Matter Physics, Institute of Physics, Chinese Academy of Sciences, Beijing 100190, China}
\affiliation{School of Physical Sciences, University of Chinese Academy of Sciences, Beijing 100049, China}
\author{Haiping Hu}
\email{hhu@iphy.ac.cn }
\affiliation{Beijing National Laboratory for Condensed Matter Physics, Institute of Physics, Chinese Academy of Sciences, Beijing 100190, China}
\author{Shu Chen}
\email{schen@iphy.ac.cn }
\affiliation{Beijing National Laboratory for Condensed Matter Physics, Institute of Physics, Chinese Academy of Sciences, Beijing 100190, China}
\affiliation{School of Physical Sciences, University of Chinese Academy of Sciences, Beijing 100049, China}
\affiliation{Yangtze River Delta Physics Research Center, Liyang, Jiangsu 213300, China }
\onecolumngrid

\twocolumngrid \clearpage \newpage
\begin{abstract}
Recent experimental advances in Floquet engineering and controlling dissipation in open systems have brought 
about unprecedented flexibility in tailoring novel phenomena without any static and Hermitian 
analogues. It can be epitomized by the various Floquet and non-Hermitian 
topological phases. Topological classifications of either static/Floquet Hermitian or 
 static non-Hermitian systems based 
on the underlying symmetries have been 
well established in the past several years. However, a coherent understanding and classification of Floquet 
non-Hermitian (FNH) topological phases have not been achieved yet. Here we systematically classify FNH 
topological bands for 54-fold generalized Bernard-LeClair (GBL)
symmetry classes and arbitrary spatial dimensions using $K$-theory.
 The classification distinguishes two different scenarios of  
the Floquet operator's spectrum gaps [dubbed as Floquet operator (FO) angle-gapped and 
FO angle-gapless]. The results culminate into two periodic tables, 
each containing 54-fold GBL symmetry
classes. Our scheme reveals FNH topological phases without
 any static/Floquet Hermitian and static non-Hermitian counterparts. And our results naturally produce the periodic tables of Floquet Hermitian 
topological insulators and Floquet unitaries. 
The framework can also be applied to characterize the topological phases of bosonic systems. 
We provide concrete examples of one and two-dimensional
fermionic/bosonic systems. And we elucidate 
the meaning of the topological invariants and their physical consequences. Our paper lays the foundation
 for a comprehensive exploration of FNH topological bands. And it 
 opens a broad avenue toward uncovering unique 
 phenomena and functionalities emerging from the synthesis of periodic driving, non-Hermiticity, 
 and band topology.
\end{abstract}
\maketitle
\section{Introduction}
Over the past decades,  topological phase of matter \cite{review1,review2,BernevigHughs,AsbothOraszlanyPalyi,shunqingshen} has become one of the major research fields in the interdisciplinary areas of condensed matter physics, photonics, cold-atom physics, electrical circuits, acoustics, etc. In topological matter, symmetry plays a central role: the topological matter can be categorized into distinct classes based on its underlying symmetries, and the appearance of gapless surface states is protected against symmetry-preserving perturbations. Quite recently, the topological phases well-studied in isolated quantum systems described by Hermitian Hamiltonians have been extended to the non-Hermitian regime \cite{coll1,coll2,coll3,coll4,coll4,coll5,coll6,colladd1,colladd2,colladd3}, partially fueled by the explosive research advancements in a 
diverse set of, e.g. atomic, molecular and optical platforms \cite{amo1,amo2,amo3,amo4,amo5,amo6,amo7}. Non-Hermitian Hamiltonian emerges as an effective description of a variety of quantum and classical systems, ranging from condensed matter materials with finite-lifetime quasiparticles \cite{finite1,finite2,finite3,finite4,finite5,finite6,finite7,finite8,BorgniaKruchkovSlager}, bosonic particles governed by Bogoliubov-de-Gennes (BdG) equations \cite{Yang,Lieu,YokomizoMurakami,XFACVO,ZapataSols,McDonaldPeregBarneaClerk,WuNiuA,WuNiuNJP,Barnett,GaliloLeeBarnett,Derezinski,GohbergLancasterRodman,KawaguchiUeda,BlonderTinkhamKlapwijk,KatsuraNagaosaLee,KondoAkagiKatsura},  open quantum systems dictated by quantum master equation \cite{Lindblad,LiuZhangYangChen,SongYaoWang,DiracDampingLiuChen}, 
to photonic setups with gain and loss \cite{op1,op2,op3,op4,op5,op6,xuepengep,xuepengep2,op8,op9,op10,op11,op12,op13,coll4,nhwsmhhp,James}. Compared to the Hermitian case, non-Hermitian Hamiltonians generally have complex eigenenergies, giving rise to a plethora of intriguing phenomena without any Hermitian analogues \cite{nhwsmhhp,etopo1,etopo2,etopo3,pointtopo4,etopoadd1,etopoadd2,etopoadd3,epknot,ne1,ne2,ne3,ne4,ne5,ne6,ne7,ne8,ne9,ne10,nhsee1,nhsee2,nhsee3,nhsee4,neadd1}.

Floquet engineering is the control of a system through the periodic drive,
 which has been widely utilized in photonic systems and ultracold 
 atoms \cite{fti1,bukovreview,fphotoexp1,fphotoexp2,fphotoexp3,fphotoexp4,fcoldexp1,fcoldexp2} and provides a powerful tool for tailoring 
 topologically nontrivial band structures. In Floquet systems, the Hamiltonian is periodic in time $H(t+\tau)=H(t)$, with $\tau$ the driving period \cite{Floquet,jiangliang,lindner,kitagawa}, and $\omega=2\pi/\tau$ the driving frequency. The topological phases in static Hermitian systems have also been generalized to the Floquet Hermitian systems \cite{ThoulessB,fclass1,fclass2,fclass3}. Stroboscopically, the time evolution over one period is effectively described by the so-called Floquet Hamiltonian (FH), with its spectra called quasienergies. The quasienergies are only well-defined up to integer multiples of the driving frequency $\omega$. Due to such periodicity of the quasienergy, the topological properties of Floquet systems turn out to be much richer than static cases. Typical examples include the anomalous Floquet topological phases \cite{ano1,anoexp,fhotihu,fhoti1,fhoti2,fhoti3,fhoti4,fhoti5}, with the appearance of boundary states even when the bulk quasienergy bands are trivial. The anomalous Floquet phases are intrinsically dynamical without static counterparts, and their topological characterization requires a scrutinization of the time-evolution operator inside the whole driving period\cite{ano1}.

For a comprehensive understanding of the various topological phases and their associated unique features, a coherent topological classification according to basic symmetry classes is the key step. 
The topological classifications for either 
the static (time-independent) Hermitian or 
Floquet Hermitian systems have previously been obtained. 
For the static case, taking into account the three fundamental internal symmetries: 
time-reversal, particle-hole, and chiral symmetry, 
yields the famous Altland-Zirnbauer (AZ) tenfold way \cite{AltlandZirnbauer,Chiu,ChiuStone,ChiuYaoRyu,MorimotoFurusaki,Kitaev,Ludwig}. For 
example, 
 the Chern insulator and quantum spin Hall insulator belong respectively to class A and class AII. They are 
described by the $\mathbb{Z}$ and $\mathbb{Z}_2$ invariants, respectively. 
For the FH real gapped Floquet case, the dynamics of the system 
are dictated by its 
 unitary time-evolution operator $U({\bf k},t)$. From a homotopic point of view, the unitary operator $U({\bf k},t)$ can be decomposed into a unitary loop operator followed by a constant evolution of the effective Floquet Hamiltonian. A complete understanding of the bulk topology involves the analysis of the effective Floquet Hamiltonian and the classification of the 
loop unitary through $K$-theory. The results are listed in the Floquet AZ periodic table \cite{fclass1,fclass2}.

\begin{figure}[t]  
\centerline{\includegraphics[width=3.35in]{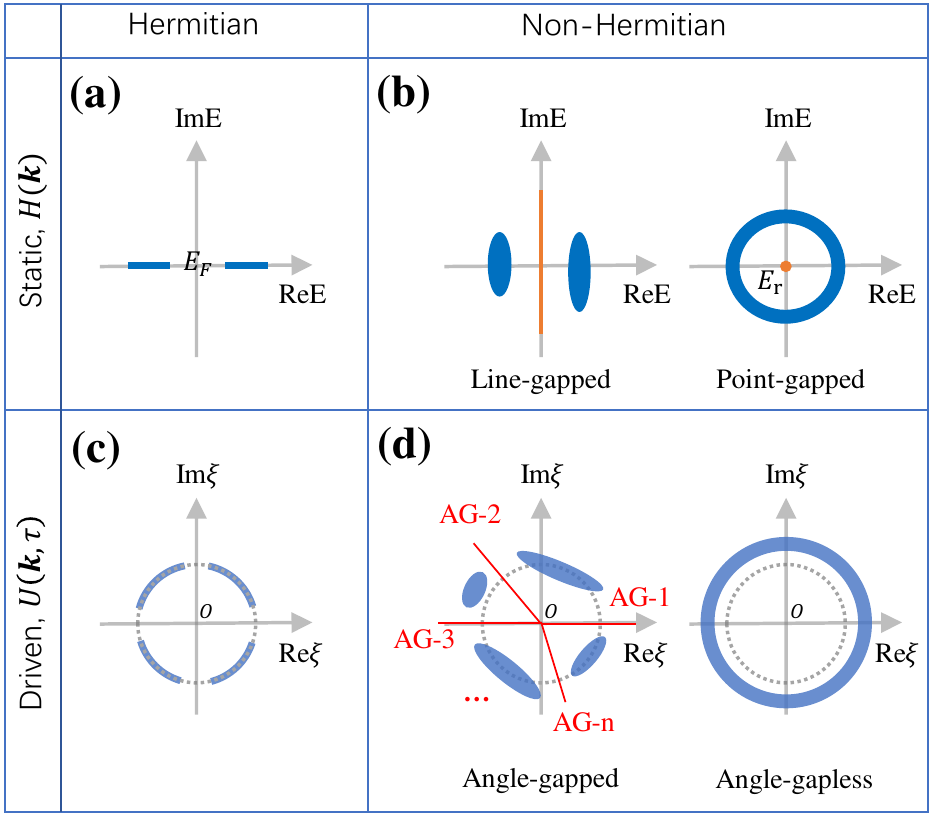}}
\caption{Schematics of the spectrum gaps for the four different types (static Hermitian/non-Hermitian and Floquet driven Hermitian/non-Hermitian) of systems. (a) The energy gap of 
a static Hermitian Hamiltonian. 
A topological transition happens through band touching at the Fermi energy $E_F$ accompanied by gap closing. (b) Energy gaps for a static non-Hermitian Hamiltonian. As the eigenenergies are complex, the band gaps can be line-like (left panel) or point-like (right panel). The line gap (orange line) separates the bands (blue) into two distinct regions. For the point-gap case, the complex bands (blue) do not touch a reference point $E_r$ (orange, here set to be at zero). (c) The spectrum gaps of the Floquet operator $U({\bf k},\tau)$
 for a Floquet driven Hermitian system. The spectra (blue) lie on the unit circle (dotted gray) as the Floquet operator is unitary. (d) The spectra (blue) of the Floquet operator for a Floquet driven non-Hermitian system. Left panel is the FO angle-gapped case. The spectra are split into several disjoint pieces around the origin by the radial lines (red), with the FO angle gaps denoted as AG-1, AG-2, ..., AG-n. Unlike the Floquet Hermitian system, the spectra of $U({\bf k},\tau)$ do not necessarily lie on the unit circle as the Floquet operator is non-unitary. Right panel is the FO angle-gapless case. The Floquet operator spectra encircle the origin without any FO angle gaps. \label{FA1}}
\end{figure}

The extension of the AZ tenfold way of the Hermitian Hamiltonians to the undriven non-Hermitian systems accomplished during the past several years is highly nontrivial. First, the complex eigenenergies of a generic non-Hermitian Hamiltonian 
bring more possibilities for the energy gap. Generally speaking, the non-Hermitian Hamiltonians feature 
either point-like or line-like gaps, as sketched in Fig. \ref{FA1} or separable bands without any band singularities \cite{fuliang}. Second, the non-Hermiticity ramifies the celebrated AZ symmetry classes to the Bernard-LeClair (BL) symmetry classes \cite{BernardLeClair,BLCclass2} due to the nonequivalence between complex conjugation and transposition for non-Hermitian operators. For separable band structures without any symmetries, a purely homotopical classification \cite{class1,class2,knotPRL} by taking into account the band braidings is carried out. 
 Further refining to spectra to possess point-like or line-line gaps, the 38 fold BL classifications and 54 fold generalized BL (GBL) classifications have been obtained, respectively \cite{GAKTHU,KawabataShiozakiUedaSato,LeeZhou,LiuChen,LiuJiangChen,colladd3}. The last quarter of the whole 
classification map--- the Floquet non-Hermitian (FNH) system, is not yet touched upon. The theoretical explorations of the richness of FNH topological phases are still in their early stages. A complete topological classification of FNH systems based on their underlying symmetries is not only of fundamental significance but also experimentally relevant. A periodically driven non-Hermitian system features
 non-unitary quantum dynamics. And it describes a variety of physical settings, e.g., photonic systems with gain and loss or nonreciprocal effects under Floquet driving \cite{fnh1,fnh2,fnh3,fnh4,fnh5,fnhano,fnhpumping,ZhouLW1,ZhouLW2,AnJH}, ultracold atoms with dissipation interacting with an external electromagnetic field \cite{drivendis1,finite6,drivendis3,drivendis4,drivendis5,drivendis6}, and non-unitary quantum walks \cite{nhsee3,op11,op12,xuepengep2,QuantumWorks5}.

In this paper, we develop a systematic topological classification of FNH systems according to the internal symmetries based on $K$-theory. 
We demonstrate that there exist 54-fold distinct GBL classes for 
time-dependent non-Hermitian systems. In contrast to the previous classification frameworks of static point gap non-Hermitian Hamiltonians \cite{GAKTHU,KawabataShiozakiUedaSato,LeeZhou,LiuChen,LiuJiangChen,colladd3}, a more natural spectrum gap is defined directly through the Floquet operator (FO) $U({\bf k},\tau)$. We distinguish two different scenarios of the spectra of $U({\bf k},\tau)$ on the
 complex plane and dub them FO angle-gapped and FO angle-gapless, respectively. As depicted in the left panel Fig. \ref{FA1}(d), the FO angle-gapped spectra manifest as the appearance of angle gaps in the spectra of Floquet operator. 
From a well-defined loop operator and a well-defined static Hamiltonian, we obtain the periodic table containing each symmetry class and its associated topological invariants, summarized in Table \ref{tab:tableI}. 
For the FO angle-gapless case, the spectra of $U({\bf k},\tau)$ enclose the origin, leaving no additional constraints on the Floquet Hamiltonian.
 The topological classification is to find all the nonequivalent homotopy classes of the Floquet operator, culminating in the second periodic table summarized in Table \ref{tab:tableII}. 
A part of our classification reproduces/consistents of Roy and Harper's periodic table of Floquet Hermitian topological insulators \cite{fclass1,fclass2,fclass3} and Higashikawa, Nakagawa, and Ueda's periodic table of Floquet unitaries \cite{HigashikawaNakagawaUeda}. Our frameworks can be directly applied to characterize the Floquet dynamics of bosonic systems. We consider concrete examples in one and 
 two-dimensional fermionic/bosonic systems and elucidate the meaning of the topological invariants and their physical consequences.

The remainder of this paper is organized as follows. Sec. \ref{secii} illustrates the classification scheme of generic time-dependent non-Hermitian systems, which contains three subsections. Sec. \ref{seciia} explores all the symmetry classes
 of time-dependent non-Hermitian systems. Consider 
 time-reversal symmetry, chiral symmetry, and particle-hole symmetry as primary symmetries, the 
 combinations of these primary symmetries produce all the 54 non-Hermitian GBL classes. Sec. \ref{seciib} deals with the gap conditions of the non-unitary Floquet operator, whose spectra on the complex plane can be either FO angle-gapped or FO angle-gapless. Sec. \ref{seciic} introduces the composition of the time-evolution
  operator and the concepts of homotopy and homeomorphic. These concepts are widely used in identifying topologically equivalent operators or spaces for later discussions. In Sec. \ref{seciii}, we provide a complete classification of FNH systems for all the 54 GBL classes based on 
 $K$-theory. We explicitly work out the extension problem of Clifford algebra in each symmetry class.
  Then we obtain periodic Table \ref{tab:tableI} and Table \ref{tab:tableII}  corresponding 
  to the FO angle-gapped and FO angle-gapless cases, respectively.
  In Sec. \ref{SecFermion}, we illustrate the FNH topology in fermionic systems through two simple examples, corresponding to the FO angle-gapped and FO angle-gapped cases, 
 respectively. And we explain the physical meanings of the topological invariants. Sec. \ref{secv} applies the classification scheme developed in Sec. \ref{seciii} and Sec. \ref{SecFermion} to the bosonic systems. Instead of the Hamiltonian itself, the dynamics of bosonic systems are governed by the M matrix. We conclude in Sec. \ref{secsum} and leave the technical details of derivations/calculations in the Appendices.

\section{Non-unitary time-evolution and symmetry classes} \label{secii}
In this section, we provide the classification scheme for generic FNH systems. The classification involves two basic ingredients, i.e., to find out all the possible symmetry classes and to identify the spectrum gaps of the Floquet operator. We start by considering a time-dependent non-Hermitian system with Hamiltonian $H(t)$. In general, $H(t)\neq H^{\dag}(t)$. The dynamics of the system is governed by 
the Schr{\"o}dinger equation:
\begin{eqnarray}
i \frac{d}{dt} |\Psi(t)\rangle= H(t)|\Psi(t)\rangle, \label{nonHerEvolu}
\end{eqnarray}
with $|\Psi(t)\rangle$ the time-dependent wave function. Suppose the initial state at $t=t_i$ is $|\Psi(t_i)\rangle$ and the time-evolved state at $t=t_f$($t_f>t_i$) is $|\Psi(t_f)\rangle$. The time-evolution can be formally represented as $|\Psi (t_f)  \rangle= U(t_f,t_i) |\Psi (t_i) \rangle$, where
\begin{eqnarray}
U(t_f,t_i):=\mathcal{T}\exp[-i\int_{t_i}^{t_f}dt~H(t)]
\end{eqnarray}
is the time-evolution operator. Here $\mathcal{T}$ takes the time-ordering product. For $t_b>t_a$, we define $U(t_a,t_b):=U^{-1}(t_b,t_a)$. We denote $U(t):=U(t,0)$ for brevity. As the Hamiltonian is non-Hermitian, the time-evolution operator $U(t)$ is non-unitary.

\subsection{Symmetry classes}\label{seciia}
In this subsection, we build all the internal (non-spatial) symmetry classes of the time-dependent non-Hermitian systems. We begin with the primary symmetries that relate the dynamics at time $t$ and $-t$:
\begin{align}
U_{T_1}H^*(-t)U_{T_1}^{-1}=H(t)&, ~~U_{T_1}U_{T_1}^*=\pm \mathbb{I}\label{eq:syms1}, \\
U_{T_2}H(-t) U_{T_2}^{-1}=-H(t)&, ~~U_{T_2}^2=\mathbb{I} \label{eq:syms2},
\end{align}
For $t=0$, Eq. (\ref{eq:syms1}) and Eq. (\ref{eq:syms2}) reduce to the time-reversal symmetry and chiral symmetry of Hermitian Hamiltonian, respectively. Here in the time-dependent settings, we dub them time-reversal symmetry (TRS) and chiral symmetry (CS), respectively. The TRS and CS keep the 
Schr{\"o}dinger equation Eq. (\ref{nonHerEvolu}) unchanged under the transformation $t\rightarrow -t$. 
Besides the TRS and CS, we need to consider the particle-hole symmetry (PHS) as another primary symmetry. 
The PHS does not flip time and is 
 preserved if the system has superconducting pairing or is described by a BdG-type Hamiltonian. The symmetry reads:
\begin{eqnarray}
U_{P}H^{\bf T}(t)U_{P}^{-1}=-H(t)&, ~~U_{P}U_{P}^*=\pm \mathbb{I}.\label{eq:syms3}
\end{eqnarray}
Starting from the above three primary symmetries, we can generate all the possible symmetry classes. For example, the combination of TRS and  CS produces a $K$-type symmetry $H(t)=-kH^*(t)k^{-1}$; the combination of CS and PHS produces a $C$-type symmetry $H(t)=cH^{\bf T}(-t)c^{-1}$ in the GBL class. All the possible symmetries 
generated from TRS, CS, and PHS are listed below:
\begin{align}
H(t)=\epsilon_kkH^*(-\epsilon_k t)k^{-1}&, ~~kk^*=\eta_k \mathbb{I}, &K \textrm{ sym.} \label{eq:symshpr}\\
H(t)=\epsilon_qqH^\dagger(\epsilon_q t) q^{-1}&, ~~q^2=\mathbb{I}, &Q \textrm{ sym.}\label{eq:symshqr}\\
H(t)=\epsilon_c cH^{\bf T}(-\epsilon_c t)c^{-1}&, ~~cc^*=\eta_c \mathbb{I},&C \textrm{ sym.}\label{eq:symshcr}\\
-H(t)=pH(-t)p^{-1}&, ~~p^2=\mathbb{I}, &P \textrm{ sym.}\label{eq:symshkr}
\end{align}
Fourier transforms Eqs. (\ref{eq:symshpr})-(\ref{eq:symshkr}) into momentum space, we get that,
\begin{align}
H({\bf k},t)=\epsilon_kkH^*(-{\bf k},-\epsilon_k t)k^{-1}&, ~~kk^*=\eta_k \mathbb{I}, &K \textrm{ sym.} \label{eq:symshp}\\
H({\bf k},t)=\epsilon_qqH^\dagger({\bf k},\epsilon_q t) q^{-1}&, ~~q^2=\mathbb{I}, &Q \textrm{ sym.}\label{eq:symshq}\\
H({\bf k},t)=\epsilon_c cH^{\bf T}(-{\bf k},-\epsilon_c t)c^{-1}&, ~~cc^*=\eta_c \mathbb{I},&C \textrm{ sym.}\label{eq:symshc}\\
-H({\bf k},t)=pH({\bf k},-t)p^{-1}&, ~~p^2=\mathbb{I}, &P \textrm{ sym.}\label{eq:symshk}
\end{align}
where $\eta_k,\eta_c,\epsilon_k,\epsilon_q,\epsilon_c=\pm1$. $k, q, c, k$ are four unitary matrices, satisfying
\begin{equation}
c=\epsilon_{pc}pcp^{\bf T}, ~k=\epsilon_{pk}pkp^{\bf T},~c=\epsilon_{qc}qcq^{\bf T},~p=\epsilon_{pq}qpq^\dagger, \label{cr}
\end{equation}
with $\epsilon_{pc}, \epsilon_{pk}, \epsilon_{qc}, \epsilon_{pq}=\pm 1$. By a full counting of the four types of symmetries $P,~Q,~C,~K$, the signs of time flipping $\epsilon_k,~\epsilon_q,~\epsilon_c$, the signs of symmetry-operator involution $\eta_k,~\eta_c$, and the signs in the commutation relations $\epsilon_{pc}, \epsilon_{pk}, \epsilon_{qc}, \epsilon_{pq}$, 
we obtain 54-fold nonequivalent symmetry classes (Details in Appendix \ref{ap0}). Each class is specified by a definite 
choice of the symmetries and signs. To avoid confusion, we stress that these symmetries are for 
time-dependent Hamiltonians. 
Yet we still utilize the convention for static line gap non-Hermitian Hamiltonians 
and call the 54-fold classes the GBL classes \cite{LiuChen}. They are labeled as \cite{LiuChen} 
Non, P, Qa-b, K1-2a-b, C1-4, PQ1-2, PK1-2, PK3a-b, PC1-4, QC1-8a-b, PQC1-8, PQC9-12a-b in 
Table \ref{tab:tableI} and Table \ref{tab:tableII}.

The 54-fold GBL classes were first constructed in Ref. \cite{LiuChen} to 
get a consistent description of line gap static non-Hermitian systems. And in the Appendix 
E of Ref. \cite{colladd3}, there is a detailed review of the 54-fold GBL classes. 
However, Ref. \cite{colladd3} still call it Bernard-LeClair class. The 
derivation of 38-fold Bernard-LeClair (BL) classes needs to use the $H\rightarrow iH$ 
as an equivalent transformation, and the transformation needs to relate two symmetry classes for some cases. 

For static point gap non-Hermitian topology, we set the point gap at $0$. 
$H\rightarrow iH$ transformation didn't change the gap, and we can 
regard it as an equivalent transformation. And 
$H\rightarrow iH$ transformation relate two static GBL symmetries, for example 
$H=kH^*k^{-1}$ ($kk^*=\mathbb{I}$) and $H=-kH^*k^{-1}$ ($kk^*=\mathbb{I}$). 
Thus, part of 54-fold GBL classes can be regarded as equivalent if we only
consider point gap topology. By subtracting the redundant equivalence classes, 
we get 38-fold BL classes. However, static point gap non-Hermitian topology also
can be described by 54-fold GBL classes. 
It is due to the redundant equivalence classes didn't lead to any inconsistent 
conclusion.

We consider static line gap non-Hermitian topology, 
$H\rightarrow iH$ transformation changes the gap, and we can't regard it as an equivalent transformation. 
Thus, static line gap non-Hermitian topology is described by 54-fold GBL classes. 
If we use the 38-fold BL classes to describe the static line gap non-Hermitian topology,
 it leads to inconsistent conclusions. For example, a non-Hermitian system with $H=kH^*k^{-1}$
 ($kk^*=\mathbb{I}$) symmetry and a non-Hermitian system with $H=-kH^*k^{-1}$ ($kk^*=\mathbb{I}$) symmetry 
 belong to the same class in the 38-fold BL classes description (BL class). 
 For a 1-dimensional real line gap system with $H=kH^*k^{-1}$ ($kk^*=\mathbb{I}$) symmetry, 
 the topological classification is $0$ \cite{KawabataShiozakiUedaSato,LiuChen}. 
 For a 1-dimensional real line gap system with $H=-kH^*k^{-1}$ ($kk^*=\mathbb{I}$) symmetry, 
 the topological classification is $\mathbb{Z}_2$ \cite{KawabataShiozakiUedaSato,LiuChen}.
 These conclusions contradict the uniqueness of the topological 
 classification in a certain dimension and symmetry class.

For time-dependent non-Hermitian systems, 
two different GBL symmetry classes didn't relate to each other by the
$H\rightarrow iH$ transformation. For example, 
 type-K symmetry takes the form of Eq. (\ref{eq:symshpr}) in the time-dependent system. 
 The $H\rightarrow iH$ transformation didn't transform type-K symmetry with $\epsilon_k=1$ and $\eta_k=1$
 into type-K symmetry with $\epsilon_k=-1$ and $\eta_k=1$. However, the $H\rightarrow iH$ transformation did
  transform type-K symmetry with $\epsilon_k=1$ and $\eta_k=1$
 into type-K symmetry with $\epsilon_k=-1$ and $\eta_k=1$ in static limit (hint: let $t=0$).

 54 is the total number of group structures generated by Eqs. (\ref{eq:symshpr})-(\ref{eq:symshkr}).
 Thus, there is no inconsistent conclusion if we use 54-fold GBL classes to 
 describe the general non-Hermitian system. If we use 38-fold BL classes to 
 describe non-Hermitian systems, sometimes,  it leads to inconsistent conclusions.

The four types of symmetries of Eqs. (\ref{eq:symshp})-(\ref{eq:symshk}) on $H({\bf k},t)$ induce the symmetries on the time-evolution operator $U({\bf k},t)$ as follows:
\begin{align}
U^*(-{\bf k},-t)=k^{-1}U({\bf k},\epsilon_k t)k&, ~~kk^*=\eta_k \mathbb{I}, &K \textrm{ sym.} \label{eq:symsup} \\
[U^{\dagger}({\bf k},t)]^{-1}=q^{-1}U({\bf k},\epsilon_q t) q&, ~~q^2=\mathbb{I}, &Q \textrm{ sym.}\label{eq:symsuq}\\
[U^{\bf T}(-{\bf k},t)]^{-1}=c^{-1}U({\bf k},-\epsilon_c t)c&, ~~cc^*=\eta_c \mathbb{I},&C \textrm{ sym.}\label{eq:symsuc}\\
U({\bf k},-t)=p^{-1}U({\bf k},t)p&, ~~p^2=\mathbb{I}. &P \textrm{ sym.}\label{eq:symsuk}
\end{align}
The derivation of  Eqs. (\ref{eq:symsup})-(\ref{eq:symsuk}) is given in Appendix \ref{apa}. 
The above discussions in this section are for generic time-dependent systems. Now we restrict to the Floquet system with a time-periodic Hamiltonian
\begin{equation}
H({\bf k},t+\tau)=H({\bf k},t),
\end{equation}
where $\tau$ is the driving period. The Floquet operator is defined as the time-evolution operator over one period. From now on, 
we concisely denote the Floquet operator as 
$U({\bf k}):=U({\bf k},\tau)$. Starting from Eqs. (\ref{eq:symsup})-(\ref{eq:symsuk}), we can obtain the symmetry operations on the Floquet operator:
\begin{align}
[U^*(-{\bf k})]^{-\epsilon_k}=k^{-1}U({\bf k})k&, ~~kk^*=\eta_k \mathbb{I}, &K \textrm{ sym.} \label{eq:symsutp} \\
[U^{\dagger}({\bf k})]^{-\epsilon_q}=q^{-1}U({\bf k}) q&, ~~q^2=\mathbb{I}, &Q \textrm{ sym.}\label{eq:symsutq}\\
[U^T(-{\bf k})]^{\epsilon_c}=c^{-1}U({\bf k})c&, ~~cc^*=\eta_c \mathbb{I},&C \textrm{ sym.}\label{eq:symsutc}\\
[U({\bf k})]^{-1}=p^{-1}U({\bf k})p&, ~~p^2=\mathbb{I}. &P \textrm{ sym.}\label{eq:symsutk}
\end{align}
The derivation of  Eqs. (\ref{eq:symsutp})-(\ref{eq:symsutk}) is given in Appendix \ref{apb}.

\subsection{Gap condition of the Floquet operator}\label{seciib}
The gap condition is an essential ingredient in the classification theory. A spectrum gap means 
a region without any spectrum. Two Hamiltonian operators 
(or time-evolution operators in the driven case) are equivalent if they can continuously 
transform into each other while keeping the gap open and preserving their corresponding symmetries. 
Topological transition happens accompanied by gap closings. Here we compare the spectrum gaps for the 
four different cases: static Hermitian/non-Hermitian Hamiltonian and 
Floquet Hermitian/non-Hermitian Hamiltonian. For a static Hermitian Hamiltonian, all eigenvalues are real, and the 
spectrum gap is defined on the real-energy axis when the energy bands do not touch the Fermi energy $E_F$, as depicted in Fig. \ref{FA1}(a). For a static non-Hermitian Hamiltonian, its eigenvalues are complex. 
 As shown in Fig. \ref{FA1}(b), the spectrum gap can be either a line-like region or a point-like region on the complex-energy plane \cite{LiuChen,KawabataShiozakiUedaSato,LeeZhou}. 
Correspondingly, the spectrum of the non-Hermitian Hamiltonian possesses a line-like gap or point-like gap.

For Floquet Hermitian/non-Hermitian systems, 
we consider the spectra of the Floquet operator (FO) $U({\bf k})$. We denote 
the spectra of FO as $\xi_n({\bf k})$, and $n$ is the band index. 
A {\it FO angle gap at $\theta$ ($\theta\in R$)} is defined as $\forall \rho>0, \forall n \quad s. t. \quad \xi_n({\bf k})\ne \rho e^{-i\theta}$.
If a Floquet Hermitian/non-Hermitian system has a FO angle gap, 
we call such system {\it FO angle-gapped}. If a Floquet Hermitian/non-Hermitian system doesn't have any FO angle gap, 
we call such system {\it FO angle-gapless}. 

We can define a widely used concept---the Floquet Hamiltonian (FH) as:
\begin{eqnarray}\label{FH}
 H_F:=\frac{i}{\tau}\ln(H({\bf k})).
  \end{eqnarray}
   The Floquet Hamiltonian's definition Eq. (\ref{FH}) is the same as previous  
  articles.  We denote 
  the spectra of FH as $\varepsilon_n({\bf k})$. 
  $\varepsilon_n({\bf k})$ usually be called quasienergies. The quasienergies 
  are only well-defined up to integer multiples of the driving frequency $\omega=2\pi/\tau$ since $\ln()$ 
  is a multivalued function. 
  A {\it FH real gap at $E_0$ ($E_0\in R$)} defined as: $\forall j\in \mathbb{Z}, \forall n \quad s. t. \quad \text{Re}(\varepsilon_n({\bf k}))\ne E_0+2\pi j/\tau$. Here, $\text{Re}(\varepsilon_n({\bf k}))$ takes the real part of $\varepsilon_n({\bf k})$.

   It is worth stressing that: A Floquet system has a FO angle gap at $\theta$ ($\theta\in R$) is equivalent to 
   the Floquet system has an FH real gap at $\theta/\tau$. 
   For the Floquet Hermitian system, $U({\bf k})$ is unitary, giving rise to real 
   quasienergies. The spectra of $U({\bf k})$ lie on the unit circle. Fig. \ref{FA1}(c) is the 
   schematic diagram of the spectrum of a Floquet Hermitian system, and it is FO angle-gapped. We can 
   directly extend this scenario to the driven non-Hermitian systems. Note that the Floquet operator 
   $U({\bf k})$ is non-unitary in FNH systems, and its spectra do not necessarily lie on the unit circle. 
   Fig. \ref{FA1}(d) is the schematic diagram of the spectrum of an FNH system, 
   Fig. \ref{FA1}(d) (left panel) is FO angle-gapped, and Fig. \ref{FA1}(d) (right panel) is FO angle-gapless.

\subsection{Composition of evolution operators, homotopy and homeomorphic}\label{seciic}
For the classification problem, a widely used concept is the composition of
 two time-evolution operators \cite{fclass2}. Given that $U_1$ is the time-evolution
  operator generated by Hamiltonian $H_1(t)$ and $U_2$ is the time-evolution operator generated by 
  Hamiltonian $H_2(t)$, we define the composition of $U_1$ and $U_2$ as $U_1*U_2$, which is the 
  time-evolution generated by the following Hamiltonian:
\begin{eqnarray}
H(t)&=&\left\{\renewcommand\arraystretch{1.2}
\begin{array}{ccc}
H_2({\bf k},2t) && 0\leq t\leq \tau/4;\\
H_1({\bf k},2t-\tau/2) && \tau/4< t< 3\tau/4;\\
H_2({\bf k},2t-\tau) && 3\tau/4\leq t\leq \tau.
\end{array}\right.\label{eq:composition_TRS}
\end{eqnarray}
The above operator composition is consistent with all the GBL symmetries. In fact, we have:
\begin{lemma}
  If $H_1(t)$ and $H_2(t)$ belong to the same GBL class defined in Eqs. (\ref{eq:symshp})-(\ref{eq:symshk}) and the first two columns of Table \ref{tab:tableI}, then the composed Hamiltonian $H(t)$ belongs to the same GBL class of  $H_1(t)$ and $H_2(t)$.
\end{lemma}

The proof of Lemma 1 is provided in Appendix \ref{apc}. Besides the operator composition, another two widely used mathematical concepts in the topological classification of Floquet systems are homotopy and 
homeomorphic \cite{fclass2}. They are defined below: (1) {\it Homotopy:} Suppose that the two operators $g$ and $h$ satisfy the 
same symmetry condition and constraint, and the operator $g$ is homotopic to $h$ if and only if there exists a continuous operator function $f_t$ ($t\in [0,1]$) with $f_0=g$ and $f_1=h$. And $f_t$ satisfies the same symmetry condition
 and constraint as $g$ or $h$. (2) {\it Homeomorphic:} Space $A$ is homeomorphic to space $B$, if and only if there exists a continuous function $F: A\rightarrow B$. $F$ is a one-to-one, onto function and has a continuous inverse. The mapping $F$ preserves all the topological properties of a given space. Two homeomorphic spaces are the same from a topological point of view.

\section{Topological classification of Floquet non-Hermitian systems} \label{seciii}
This section is devoted to a complete classification of the FNH band topology, which contains both the FO angle-gapped and FO angle-gapless cases. 
We have demonstrated that for the FO angle-gapped case,
 there exists FH real gap. 
Thus, we also call the FO angle-gapped topological 
classification as FH real gapped topological classification. 
For the FO angle-gapped case, two time-evolution operators are considered topologically equivalent 
if they can be continuously transformed to each other while preserving the FO angle gaps and corresponding 
symmetries. It turns out the band topology of a FO angle-gapped system is not fully 
encoded in the Floquet Hamiltonian itself \cite{fclass1,fclass2,ano1,anoexp,fhotihu,fhoti1,fhoti2,fhoti3,fhoti4,fhoti5} 
due to the periodicity of the quasienergy zone. In fact, a FO angle gap naturally defines the branch cut 
of the Floquet Hamiltonian. 
To complete the topological classification, both the branch-cut-dependent Floquet Hamiltonian
 and its associated loop operator should be taken into account. While for 
the FO angle-gapless system, there is no additional spectrum restriction on the 
quasienergies. And the Floquet Hamiltonian is not always continuous in 
 the Brillouin zone (BZ) for any chosen branch cut in 
the absence of the FO angle gap. For the FO angle-gapless systems, the band topology is 
only extracted from the Floquet operator \cite{kitagawa,HigashikawaNakagawaUeda,UTtopology3}.
Thus, we also call the FO angle-gapless topological 
classification as the Floquet operator's topological classification.

To proceed, we expand the Floquet operator $U({\bf k})$ according to its eigenenergy spectra:
\begin{equation}
U({\bf k})=\sum_n \xi_n({\bf k})|\psi_{n,R}\rangle \langle \psi_{n,L}|.
 \end{equation}
Here $n$ is the band index. As $U({\bf k})$ is non-unitary, the expansion involves both the left and right eigenvectors \cite{biqm}, which are defined as $U({\bf k})|\psi_{n,R}\rangle=\xi_n({\bf k})|\psi_{n,R}\rangle$, and $U^{\dag}({\bf k})|\psi_{n,L}\rangle=\xi^{*}_n({\bf k})|\psi_{n,L}\rangle$. 
We further define the logarithm function $\ln_{\alpha}$ as:
\begin{equation}\label{loga}
e^{\ln_{\alpha}(z)}:=z,~~\textrm{and}~~\alpha<\text{Im}[\ln_{\alpha}(z)]< \alpha+2\pi.
\end{equation}
Here, $\text{Im}(z)$ takes the imaginary part of $z$. 
We have set the branch cut of the logarithm $\ln_{\alpha}$  at $\alpha$, and it is 
a single-valued function. For example, let us suppose 
$\phi_1,\phi_2\in R$ ($R$ is real number field)
 and $\alpha<\phi_1< \alpha+2\pi$. According to our definition, 
$\ln_{\alpha}[e^{i(\phi_1+2\pi N_1)+\phi_2}]=i\phi_1+\phi_2$, for any $N_1\in \mathbb{Z}$. Using the above 
logarithm function with a definite branch cut $\alpha$, 
we define the effective Floquet Hamiltonian at a 
branch cut $\theta$ as follows:
\begin{equation}\label{HFa}
H_{F,\theta}:= \frac{i}{\tau}\ln_{-\theta}[U({\bf k})]=\sum_n \frac{i}{\tau}\ln_{-\theta}(\xi_n)|\psi_{n,R}\rangle \langle \psi_{n,L}|.
\end{equation}
The Floquet Hamiltonian defined above explicitly depends on the branch cut $\theta$. A suitable choice of 
the branch cut is important when we consider the topological equivalence in the following discussions.
According to Eqs. (\ref{loga}) and (\ref{HFa}), the definition of $H_{F,\theta}$ requires 
a FO angle gap at $\theta$.  And $H_{F,\theta}$ is a continuous function in BZ. 
With a bit of abuse of notation, 
we still call the spectrum of $H_{F,\theta}$ as the quasienergy spectrum. From 
Eqs. (\ref{loga}) and (\ref{HFa}), it is easy to see the real part of the quasienergy spectrum of 
$H_{F,\theta}$ lies inside 
$((\theta-2\pi)/\tau,\theta/\tau)$.

Correspondingly, from $H_{F,\theta}$ and the time-evolution operator $U({\bf k},t)$ (Note that in general, $U({\bf k},t)\neq U({\bf k},t+\tau)$), we can define a time-periodic evolution operator
\begin{equation}
U_{l,\theta}({\bf k},t):=U({\bf k},t)*e^{iH_{F,\pi}({\bf k})t}.
\end{equation}
$U_{l,\theta}({\bf k},t)$ satisfies $U_{l,\theta}({\bf k},t+\tau)=U_{l,\theta}({\bf k},t)$ and is usually 
called as the \textit{loop operator} \cite{fclass1,fclass2,fclass3}. 
The definition of $H_{F,\theta}$ and $U_{l,\theta}({\bf k},t)$ both require 
a FO angle gap at $\theta$. And $U_{l,\theta}({\bf k},t)$ is a continuous function of (${\bf k},t$). 
For generic settings of the 
time evolution (It may not have a FO angle gap), we can't define $H_{F,\theta}$ which is a well-defined 
single-value continuous function in BZ.

It is worth stressing that: A Floquet system haven $n$ FO angle gaps at $\tilde{\theta}_1,\tilde{\theta}_2,...,\tilde{\theta}_n$ ($\tilde{\theta}_j\in [\tilde{\theta}_1-2\pi,\tilde{\theta}_1]$, $j=1,2,...,n$) is equivalent to 
   the Floquet system haven $n$ FH real gaps at $\tilde{\theta}_1/\tau,\tilde{\theta}_2/\tau,...,\tilde{\theta}_n/\tau$. It is also 
   is equivalent to the $H_{F,\tilde{\theta}_1}$ of this system haven $n-1$ real gaps at $\tilde{\theta}_2/\tau,\tilde{\theta}_3/\tau,...,\tilde{\theta}_n/\tau$. The
   $\tilde{\theta}_1$ is chosen as the branch cut and doesn't contribute to a $H_{F,\tilde{\theta}_1}$ real gap.

\subsection{FO angle-gapped case}\label{seciiia}
In this subsection, we consider the topological classification for the FO 
angle-gapped FNH systems. We first elaborate on the simplest case 
with only 
 a FO angle gap at $\theta_0=\pi$ in the Floquet operator spectra and explicitly 
work out the Clifford algebra's extension problem for all the symmetry classes. The discussions are then 
extended to the generic cases with more real gaps allowed by their underlying 
symmetries. The results are listed in the 
topological classification Table \ref{tab:tableI} for all the FO angle-gapped FNH topological phases.

\subsubsection{Only one FO angle gap at $\theta_0=\pi$}\label{seciiia1}
For the simple case, when the quasienergy spectra possess one 
 FO angle gap at $\theta_0=\pi$, we have
\begin{lemma}
The loop operator $U_{l,\pi}({\bf k},t)$ and time-evolution operator $U({\bf k},t)$ have the same symmetries. The Floquet Hamiltonian $H_{F,\pi}$ and the initial Hamiltonian $H({\bf k}, t=0)$ also have the same symmetries.
\end{lemma}

The proof of Lemma 2 is provided in Appendix \ref{apd} for mathematical rigorousness. 
Following Lemma 2, we can decompose the time-evolution operator $U({\bf k},t)$ into two separate parts, which is described below
\begin{theorem}
  The time-evolution operator $U({\bf k},t)$ with 
 FO angle gap at $\pi$ can be continuously 
  transformed to $U_f({\bf k},t)=L * C$, where $L$ is a loop operator satisfying $L(t)=L(t+\tau)$ and $C$ 
  is a constant evolution generated by a time-independent Hamiltonian. Here $L$ and $C$ are unique up to 
  homotopy. The loop operator can be chosen as $L=U_{l,\pi}({\bf k},t)$. Correspondingly, $C$ is the 
  constant evolution of Hamiltonian $H_{F,\pi}$. The continuous transformation preserves all the GBL symmetries.
\end{theorem}

Theorem 1 can be regarded as a non-unitary 
and GBL symmetries generalization of the Theorem III. 1 in Ref. \cite{fclass2}. The detailed proof of Theorem 1 is provided in Appendix \ref{ape}.  
Theorem 1 indicates that the topological classification for FNH systems with FO angle gap at 
$\pi$ reduces to the topological classification of $U_{l,\pi}({\bf k},t)$ and $H_{F,\pi}$.
The latter has been previously investigated for static non-Hermitian Hamiltonians 
\cite{LiuChen,KawabataShiozakiUedaSato,LeeZhou}. For the simplest case with only a single FO angle gap at $\theta_0=\pi$ (it is equivalent to 
only a single FH real gap at $E_0=\pi/\tau$),
the spectra of $H_{F,\pi}$ can be 
continuously contracted to a single point since there is no FH real gap that 
separates them. Therefore we only need to consider the
 topological classification of the loop operator $U_{l,\pi}$.

The topological classification of $U_{l,\pi}({\bf k},t)$ can be obtained through a ``Hermitianization" procedure. We consider the following Hermitian operator:
\begin{equation}
\tilde {H}({\bf k}, t)=\left[ \begin{array}{cc}
0 & U_{l,\pi}({\bf k}, t)\\
U_{l,\pi}({\bf k}, t)^\dagger & 0
\end{array}
\right ]. \label{H}
\end{equation}
The mapping from $U_{l,\pi}({\bf k}, t)$ to $\tilde {H}({\bf k}, t)$ is homeomorphic since $\det(U_{l,\pi}({\bf k}, t))\ne 0$ implies $\det(\tilde {H}({\bf k}, t))\ne 0$. The Hermitianization procedure greatly simplifies the classification problem, e.g., we can continuously transform $\tilde {H}({\bf k}, t)$ into a band-flattened Hamiltonian [i.e., $(\tilde {H}({\bf k},t))^2=\mathbb{I}$] without altering any symmetries \cite{Kitaev}. 
According to Lemma 2, 
$U_{l,\pi}({\bf k}, t)$ and $U({\bf k}, t)$ 
belong to the same GBL symmetry class. From the 
symmetries of the loop operator $U_{l,\pi}({\bf k}, t)$ as described in 
Eqs. (\ref{eq:symsup})-(\ref{eq:symsuk}), we obtain the symmetries:
\begin{align}
\tilde {H}(-{\bf k}, -\epsilon_kt)=\tilde{K}\tilde {H}({\bf k}, t)\tilde{K}^{-1}&, ~~kk^*=\eta_k \mathbb{I}, &K \textrm{ sym.} \label{eq:symshtp} \\
\tilde {H}({\bf k}, \epsilon_q t)=\tilde{Q}\tilde {H}({\bf k}, t) \tilde{Q}^{-1}&, ~~q^2=\mathbb{I}, &Q \textrm{ sym.}\label{eq:symshtq}\\
\tilde {H}(-{\bf k}, -\epsilon_c t)=\tilde{C}\tilde {H}({\bf k}, t)\tilde{C}^{-1}&, ~~cc^*=\eta_c \mathbb{I},&C \textrm{ sym.}\label{eq:symshtc}\\
\tilde {H}({\bf k}, - t)=\tilde{P}\tilde {H}({\bf k}, t)\tilde{P}^{-1}&, ~~p^2=\mathbb{I}, &P \textrm{ sym.}\label{eq:symshtk}
\end{align}
satisfied by the Hamiltonian $\tilde {H}({\bf k}, t)$. Here $\tilde{K}=\sigma_0 \otimes k\mathcal{K}$, $\tilde{Q}=\sigma_0\otimes q$, $\tilde{P}=\sigma_0 \otimes p$ and $\tilde{C}=\sigma_0\otimes c\mathcal{K}$. $\mathcal{K}$ is the complex conjugate. The derivation of  Eqs. (\ref{eq:symshtp})-(\ref{eq:symshtk}) is given in Appendix \ref{apf}. Besides, the Hamiltonian $\tilde {H}({\bf k},t)$ has an additional chiral symmetry $\Sigma=\sigma_z\otimes\mathbb{I}$, with
\begin{equation}
\Sigma \tilde {H}({\bf k})=-\tilde {H}({\bf k})\Sigma.
\end{equation}

We then utilize $K$-theory to deal with the topological classification of $\tilde {H}({\bf k}, t)$ for each of the 54 GBL symmetry classes. According to the standard classification scheme in terms of Clifford algebra \cite{ChiuYaoRyu}, we represent $\tilde {H}({\bf k}, t)$ as
\begin{equation}
\tilde {H}({\bf k}, t)=m\tilde{\gamma}_0+ k_1\tilde{\gamma}_1+...+k_d \tilde{\gamma}_d +t\tilde{\gamma}_t.
\end{equation}
Here, $\tilde{\gamma}_0,~\tilde{\gamma}_i~(i=1,...,d)$ and $\tilde{\gamma}_t$ are the basis of the Clifford algebra. They anticommute with each other and square to the identity. $m\tilde{\gamma}_0$ is the mass term. From the commutation relations of the symmetry operators and Hamiltonian, we can construct the Clifford algebra's extension for each symmetry class. The space of the mass term is obtained through its correspondence with Clifford algebra's extension. (See Table I and Table VI of Ref. \cite{LiuChen}). Once the space of the mass term is obtained, we finalize the topological classification by calculating its
 0th homotopy group. We illustrate the above procedure by an explicit example of the class Non in Table \ref{tab:tableI}. 
 The generators of this class are $\left\{\tilde{\gamma}_0,\tilde{\gamma}_1,...,\tilde{\gamma}_d,\tilde{\gamma}_t,\Sigma\right\}$. The Clifford algebra's extension of this class is $\left\{\tilde{\gamma}_1,...,\tilde{\gamma}_d,\tilde{\gamma}_t,\Sigma\right\} \rightarrow\left\{\tilde{\gamma}_0,\tilde{\gamma}_1,...,\tilde{\gamma}_d,\tilde{\gamma}_t,\Sigma\right\}=Cl_{d+2}\rightarrow Cl_{d+3}$. The space of the mass term follows as $C_{d+2}$, and the topological classification of the Non class is given by the homotopy group: 
 $\pi_0(C_{d+2})=\mathbb{Z} ~(0)$ for even (odd) $d$ ($d$ is the spatial dimension).  
 The classifying space is equal to the space of mass term at $d=0$. In a similar vein, we can construct the Clifford algebra's extension for all the other GBL classes, as summarized in Table \ref{tab:tableIII}.

\subsubsection{Complete classification for classes without P, Q ($\epsilon_q=-1$), C ($\epsilon_c=-1$) and K ($\epsilon_k=-1$) symmetry}\label{seciiia2}
In this part, we discuss the classification for classes without any of the $P$, $Q$ ($\epsilon_q=-1$), $C$ ($\epsilon_c=-1$), or $K$ ($\epsilon_k=-1$) symmetry. For these classes, 
the FO angle gap could be along any radial line emitted from the origin, with the spectra 
satisfying the full symmetry of the system. Let us consider a FO angle gap at $\theta_0$ and a constant 
Hamiltonian evolution $U_c({\bf k},t)=e^{-it\theta_c\mathbb{I}/\tau}$, with $\theta_c$ a real constant. The 
$U_c({\bf k},t)$ preserves the full symmetry of the system. The composition 
$\bar{U}({\bf k},t)=U({\bf k},t)*U_c({\bf k},t)$, which  is homeomorphic to $U({\bf k},t)$. $\bar{U}({\bf k},t)$ 
preserves the full symmetry of $U({\bf k},t)$ and has a gap located at $\tilde{\theta}_0=\theta_0+\theta_c$. In general, there may exist 
multiple FO angle gaps. For the Floquet spectra with $n$
 FO angle gaps, we can always continuously shift one of the FO angle gaps to $\pi$ by adjusting $\theta_c$ without altering any symmetry of
$\bar{U}({\bf k},t)$. Thus, we can focus on the case that 
$U({\bf k})$ has 
 a FO angle gap at $\pi$ and $n-1$ FO angle gaps at other angles. 
From Theorem 1, we conclude a complete
 topological classification of $U({\bf k},t)$ is equivalent to the topological classification of $H_{F,\pi}$ with $n-1$ 
 real 
 gaps in  $(-\pi/\tau,\pi/\tau)$ \cite{KawabataShiozakiUedaSato} plus the 
 topological classification of loop operator $U_{l,\pi}$, as demonstrated in Sec. \ref{seciiia1}.

\subsubsection{Complete classification for classes with P or Q ($\epsilon_q=-1$) or C ($\epsilon_c=-1$) or K ($\epsilon_k=-1$) symmetry}\label{seciiia3}
In this part, we demonstrate the complete classification for classes when any of the $P$ or $Q$ 
($\epsilon_q=-1$) or $C$ ($\epsilon_c=-1$) or $K$ ($\epsilon_k=-1$) symmetry exists. For these classes, 
the FO angle gaps are pinned at $\theta_0=0$ or 
$\pi$ or a pair of FO angle gaps located at ($\theta_i$, $-\theta_i$) to make the 
spectra have the full symmetry of the system \cite{KawabataShiozakiUedaSato}. In the following, we 
elaborate on the different situations of the FO angle gap configurations.

{\bf S1:} There is only one 
FO angle gap at $\pi$ or $0$. The topological classification for the former is given 
in Section \ref{seciiia1}. For the latter, we can shift the 
FO angle gap at $0$ to $\pi$ through a 
homeomorphic mapping. As a representative example, we consider the class $P$ with $p=\sigma_z$. The 
constant evolution $U_0({\bf k},t)=e^{-i\pi\sigma_x t/\tau}$ fulfills the $P$ symmetry. The 
composition $\bar{U}({\bf k},t)=U_0({\bf k},t)*U({\bf k},t)$ is homeomorphic to $U({\bf k},t)$ and 
preserves the symmetry of $U({\bf k},t)$, 
while its associated Floquet spectra possess a FO angle gap at $\pi$. It follows 
that the system with only one FO angle gap at $0$ has the same topological classification as that with only 
one FO angle gap at $\pi$.
 Similarly, we can shift the 
 FO angle gap from $0$ to $\pi$ for all
 the other GBL classes.

{\bf S2:} There are two 
FO angle gaps at $0$ and $\pi$. According to Theorem 1, the topological 
classification is reduced to the topological classification of $H_{F,\pi}$ 
with an real gap at $0$ 
\cite{KawabataShiozakiUedaSato} plus the topological classification of 
 $U_{l,\pi}$, which is demonstrated in 
Sec. \ref{seciiia1}.

{\bf S3:} There are one FO angle gap at $\pi$ or $0$ and $n_q$ pairs of FO angle gaps at 
($\theta_m$, $-\theta_m$) ($m=1,2,...,n_q$). Here $0< \theta_1<\theta_2<...<\theta_{n_q}< \pi$. 
Thus, there are one FH real gap at $\pi/\tau$ or $0$ and $n_q$ pairs of FH real gaps at 
$(\theta_m/\tau, -\theta_m/\tau)$ ($m=1,2,...,n_q$). 
For the former case, when one of the FO angle gaps is located at $\pi$ and $n_q$ pairs of FO angle gaps at 
($\theta_m$, $-\theta_m$) ($m=1,2,...,n_q$), 
$H_{F,\pi}$ possesses $n_q$ pairs of real gap at ($\theta_m/\tau$, $-\theta_m/\tau$), $\pi$ is chosen as the branch cut and doesn't contribute to a $H_{F,\pi}$ real gap. $H_{F,\pi}$ is formally written as $H_{F,\pi}=\sum_j\mathcal{E}_j|\psi_j^R\rangle 
\langle \psi_j^L|$. According to Theorem 1, the
topological classification is reduced to the classification of $H_{F,\pi}$ plus 
$U_{l,\pi}$. We rearrange the spectra of $H_{F,\pi}$ according to it's real gaps and decompose $H_{F,\pi}$
 into $n_q+1$ sub-Hamiltonians $H_{F,\pi}=H_1+H_2+...+H_{n_q+1}$. Here $H_1=\sum_j
 \mathcal{E}_j|\psi_j^R\rangle \langle \psi_j^L|$ with $-\theta_1/\tau<\text{Re}(\mathcal{E}_j)<\theta_1/\tau$, $H_{n_q+1}=\sum_j
 \mathcal{E}_j|\psi_j^R\rangle \langle \psi_j^L|$ with $\text{Re}(\mathcal{E}_j)\in (-\pi/\tau,-\theta_{n_q}/\tau)
 \cup(\theta_{n_q}/\tau,
 \pi/\tau)$, and $H_m=\sum_j\mathcal{E}_j|\psi_j^R\rangle \langle \psi_j^L|$ with $\text{Re}(\mathcal{E}_j)\in 
 (-\theta_{m}/\tau,-\theta_{m-1}/\tau)\cup(\theta_{m-1}/\tau,\theta_m/\tau)$ for $1<m<n_q+1$. Each sub-Hamiltonian belongs to the same symmetry 
 class as $H_{F,\pi}$. And $H_n$ $(n=1,2,...,n_{q}+1)$ possesses an real gap at $0$ except for $H_1$. Thus $H_1$ is contractible
  to a trivial constant Hamiltonian. Combine the classification of $H_n$ $(n=2,3,...,n_{q}+1)$ with an real gap at
   $0$ \cite{KawabataShiozakiUedaSato} and the topological classification of loop operator $U_{l,\pi}$ in 
   Sec. \ref{seciiia1}, we obtain the full classification of the system. 
   
   The discussion for the latter case
    when one of the FO angle gaps is located at $0$ and $n_q$ pairs of FO angle gaps at 
    ($\theta_m$, $-\theta_m$) ($m=1,2,...,n_q$) is similar to that in {\bf S1}. We still take class 
    $P$ with $p=\sigma_z$ as an example. The homeomorphic mapping
$\bar{U}({\bf k},t)=U_0({\bf k},t)*U({\bf k},t)$ does not alter any symmetry of $U({\bf k},t)$; 
while its associated 
Floquet spectra possess FO angle gap at $\pi$, $\theta_m-\pi$ and $\pi-\theta_m$ ($m=1,2,...,n_q$). 
It follows the topological classification is the same as the former case. Such a homeomorphic mapping 
works for all the other GBL classes.

{\bf S4:} 
  There are both FO angle gaps at $0$ and $\pi$ and
   $n_q$ pairs of FO angle gaps at ($\theta_m, -\theta_m$), where $m=1,2,...,n_q$ and 
   $0< \theta_1<\theta_2<...<\theta_{n_q}< \pi$. It is equivalent to 
   there are both FH real gaps at $0$ and $\pi/\tau$ and
   $n_q$ pairs of FH real gaps at ($\theta_m/\tau, -\theta_m/\tau$), where $m=1,2,...,n_q$.
   According to Theorem 1, the
   topological classification is reduced to the classification of $H_{F,\pi}$ plus 
   $U_{l,\pi}$.
   For this case, the $H_{F,\pi}$ has one real
    gap at $0$, and $n_q$ pairs of real gaps at 
    ($\theta_m/\tau, -\theta_m/\tau$), $\pi$ is chosen as the branch cut and doesn't contribute to a real gap. We can decompose the Floquet Hamiltonian into $n_q+1$ constant Hamiltonians according to 
    quasienergy gaps: $H_{F,\pi}=H_1+H_2+...+H_m+...+H_{n_q+1}$. Here $H_1=\sum_j\mathcal{E}_j|\psi_j^R\rangle
     \langle \psi_j^L|$ with $\text{Re}(\mathcal{E}_j)\in (-\theta_{1}/\tau,0)\cup(0,\theta_1/\tau)$; $H_{n_q+1}=\sum_j\mathcal{E}_j
     |\psi_j^R\rangle \langle \psi_j^L|$ with $\text{Re}(\mathcal{E}_j)\in (-\pi/\tau,-\theta_{n_q}/\tau)\cup(\theta_{n_q}/\tau,\pi/\tau)$; 
     and $H_m=\sum_j\mathcal{E}_j|\psi_j^R\rangle \langle \psi_j^L|$ with $\text{Re}(\mathcal{E}_j)\in (-\theta_{m}/\tau,
     -\theta_{m-1}/\tau)\cup(\theta_{m-1}/\tau,\theta_m/\tau)$ for $1<m<n_q+1$.  All the sub-Hamiltonians $H_m$ belong to 
     the same symmetry class as $H_{F,\pi}$ and have a real gap located at $0$. Combine the classification 
     of $H_n$ $(n=1,2,...,n_{q}+1)$ with a real gap at $0$ \cite{KawabataShiozakiUedaSato} and the classification of
     $U_{l,\pi}$ in Sec. \ref{seciiia1}, we obtain a full topological classification for 
     the system.

The discussions in Sec. \ref{seciiia} 1-3 cover all the 54 GBL symmetry classes and the possible 
FO angle gap conditions of the Floquet
 spectra, culminating in a complete topological classification of all the FO angle-gapped FNH topological phases as summarized in the periodic Table \ref{tab:tableI}.

\subsection{FO angle-gapless case}\label{seciiib}
For the FO angle-gapless case, the spectra of the Floquet operator winds 
around the origin without any FO angle gaps [See Fig. \ref{FA1}(d)]. We remind that 
the FO angle gap is not allowed to be closed in the topological classification in the 
previous FO angle-gapped case. If there is a FO angle gap at $\theta$, 
 we can define $H_{F,\theta}$, which is a continuous function. In the FO angle-gapless 
case, there is no such spectrum restriction, 
and we can't define $H_{F,\theta}$. But Floquet 
operator $U({\bf k})$ is a continuous function. To extract the band topology of the Floquet 
operator $U({\bf k})$, we consider the following Hermitian operator:
\begin{equation}
\tilde {H}({\bf k})=\left[ \begin{array}{cc}
0 & U({\bf k})\\
U({\bf k})^\dagger & 0
\end{array}
\right ]. \label{Hk}
\end{equation}
As $\det(U({\bf k}))\ne 0$ is equivalent to $\det(\tilde {H}({\bf k}))\ne 0$, the mapping from $U({\bf k})$ to $\tilde {H}({\bf k})$ is homeomorphic, yielding that $U({\bf k})$ has the same topology as $\tilde {H}({\bf k})$.              
We can continuously transform the Hermitian operator $\tilde {H}({\bf k})$ into a band-flattened Hamiltonian $\tilde {H}({\bf k})$ [$(\tilde {H}({\bf k}))^2=1$] without altering any symmetry \cite{Kitaev}. When $U({\bf k})$ satisfies the symmetries described by Eqs. (\ref{eq:symsutp})-(\ref{eq:symsutk}), the corresponding symmetries satisfied by the band-flattened Hamiltonian $\tilde {H}({\bf k})$ are:
\begin{align}
\tilde {H}(-{\bf k})=\bar{K}\tilde {H}({\bf k})\bar{K}^{-1}&, ~~kk^*=\eta_k \mathbb{I}, &K \textrm{ sym.} \label{eq:symshttp} \\
\tilde {H}({\bf k})=\bar{Q}\tilde {H}({\bf k}) \bar{Q}^{-1}&, ~~q^2=\mathbb{I}, &Q \textrm{ sym.}\label{eq:symshttq}\\
\tilde {H}(-{\bf k})=\bar{C}\tilde {H}({\bf k})\bar{C}^{-1}&, ~~cc^*=\eta_c \mathbb{I},&C \textrm{ sym.}\label{eq:symshttc}\\
\tilde {H}({\bf k})=\bar{P}\tilde {H}({\bf k})\bar{P}^{-1}&. ~~p^2=\mathbb{I}. &P \textrm{ sym.}\label{eq:symshttk}
\end{align}
Here $\bar{K}=\sigma_x \otimes k\mathcal{K}$ if $\epsilon_k=1$ and $\bar{K}=\sigma_0 \otimes k\mathcal{K}$ if $\epsilon_k=-1$; $\bar{Q}=\sigma_0\otimes q$ if $\epsilon_q=1$ and $\bar{Q}=\sigma_x\otimes q$ if $\epsilon_q=-1$; $\bar{P}=\sigma_x \otimes p$; $\bar{C}=\sigma_x\otimes c\mathcal{K}$ if $\epsilon_c=1$ and $\bar{C}=\sigma_0\otimes c\mathcal{K}$ if $\epsilon_c=-1$. The derivation of  Eqs. (\ref{eq:symshttp})-(\ref{eq:symshttk}) is given in Appendix \ref{apg}. Besides, the Hamiltonian $\tilde {H}({\bf k})$ possesses an additional chiral symmetry ($\Sigma=\sigma_z$):
\begin{eqnarray}
\Sigma \tilde {H}({\bf k})=-\tilde {H}({\bf k})\Sigma.
\end{eqnarray}

We then use the $K$-theory to obtain the topological classification of $\tilde {H}({\bf k})$ for each of the 54 GBL classes. Following the standard classification scheme \cite{ChiuYaoRyu} in terms of Clifford algebra, we represent $\tilde {H}({\bf k})$ as
\begin{equation}
\tilde {H}({\bf k})=m\gamma_0+ k_1\gamma_1+...+k_d \gamma_d.
\end{equation}
Here, $\gamma_0,\gamma_i~(i=1,...,d)$ anticommute with each other and square to identity. Using the commutation relations of the symmetry operators and $\tilde {H}({\bf k})$, we construct the Clifford algebra's extension for 
each symmetry class. We 
can get the space of mass term from the Clifford aglebra's extension. (See Table I and Table IV of Ref. \cite{LiuChen}). By calculating the 0th homotopy group of the space of the mass term, we obtain the topological classification of the FO angle-gapless FNH topological phase for all the GBL classes. We still take the Non class as an example, 
with the generators of this class given by $\left\{\gamma_0,\gamma_1,...,\gamma_d,\Sigma\right\}$. The Clifford algebra's extension of this class is $\left\{\gamma_1,...,\gamma_d,\Sigma\right\} \rightarrow\left\{\gamma_0,\gamma_1,...,\gamma_d,\Sigma\right\}=Cl_{d+1}\rightarrow Cl_{d+2}$. The space of the mass term is $C_{d+1}$, and the classifying space is $C_1$. The topological classification for the Non class is then $\pi_0(C_{d+1})=0~(\mathbb{Z})$ for even (odd) $d$. 
We note that it is different from the FO angle-gapped case (See Sec. \ref{seciiia1}), for example, the additional time dimension in FO angle-gapped case. 
We can similarly work out the Clifford algebra's extension for all the other
 GBL classes, as demonstrated in Table. \ref{tab:tableIV} of Appendix \ref{aph}. The topological 
 classification for all the GBL classes in the FO angle-gapless case is summarized in 
 Table \ref{tab:tableII}.

 \subsection{Relations with Floquet Hermitian topological phases}\label{seciiic}
 Our classification for the FNH topological phases fully covers the previous 
 topological classifications of Floquet Hermitian topological phases \cite{fclass2,HigashikawaNakagawaUeda}
  as the special cases. In fact, the Hermitian constraint $H({\bf k},t)=H^{\dag}({\bf k},t)$ can be 
  represented by a type-$Q$ ($\epsilon_q=1$, $\eta_q=1$, and $q=\mathbb{I}$) symmetry 
   of the GBL classes in Eq. (\ref{eq:symshq}). Thus,
  the Hermitian A, AIII, AI, AII, C, D, CI, CII, BDI, and DIII classes of the AZ tenfold way correspond to the
   non-Hermitian Qa, PQ1, QC1a, QC3a, QC7a, QC5a, PQC9a, PQC5, PQC1, and PQC11a classes of the GBL 54-fold way, respectively.

   Here we derive this conclusion. The Hermitian constraint is a type-$\eta$ (pseudo-Hermitian) symmetry in 
   the paper of Kawabata et al. \cite{KawabataShiozakiUedaSato}. Due to the Hermitian operator commute with TRS, PHS, and CS operators, 
   the Hermitian A, AIII, AI, AII, C, D, CI, CII, BDI, and DIII classes of the AZ tenfold way corresponds to the
   non-Hermitian $\eta_+$A, $\eta_+$AIII, $\eta_+$AI, $\eta_+$AII, $\eta_{++}$C, $\eta_{++}$D, $\eta_{++}$CI, 
   $\eta_{++}$CII, $\eta_{++}$BDI, and $\eta_{++}$DIII classes of the GBL 54-fold way in Kawabata-Shiozaki-Ueda-Sato's notation, 
   respectively. 
   According to the correspondence relations between Kawabata-Shiozaki-Ueda-Sato notation and Ref. \cite{LiuChen}'s notation in  
   Table VIII of Ref. \cite{LiuChen}, the non-Hermitian $\eta_+$A, $\eta_+$AIII, $\eta_+$AI, $\eta_+$AII, $\eta_{++}$C, 
   $\eta_{++}$D, $\eta_{++}$CI, $\eta_{++}$CII, $\eta_{++}$BDI, and $\eta_{++}$DIII classes of the GBL 
   54-fold way in Kawabata-Shiozaki-Ueda-Sato's notation correspond to the
   non-Hermitian Qa, PQ1, QC1a, QC3a, QC7a, QC5a, PQC9a, PQC5, PQC1, and PQC11a classes of the GBL 54-fold way in 
   Ref. \cite{LiuChen}'s notation, respectively. 
 
    For the above 10 classes, our classification for the FO angle-gapped case 
    in Table \ref{tab:tableI} reproduces Roy and Harper's periodic table \cite{fclass2} of Floquet Hermitian topological insulator 
    with FH real gaps (or FO angle gaps). While for the FO angle-gapless case, our classification in Table \ref{tab:tableII} reproduces 
    Higashikawa, Nakagawa, and Ueda's periodic table \cite{HigashikawaNakagawaUeda} of unitary Floquet operator. The superficial doubling of 
    the topological invariants $\mathbb{Z}^{\times 2n}$, $\mathbb{Z}^{\times 2n_p}$, $\mathbb{Z}_2^{\times 2n}$, or 
    $\mathbb{Z}_2^{\times 2n_p}$ appeared in Table \ref{tab:tableI} and \ref{tab:tableII} comes from the existence of the $Q$ 
    ($\epsilon_q=1$) symmetry, which enforces the $\tilde {H}({\bf k}, t)$ in Eq. (\ref{H}) and $\tilde {H}({\bf k})$ 
    in Eq. (\ref{Hk}) to be diagonalized into two irreducible blocks. Each block either corresponds to the $1$ or the $-1$ eigenvalue 
    of $\tilde{Q}$ (note that $\tilde{Q}^2=1$), respectively. 
    Consider Hermitian as a type-Q symmetry, we have $\epsilon_q=1$, $\eta_q=1$, and $q=\mathbb{I}$, the $-1$ eigensubspace of $\tilde{Q}$ is vanish. By kicking 
    out the half topological numbers that correspond to -1 eigensubspace, 
    we are left with only the half topological numbers that correspond to the 1 eigensubspace and fully recover previous 
    results for Floquet Hermitian topological phases.
 
    We stress that the classification of Floquet topological insulator is equivalent to the topological classification of 
    FO angle gapped Hermitian system. And 
    unitary Floquet operator's topological classification is equivalent to the topological classification of 
    FO angle gapless Hermitian system. They are different names for the same thing.
 
    For example, according to Table \ref{tab:tableI},
     the topological classification of the non-Hermitian 1-dimensional GBL PQC1 class FO angle gapped system is $\mathbb{Z}^{2n_p}$. 
     Consider Hermitian as a type-Q symmetry, and we have $q=\mathbb{I}$. Thus, the $-1$ eigensubspace of $\tilde{Q}$ vanish, we should kick out the half topological number (they correspond to $-1$ the eigensubspace and must be trivial). And the topological classification is $\mathbb{Z}^{n_p}$. 
    According to Ref. \cite{fclass2} the topological classification of Hermitian 1-dimensional AZ BDI class Floquet topological insulator
     is $\mathbb{Z}^{n_p}$. Thus, our result is consistent with Roy and Harper's result \cite{fclass2}.
    Similarly, it is not difficult to verify that our results are consistent with Roy and Harper's results \cite{fclass2} 
    for any dimensional and symmetry class. 
    According to Table \ref{tab:tableII},
    the topological classification of the non-Hermitian 1-dimensional GBL Qa class FO angle gapless system is $\mathbb{Z}^{2}$. 
    Consider Hermitian as a type-Q symmetry, and we have $q=\mathbb{I}$. Thus, the $-1$ eigensubspace of $\tilde{Q}$ vanish, we should kick out the half topological number (they correspond to $-1$ the eigensubspace and must be trivial). And the topological classification is $\mathbb{Z}$. 
   According to Ref. \cite{HigashikawaNakagawaUeda}, the topological classification of Hermitian 1-dimensional A class unitary Floquet operator 
   is $\mathbb{Z}$. Thus, our result is consistent with Higashikawa, Nakagawa, and Ueda's 
   result \cite{HigashikawaNakagawaUeda}.
   Similarly, it is not difficult to verify that our results are consistent with Higashikawa, Nakagawa, and Ueda's results \cite{HigashikawaNakagawaUeda} 
   for any dimensional and symmetry class.

\onecolumngrid \clearpage \newpage

\begin{table*}[htbp]\footnotesize
\begin{center}
\caption{\label{tab:tableI}
 Periodic table of Floquet non-Hermitian topological phases for FO angle gapped case. 
$d$ is the spatial dimension. Each row corresponds to a specific generalized Bernard-LeClair (GBL) symmetry
 class, labeled by its name in the first column. The second column lists the symmetry-generator relations,
  including the signs of time-flipping, operator involution, and commutation relations. For the classes with $P$ and
   at least one of the $Q$ and $K$ symmetries, only the cases of $\epsilon_q=1$ or $\epsilon_k=1$ are 
   listed in the table. This is because the classes of $\epsilon_q=-1$ or $\epsilon_k=-1$ are equivalent to
    the corresponding classes of $\epsilon_q=1$ or $\epsilon_k=1$ in the presence of $P$ symmetry, as can 
    be seen from Eqs. (\ref{eq:symshp})-(\ref{eq:symshk}). The third column gives the classifying space of each symmetry class. In the table, $n\in \mathbb{Z}^+$
      is the total number of relevant spectrum gaps. $n_p=n_q+n_r$. And $n_r\in \left\{1,2\right\}$ counts
       the relevant 
       FO angle gaps at $0$ and $\pi$. $n_q\in \mathbb{Z}^+\cup \left\{0\right\}$ is the
        number of pairs of FO angle gaps at ($\theta_m$, $-\theta_m$) ($m=1,2,...,n_q$, 
        $\theta_m\ne 0, 
        \pi$). The topological numbers in the table are stable strong topological numbers.}

\begin{tabular}{|c|c|c|cccccccc|}
  \hline
    GBL&Gen. Rel.&Cl&$d=0$&1&2\;\;\;&\;\;\;3\;\;\;&\;\;\;4\;\;\;&\;\;\;\;5\;\;\;\;&\;\;\;\;6\;\;\;\;&\;\;\;\;7\;\;\;\\
  \hline
Non &  &$C_0^{\times n}$&$\mathbb{Z}^{\times n}$ & $0$ & $\mathbb{Z}^{\times n}$ & $0$ & $\mathbb{Z}^{\times n}$ & $0$ & $\mathbb{Z}^{\times n}$ & $0$\\
  \hline
P&   &$C_1^{\times n_p} $&$0$ & $\mathbb{Z}^{\times n_p}$ & $0$ & $\mathbb{Z}^{\times n_p}$ & $0$ & $\mathbb{Z}^{\times n_p}$ & $0$ & $\mathbb{Z}^{\times n_p}$\\
  \hline
Qa& $\epsilon_q=1 $ & $ C_0^{\times 2n} $ & $\mathbb{Z} ^{\times 2n}$ & $0$ & $\mathbb{Z}^{\times 2n}$ & $0$ & $\mathbb{Z} ^{\times 2n}$ & $0$ & $\mathbb{Z} ^{\times 2n}$ &  $0$ \\
  \hline
  Qb& $\epsilon_q=-1 $ & $ C_1^{\times n_p}$& $0$ & $\mathbb{Z}^{\times n_p}$ & $0$ & $\mathbb{Z}^{\times n_p}$ & $0$ & $\mathbb{Z}^{\times n_p}$ & $0$ & $\mathbb{Z}^{\times n_p}$\\
  \hline
K1a& $\epsilon_k =1$, $\eta_k =1$ & $ R_0 ^{\times n}$ &  $\mathbb{Z}^{\times n}$ & $0$ & $0$ & $0$ & $2\mathbb{Z}^{\times n}$ & $0$ & $\mathbb{Z}_2^{\times n}$ & $\mathbb{Z}_2^{\times n}$\\
  \hline
  K1b& $\epsilon_k =-1$, $\eta_k =1$ & $ R_2 ^{\times n_p}$ &  $\mathbb{Z}_2^{\times n_p}$ & $\mathbb{Z}_2^{\times n_p}$ & $\mathbb{Z}^{\times n_p}$ & $0$ & $0$ & $0$ & $2\mathbb{Z}^{\times n_p}$ & $0$\\
  \hline

K2a& $\epsilon_k =1$, $\eta_k =-1$ & $ R_4 ^{\times n}$& $2\mathbb{Z}^{\times n}$ & $0$ & $\mathbb{Z}_2^{\times n}$ & $\mathbb{Z}_2^{\times n}$ & $\mathbb{Z}^{\times n}$ & $0$ & $0$ & $0$\\
  \hline
  K2b& $\epsilon_k =-1$,  $\eta_k =-1$ &$ R_6 ^{\times n_p}$& $0$ & $0$ & $2\mathbb{Z}^{\times n_p}$ & $0$ & $\mathbb{Z}_2^{\times n_p}$ & $\mathbb{Z}_2^{\times n_p}$ & $\mathbb{Z}^{\times n_p}$ & $0$\\
  \hline

C1& $\epsilon_c=1$, $\eta_c=1$ & $ R_0 ^{\times n}$& $\mathbb{Z}^{\times n}$ & $0$ & $0$ & $0$ & $2\mathbb{Z}^{\times n}$ & $0$ & $\mathbb{Z}_2^{\times n}$ & $\mathbb{Z}_2^{\times n}$\\
  \hline
C2& $\epsilon_c=1$, $\eta_c=-1$ & $ R_4 ^{\times n}$& $2\mathbb{Z}^{\times n}$ & $0$ & $\mathbb{Z}_2^{\times n}$ & $\mathbb{Z}_2^{\times n}$ & $\mathbb{Z}^{\times n}$ & $0$ & $0$ & $0$\\
  \hline
C3& $\epsilon_c=-1$, $\eta_c=1$  & $ R_2 ^{\times n_p}$& $\mathbb{Z}_2^{\times n_p}$ & $\mathbb{Z}_2^{\times n_p}$ & $\mathbb{Z}^{\times n_p}$ & $0$ & $0$ & $0$ & $2\mathbb{Z}^{\times n_p}$ & $0$\\
  \hline
C4& $\epsilon_c=-1$, $\eta_c=-1$ & $ R_6 ^{\times n_p}$& $0$ & $0$ & $2\mathbb{Z}^{\times n_p}$ & $0$ & $\mathbb{Z}_2^{\times n_p}$ & $\mathbb{Z}_2^{\times n_p}$ & $\mathbb{Z}^{\times n_p}$ & $0$\\
  \hline

PQ1& $\epsilon_q=1$, $\epsilon_{pq}=1$ & $ C_1^{\times 2n_p} $& $0$ & $\mathbb{Z}^{\times 2n_p}$ & $0$ & $\mathbb{Z} ^{\times 2n_p}$ & $0$ & $\mathbb{Z} ^{\times 2n_p}$ & $0$ & $\mathbb{Z} ^{\times 2n_p}$\\
  \hline
PQ2& $\epsilon_q=1$, $\epsilon_{pq}=-1$ & $ C_0  ^{\times n_p}$&  $\mathbb{Z} ^{\times n_p}$ & $0$ & $\mathbb{Z} ^{\times n_p}$ & $0$ & $\mathbb{Z} ^{\times n_p}$ & $0$ & $\mathbb{Z} ^{\times n_p}$ & $0$\\
  \hline
PK1& $\epsilon_k=1$, $\eta_k=1$, $\epsilon_{pk}=1$ & $ R_1  ^{\times n_p}$&  $\mathbb{Z}_2^{\times n_p}$ & $\mathbb{Z}^{\times n_p}$ & $0$ & $0$ & $0$ & $2\mathbb{Z}^{\times n_p}$ & $0$ & $\mathbb{Z}_2^{\times n_p}$\\
  \hline
PK2& $\epsilon_k=1$, $\eta_k=-1$, $\epsilon_{pk}=1$ & $ R_5 ^{\times n_p} $& $0$ & $2\mathbb{Z}^{\times n_p}$ & $0$ & $\mathbb{Z}_2^{\times n_p}$ & $\mathbb{Z}_2^{\times n_p}$ & $\mathbb{Z}^{\times n_p}$ & $0$ & $0$\\
  \hline
PK3a& \begin{tabular}{c} $\epsilon_k=1$, $\eta_k=1$, $\epsilon_{pk}=-1$  \end{tabular} & $ R_7 ^{\times n_p}$& $0$ & $0$ & $0$ & $2\mathbb{Z}^{\times n_p}$ & $0$ & $\mathbb{Z}_2^{\times n_p}$ & $\mathbb{Z}_2^{\times n_p}$ & $\mathbb{Z}^{\times n_p}$\\
  \hline
PK3b& \begin{tabular}{c} $\epsilon_k=1$, $\eta_k=-1$, $\epsilon_{pk}=-1$   \end{tabular} & $ R_3 ^{\times n_p}$&  $0$ & $\mathbb{Z}_2^{\times n_p}$ & $\mathbb{Z}_2^{\times n_p}$ & $\mathbb{Z}^{\times n_p}$ & $0$ & $0$ & $0$& $2\mathbb{Z}^{\times n_p}$ \\
  \hline
PC1& \begin{tabular}{c} $\epsilon_c=1$, $\eta_c=1$, $\epsilon_{pc}=1$ \\ \hline $\epsilon_c=-1$, $\eta_c=1$, $\epsilon_{pc}=1$ \end{tabular} & $ R_1 ^{\times n_p}$& $\mathbb{Z}_2^{\times n_p}$ & $\mathbb{Z}^{\times n_p}$ & $0$ & $0$ & $0$ & $2\mathbb{Z}^{\times n_p}$ & $0$ & $\mathbb{Z}_2^{\times n_p}$\\
  \hline
PC2& \begin{tabular}{c} $\epsilon_c=1$, $\eta_c=1$, $\epsilon_{pc}=-1$ \\ \hline $\epsilon_c=-1$, $\eta_c=-1$, $\epsilon_{pc}=-1$ \end{tabular} & $ R_7 ^{\times n_p}$& $0$ & $0$ & $0$ & $2\mathbb{Z}^{\times n_p}$ & $0$ & $\mathbb{Z}_2^{\times n_p}$ & $\mathbb{Z}_2^{\times n_p}$ & $\mathbb{Z}^{\times n_p}$\\
  \hline
PC3& \begin{tabular}{c} $\epsilon_c=1$, $\eta_c=-1$, $\epsilon_{pc}=1$ \\ \hline $\epsilon_c=-1$, $\eta_c=-1$, $\epsilon_{pc}=1$ \end{tabular} & $ R_5 ^{\times n_p}$& $0$ & $2\mathbb{Z}^{\times n_p}$ & $0$ & $\mathbb{Z}_2^{\times n_p}$ & $\mathbb{Z}_2^{\times n_p}$ & $\mathbb{Z}^{\times n_p}$ & $0$ & $0$\\
  \hline
PC4& \begin{tabular}{c} $\epsilon_c=1$, $\eta_c=-1$, $\epsilon_{pc}=-1$ \\ \hline $\epsilon_c=-1$, $\eta_c=1$, $\epsilon_{pc}=-1$ \end{tabular} & $ R_3^{\times n_p} $&  $0$ & $\mathbb{Z}_2^{\times n_p}$ & $\mathbb{Z}_2^{\times n_p}$ & $\mathbb{Z}^{\times n_p}$ & $0$ & $0$ & $0$& $2\mathbb{Z}^{\times n_p}$ \\
  \hline
QC1a& $\epsilon_q=1$, $\epsilon_c=1$, $\eta_c=1$, $\epsilon_{qc}=1$ & $ R_0^{\times 2n}$&  $\mathbb{Z}^{\times 2n}$ & $0$ & $0$ & $0$ & $2\mathbb{Z}^{\times 2n}$ & $0$ & $\mathbb{Z}_2 ^{\times 2n}$ & $\mathbb{Z}_2 ^{\times 2n}$ \\
  \hline
  QC1b& $\epsilon_q=-1$, $\epsilon_c=1$, $\eta_c=1$, $\epsilon_{qc}=1$ & $ R_1^{\times n_p} $&  $\mathbb{Z}_2^{\times n_p}$ & $\mathbb{Z}^{\times n_p}$ & $0$ & $0$ & $0$ & $2\mathbb{Z}^{\times n_p}$ & $0$ & $\mathbb{Z}_2^{\times n_p}$\\
  \hline
QC2a&  $\epsilon_q=1$, $\epsilon_c=1$, $\eta_c=1$, $\epsilon_{qc}=-1$  & $ C_0 ^{\times n}$& $\mathbb{Z}^{\times n}$ & $0$ & $\mathbb{Z}^{\times n}$ & $0$ & $\mathbb{Z}^{\times n}$ & $0$ & $\mathbb{Z}^{\times n}$ & $0$\\
  \hline
  QC2b&  $\epsilon_q=-1$, $\epsilon_c=1$, $\eta_c=1$, $\epsilon_{qc}=-1$ & $ R_7 ^{\times n_p}$&  $0$ & $0$ & $0$ & $2\mathbb{Z}^{\times n_p}$ & $0$ & $\mathbb{Z}_2^{\times n_p}$ & $\mathbb{Z}_2^{\times n_p}$ & $\mathbb{Z}^{\times n_p}$\\
  \hline
QC3a&  $\epsilon_q=1$, $\epsilon_c=1$, $\eta_c=-1$, $\epsilon_{qc}=1$  & $ R_4^{\times 2n}$& $2\mathbb{Z}^{\times 2n}$ & $0$ & $\mathbb{Z}_2 ^{\times 2n}$ & $\mathbb{Z}_2 ^{\times 2n}$ & $\mathbb{Z}^{\times 2n}$ & $0$ & $0$ & $0$\\
  \hline
  QC3b&  $\epsilon_q=-1$, $\epsilon_c=1$, $\eta_c=-1$, $\epsilon_{qc}=1$  & $ R_5^{\times n_p} $& $0$ & $2\mathbb{Z}^{\times n_p}$ & $0$ & $\mathbb{Z}_2^{\times n_p}$ & $\mathbb{Z}_2^{\times n_p}$ & $\mathbb{Z}^{\times n_p}$ & $0$ & $0$\\
  \hline
QC4a& $\epsilon_q=1$, $\epsilon_c=1$, $\eta_c=-1$, $\epsilon_{qc}=-1$  & $ C_0 ^{\times n}$&  $\mathbb{Z}^{\times n}$ & $0$ & $\mathbb{Z}^{\times n}$ & $0$ & $\mathbb{Z}^{\times n}$ & $0$ & $\mathbb{Z}^{\times n}$ & $0$\\
  \hline
  QC4b& $\epsilon_q=-1$, $\epsilon_c=1$, $\eta_c=-1$, $\epsilon_{qc}=-1$  & $ R_3 ^{\times n_p}$&  $0$ & $\mathbb{Z}_2^{\times n_p}$ & $\mathbb{Z}_2^{\times n_p}$ & $\mathbb{Z}^{\times n_p}$ & $0$ & $0$ & $0$& $2\mathbb{Z}^{\times n_p}$ \\
  \hline
QC5a& $\epsilon_q=1$, $\epsilon_c=-1$, $\eta_c=1$, $\epsilon_{qc}=1$  & $ R_2^{\times 2n_p} $&  $\mathbb{Z}_2 ^{\times 2n_p}$ & $\mathbb{Z}_2 ^{\times 2n_p}$ & $\mathbb{Z} ^{\times 2n_p}$ & $0$ & $0$ & $0$ & $2\mathbb{Z}^{\times 2n_p}$ & $0$\\
  \hline
  QC5b& $\epsilon_q=-1$, $\epsilon_c=-1$, $\eta_c=1$, $\epsilon_{qc}=1$  & $ R_1 ^{\times n_p}$&   $\mathbb{Z}_2^{\times n_p}$ & $\mathbb{Z}^{\times n_p}$ & $0$ & $0$ & $0$ & $2\mathbb{Z}^{\times n_p}$ & $0$ & $\mathbb{Z}_2^{\times n_p}$\\
  \hline
QC6a& $\epsilon_q=1$, $\epsilon_c=-1$, $\eta_c=1$, $\epsilon_{qc}=-1$  & $ C_0 ^{\times n_p}$& $\mathbb{Z}^{\times n_p}$ & $0$ & $\mathbb{Z}^{\times n_p}$ & $0$ & $\mathbb{Z}^{\times n_p}$ & $0$ & $\mathbb{Z}^{\times n_p}$ & $0$\\
  \hline
  QC6b& $\epsilon_q=-1$, $\epsilon_c=-1$, $\eta_c=1$, $\epsilon_{qc}=-1$  & $ R_3 ^{\times n_p}$&  $0$ & $\mathbb{Z}_2^{\times n_p}$ & $\mathbb{Z}_2^{\times n_p}$ & $\mathbb{Z}^{\times n_p}$ & $0$ & $0$ & $0$& $2\mathbb{Z}^{\times n_p}$ \\
  \hline
QC7a& $\epsilon_q=1$, $\epsilon_c=-1$, $\eta_c=-1$, $\epsilon_{qc}=1$  & $ R_6^{\times 2n_p}$& $0$ & $0$ & $2\mathbb{Z}^{\times 2n_p}$ & $0$ & $\mathbb{Z}_2 ^{\times 2n_p}$ &$\mathbb{Z}_2 ^{\times 2n_p}$ & $\mathbb{Z}^{\times 2n_p}$ & $0$\\
  \hline
  QC7b& $\epsilon_q=-1$, $\epsilon_c=-1$, $\eta_c=-1$, $\epsilon_{qc}=1$  & $ R_5^{\times n_p} $& $0$ & $2\mathbb{Z}^{\times n_p}$ & $0$ & $\mathbb{Z}_2^{\times n_p}$ & $\mathbb{Z}_2^{\times n_p}$ & $\mathbb{Z}^{\times n_p}$ & $0$ & $0$\\
  \hline
QC8a& $\epsilon_q=1$, $\epsilon_c=-1$, $\eta_c=-1$, $\epsilon_{qc}=-1$   & $ C_0 ^{\times n_p}$&  $\mathbb{Z}^{\times n_p}$ & $0$ & $\mathbb{Z}^{\times n_p}$ & $0$ & $\mathbb{Z}^{\times n_p}$ & $0$ & $\mathbb{Z}^{\times n_p}$ & $0$\\
  \hline
  QC8b& $\epsilon_q=-1$, $\epsilon_c=-1$, $\eta_c=-1$, $\epsilon_{qc}=-1$   & $ R_7 ^{\times n_p}$&  $0$ & $0$ & $0$ & $2\mathbb{Z}^{\times n_p}$ & $0$ & $\mathbb{Z}_2^{\times n_p}$ & $\mathbb{Z}_2^{\times n_p}$ & $\mathbb{Z}^{\times n_p}$\\
  \hline
PQC1& \begin{tabular}{c} $\epsilon_c=1$, $\eta_c=1$, $\epsilon_{pq}=1$, $\epsilon_{pc}=1$, $\epsilon_{qc}=1$ \\ \hline $\epsilon_c=-1$, $\eta_c=1$, $\epsilon_{pq}=1$, $\epsilon_{pc}=1$, $\epsilon_{qc}=1$ \end{tabular} & $ R_1^{\times 2n_p} $&  $\mathbb{Z}_2 ^{\times 2n_p}$ & $\mathbb{Z}^{\times 2n_p}$ & $0 $ & $0 $ & $0 $ & $2\mathbb{Z}^{\times 2n_p}$ & $0$ & $\mathbb{Z}_2 ^{\times 2n_p}$\\
  \hline
PQC2& \begin{tabular}{c} $\epsilon_c=1$, $\eta_c=1$, $\epsilon_{pq}=1$, $\epsilon_{pc}=1$, $\epsilon_{qc}=-1$ \\ \hline $\epsilon_c=-1$, $\eta_c=1$, $\epsilon_{pq}=1$, $\epsilon_{pc}=1$, $\epsilon_{qc}=-1$ \end{tabular} & $ C_1 ^{\times n_p}$&  $0$ & $\mathbb{Z}^{\times n_p}$ & $0$ & $\mathbb{Z}^{\times n_p}$ & $0$ & $\mathbb{Z}^{\times n_p}$ & $0$ & $\mathbb{Z}^{\times n_p}$\\
  \hline
PQC3& \begin{tabular}{c} $\epsilon_c=1$, $\eta_c=1$, $\epsilon_{pq}=-1$, $\epsilon_{pc}=-1$, $\epsilon_{qc}=1$ \\ \hline $\epsilon_c=-1$, $\eta_c=-1$, $\epsilon_{pq}=-1$, $\epsilon_{pc}=-1$, $\epsilon_{qc}=-1$ \end{tabular} & $ R_0 ^{\times n_p}$&  $\mathbb{Z}^{\times n_p}$ & $0$ & $0$ & $0$ & $2\mathbb{Z}^{\times n_p}$ & $0$ & $\mathbb{Z}_2^{\times n_p}$ & $\mathbb{Z}_2^{\times n_p}$\\
  \hline
PQC4& \begin{tabular}{c} $\epsilon_c=1$, $\eta_c=1$, $\epsilon_{pq}=-1$, $\epsilon_{pc}=-1$, $\epsilon_{qc}=-1$ \\ \hline $\epsilon_c=-1$, $\eta_c=-1$, $\epsilon_{pq}=-1$, $\epsilon_{pc}=-1$, $\epsilon_{qc}=1$ \end{tabular} & $ R_6^{\times n_p}  $& $0$ & $0$ & $2\mathbb{Z}^{\times n_p}$ & $0$ & $\mathbb{Z}_2^{\times n_p}$ & $\mathbb{Z}_2^{\times n_p}$ & $\mathbb{Z}^{\times n_p}$ & $0$\\
  \hline
   \multicolumn{11}{c}{continued on next page}
  \end{tabular}
\end{center}
\end{table*}

\begin{table*}[htbp]\footnotesize
  \begin{center}
\begin{tabular}{|c|c|c|cccccccc|}
 \multicolumn{11}{c}{TABLE I  --- continued} \\ \hline
PQC5& \begin{tabular}{c} $\epsilon_c=1$, $\eta_c=-1$, $\epsilon_{pq}=1$, $\epsilon_{pc}=1$, $\epsilon_{qc}=1$ \\ \hline $\epsilon_c=-1$, $\eta_c=-1$, $\epsilon_{pq}=1$, $\epsilon_{pc}=1$, $\epsilon_{qc}=1$ \end{tabular} &$ R_5^{\times 2n_p} $&  $0$ & $2\mathbb{Z}^{\times 2n_p}$ & $0$ & $\mathbb{Z}_2^{\times 2n_p}$ & $\mathbb{Z}_2 ^{\times 2n_p}$ & $\mathbb{Z}^{\times 2n_p}$ & $0$ & $0$\\
  \hline
PQC6& \begin{tabular}{c} $\epsilon_c=1$, $\eta_c=-1$, $\epsilon_{pq}=1$, $\epsilon_{pc}=1$, $\epsilon_{qc}=-1$ \\ \hline $\epsilon_c=-1$, $\eta_c=-1$, $\epsilon_{pq}=1$, $\epsilon_{pc}=1$, $\epsilon_{qc}=-1$ \end{tabular} & $ C_1^{\times n_p} $&   $0$ & $\mathbb{Z}^{\times n_p}$ & $0$ & $\mathbb{Z}^{\times n_p}$ & $0$ & $\mathbb{Z}^{\times n_p}$ & $0$ & $\mathbb{Z}^{\times n_p}$\\
  \hline
PQC7& \begin{tabular}{c} $\epsilon_c=1$, $\eta_c=-1$, $\epsilon_{pq}=-1$, $\epsilon_{pc}=-1$, $\epsilon_{qc}=1$ \\ \hline $\epsilon_c=-1$, $\eta_c=1$, $\epsilon_{pq}=-1$, $\epsilon_{pc}=-1$, $\epsilon_{qc}=-1$ \end{tabular} & $ R_4^{\times n_p}  $&  $2\mathbb{Z}^{\times n_p}$ & $0$ & $\mathbb{Z}_2^{\times n_p}$ & $\mathbb{Z}_2^{\times n_p}$ & $\mathbb{Z}^{\times n_p}$ & $0$ & $0$ & $0$\\
  \hline
PQC8& \begin{tabular}{c} $\epsilon_c=1$, $\eta_c=-1$, $\epsilon_{pq}=-1$, $\epsilon_{pc}=-1$, $\epsilon_{qc}=-1$ \\ \hline $\epsilon_c=-1$, $\eta_c=1$, $\epsilon_{pq}=-1$, $\epsilon_{pc}=-1$, $\epsilon_{qc}=1$ \end{tabular} & $ R_2 ^{\times n_p} $&  $\mathbb{Z}_2^{\times n_p}$ & $\mathbb{Z}_2^{\times n_p}$ & $\mathbb{Z}^{\times n_p}$ & $0$ & $0$ & $0$ & $2\mathbb{Z}^{\times n_p}$ & $0$\\
  \hline

PQC9a& \begin{tabular}{c} $\epsilon_c=1$, $\eta_c=1$, $\epsilon_{pq}=1$, $\epsilon_{pc}=-1$, $\epsilon_{qc}=1$ \\ \hline $\epsilon_c=-1$, $\eta_c=-1$, $\epsilon_{pq}=1$, $\epsilon_{pc}=-1$, $\epsilon_{qc}=1$  \end{tabular} & $ R_7^{\times 2n_p} $&  $0$ & $0$ & $0$ & $2\mathbb{Z}^{\times 2n_p}$ & $0$ & $\mathbb{Z}_2^{\times 2n_p}$ &  $\mathbb{Z}_2^{\times 2n_p}$ & $\mathbb{Z}^{\times 2n_p}$\\
  \hline
  PQC9b& \begin{tabular}{c}  $\epsilon_c=1$, $\eta_c=1$, $\epsilon_{pq}=1$, $\epsilon_{pc}=-1$, $\epsilon_{qc}=-1$ \\ \hline $\epsilon_c=-1$, $\eta_c=-1$, $\epsilon_{pq}=1$, $\epsilon_{pc}=-1$, $\epsilon_{qc}=-1$ \end{tabular} & $ C_1 ^{\times n_p}$&  $0$ & $\mathbb{Z}^{\times n_p}$ & $0$ & $\mathbb{Z}^{\times n_p}$ & $0$ & $\mathbb{Z}^{\times n_p}$ & $0$ & $\mathbb{Z}^{\times n_p}$\\
  \hline

PQC10a& \begin{tabular}{c} $\epsilon_c=1$, $\eta_c=1$, $\epsilon_{pq}=-1$, $\epsilon_{pc}=1$, $\epsilon_{qc}=1$ \\ \hline $\epsilon_c=-1$, $\eta_c=1$, $\epsilon_{pq}=-1$, $\epsilon_{pc}=1$, $\epsilon_{qc}=-1$  \end{tabular} & $ R_0 ^{\times n_p}$& $\mathbb{Z}^{\times n_p}$ & $0$ & $0$ & $0$ & $2\mathbb{Z}^{\times n_p}$ & $0$ & $\mathbb{Z}_2^{\times n_p}$ & $\mathbb{Z}_2^{\times n_p}$\\
  \hline
  PQC10b& \begin{tabular}{c}  $\epsilon_c=1$, $\eta_c=1$, $\epsilon_{pq}=-1$, $\epsilon_{pc}=1$, $\epsilon_{qc}=-1$ \\ \hline $\epsilon_c=-1$, $\eta_c=1$, $\epsilon_{pq}=-1$, $\epsilon_{pc}=1$, $\epsilon_{qc}=1$ \end{tabular} & $ R_2 ^{\times n_p}$&  $\mathbb{Z}_2^{\times n_p}$ & $\mathbb{Z}_2^{\times n_p}$ & $\mathbb{Z}^{\times n_p}$ & $0$ & $0$ & $0$ & $2\mathbb{Z}^{\times n_p}$ & $0$\\
  \hline

PQC11a& \begin{tabular}{c}$\epsilon_c=1$, $\eta_c=-1$, $\epsilon_{pq}=1$, $\epsilon_{pc}=-1$, $\epsilon_{qc}=1$ \\ \hline $\epsilon_c=-1$, $\eta_c=1$, $\epsilon_{pq}=1$, $\epsilon_{pc}=-1$, $\epsilon_{qc}=1$ \end{tabular} & $ R_3^{\times 2n_p} $& $0$ & $\mathbb{Z}_2^{\times 2n_p}$ & $\mathbb{Z}_2^{\times 2n_p}$ & $\mathbb{Z}^{\times 2n_p}$ & $0$ & $0$ & $0$& $2\mathbb{Z}^{\times 2n_p}$ \\
  \hline
  PQC11b& \begin{tabular}{c}  $\epsilon_c=1$, $\eta_c=-1$, $\epsilon_{pq}=1$, $\epsilon_{pc}=-1$, $\epsilon_{qc}=-1$ \\ \hline $\epsilon_c=-1$, $\eta_c=1$, $\epsilon_{pq}=1$, $\epsilon_{pc}=-1$, $\epsilon_{qc}=-1$ \end{tabular} & $ C_1 ^{\times n_p}$& $0$ & $\mathbb{Z}^{\times n_p}$ & $0$ & $\mathbb{Z}^{\times n_p}$ & $0$ & $\mathbb{Z}^{\times n_p}$ & $0$ & $\mathbb{Z}^{\times n_p}$\\
  \hline

PQC12a& \begin{tabular}{c} $\epsilon_c=1$, $\eta_c=-1$, $\epsilon_{pq}=-1$, $\epsilon_{pc}=1$, $\epsilon_{qc}=1$ \\ \hline $\epsilon_c=-1$, $\eta_c=-1$, $\epsilon_{pq}=-1$, $\epsilon_{pc}=1$, $\epsilon_{qc}=-1$  \end{tabular} & $ R_4 ^{\times n_p}$& $2\mathbb{Z}^{\times n_p}$ & $0$ & $\mathbb{Z}_2^{\times n_p}$ & $\mathbb{Z}_2^{\times n_p}$ & $\mathbb{Z}^{\times n_p}$ & $0$ & $0$ & $0$\\
  \hline
  PQC12b& \begin{tabular}{c} $\epsilon_c=1$, $\eta_c=-1$, $\epsilon_{pq}=-1$, $\epsilon_{pc}=1$, $\epsilon_{qc}=-1$ \\ \hline $\epsilon_c=-1$, $\eta_c=-1$, $\epsilon_{pq}=-1$, $\epsilon_{pc}=1$, $\epsilon_{qc}=1$ \end{tabular} & $ R_6 ^{\times n_p}$& $0$ & $0$ & $2\mathbb{Z}^{\times n_p}$ & $0$ & $\mathbb{Z}_2^{\times n_p}$ & $\mathbb{Z}_2^{\times n_p}$ & $\mathbb{Z}^{\times n_p}$ & $0$\\
  \hline
\end{tabular}

\end{center}
\end{table*}

\begin{table*}[htbp]\footnotesize
  \begin{center}
  \caption{\label{tab:tableII} 
  Periodic table of Floquet non-Hermitian topological phases for FO angle gapless case.  $d$ is the spatial dimension. The first two columns are the same as those of Table \ref{tab:tableI}. Each row corresponds to a specific generalized Bernard-LeClair (GBL) symmetry class, labeled by its name in the first column. The second column lists the symmetry generator relations, 
  including the signs of time-flipping, operator involution, and commutation relations. The third column gives the classifying space of each symmetry class in the FO angle-gapless case, which is different from Table \ref{tab:tableI}. The topological numbers in the table are stable strong topological numbers.}

  \begin{tabular}{|c|c|c|cccccccc|}
    \hline
    \;\;\;GBL\;\;\;&Gen. Rel.&Cl&\;\;\;\;\;$d=0$\;\;\;\;&1& \;\;\;\;\;2\;\;\;\;\;&\;\;\;\;\;3\;\;\;\;\;&\;\;\;\;\;4\;\;\;\;\;&\;\;\;\;\;5\;\;\;\;\;&\;\;\;\;\;6\;\;\;\;\;&\;\;\;\;\;7\;\;\;\;\;\\
  \hline
Non& $\;\;\qquad \qquad \qquad \qquad \qquad \qquad  \qquad \qquad \qquad \qquad$\;\;&$C_1$& \Con\\
  \hline
P&   &$ C_0 $& \Cze \\
  \hline
Qa& $\epsilon_q=1$ & $ C_1^2 $& \Conon \\
  \hline
  Qb& $\epsilon_q=-1$ & $  C_0$ & \Cze \\
  \hline
K1a& $\epsilon_k=1$ ,$\eta_k =1$ & $ R_7 $ & \Rse \\
  \hline
  K1b& $\epsilon_k=-1$ ,$\eta_k =1$ & $ R_1 $ &  \Ron \\
  \hline
K2a& $\epsilon_k=1$ ,$\eta_k =-1$ & $ R_3 $& \Rth \\
  \hline
  K2b& $\epsilon_k=-1$ , $\eta_k =-1$ & $ R_5 $& \Rfi \\
  \hline
C1& $\epsilon_c=1$, $\eta_c=1$ & $  R_7$& \Rse \\
  \hline
C2& $\epsilon_c=1$, $\eta_c=-1$ & $ R_3 $&  \Rth \\
  \hline
C3& $\epsilon_c=-1$, $\eta_c=1$ & $ R_1 $& \Ron \\
  \hline
C4& $\epsilon_c=-1$, $\eta_c=-1$ & $ R_5  $& \Rfi \\
  \hline
PQ1& $\epsilon_q=1$, $\epsilon_{pq}=1$ & $ C_0^2 $& \Czeze  \\
  \hline
PQ2& $\epsilon_q=1$, $\epsilon_{pq}=-1$ & $ C_1 $& \Con  \\
  \hline
PK1& $\epsilon_k=1$, $\eta_k=1$, $\epsilon_{pk}=1$ & $ R_0 $& \Rze \\
  \hline
PK2& $\epsilon_k=1$, $\eta_k=-1$, $\epsilon_{pk}=1$ & $ R_4 $& \Rfo \\
  \hline
PK3a& \begin{tabular}{c} $\epsilon_k=1$, $\eta_k=1$, $\epsilon_{pk}=-1$   \end{tabular} & $ R_6 $& \Rsi  \\
  \hline
  PK3b& \begin{tabular}{c} $\epsilon_k=1$, $\eta_k=-1$, $\epsilon_{pk}=-1$  \end{tabular} & $  R_2 $& \Rtw \\
  \hline
PC1& \begin{tabular}{c} $\epsilon_c=1$, $\eta_c=1$, $\epsilon_{pc}=1$ \\ \hline $\epsilon_c=-1$, $\eta_c=1$, $\epsilon_{pc}=1$ \end{tabular} & $ R_0 $& \Rze \\
  \hline
PC2& \begin{tabular}{c} $\epsilon_c=1$, $\eta_c=1$, $\epsilon_{pc}=-1$ \\ \hline $\epsilon_c=-1$, $\eta_c=-1$, $\epsilon_{pc}=-1$ \end{tabular} & $ R_6 $& \Rsi \\
  \hline
PC3& \begin{tabular}{c} $\epsilon_c=1$, $\eta_c=-1$, $\epsilon_{pc}=1$ \\ \hline $\epsilon_c=-1$, $\eta_c=-1$, $\epsilon_{pc}=1$ \end{tabular} & $ R_4 $& \Rfo \\
  \hline
PC4& \begin{tabular}{c} $\epsilon_c=1$, $\eta_c=-1$, $\epsilon_{pc}=-1$ \\ \hline $\epsilon_c=-1$, $\eta_c=1$, $\epsilon_{pc}=-1$ \end{tabular} & $ R_2 $& \Rtw \\
  \hline
QC1a& $\epsilon_q=1$ , $\epsilon_c=1$, $\eta_c=1$, $\epsilon_{qc}=1$ & $ R_7^2 $&  \Rsese  \\
  \hline
  QC1b& $\epsilon_q=-1$ , $\epsilon_c=1$, $\eta_c=1$, $\epsilon_{qc}=1$ & $ R_0 $&  \Rze  \\
  \hline
 QC2a&  $\epsilon_q=1$, $\epsilon_c=1$, $\eta_c=1$, $\epsilon_{qc}=-1$ & $ C_1 $& \Con \\
  \hline
QC2b&  $\epsilon_q=-1$, $\epsilon_c=1$, $\eta_c=1$, $\epsilon_{qc}=-1$  & $ R_6 $& \Rsi  \\
  \hline
\multicolumn{11}{c}{continued on next page}
  \end{tabular}
\end{center}
\end{table*}

\begin{table*}[htbp]\footnotesize
  \begin{center}
\begin{tabular}{|c|c|c|cccccccc|}
 \multicolumn{11}{c}{TABLE II --- continued} \\ \hline
  QC3a&  $\epsilon_q=1$, $\epsilon_c=1$, $\eta_c=-1$, $\epsilon_{qc}=1$  & $ R_3^2 $& \Rthth \\
  \hline
QC3b&  $\epsilon_q=-1$, $\epsilon_c=1$, $\eta_c=-1$, $\epsilon_{qc}=1$  & $ R_4 $& \Rfo  \\
  \hline
  QC4a& $\epsilon_q=1$, $\epsilon_c=1$, $\eta_c=-1$, $\epsilon_{qc}=-1$  & $ C_1 $& \Con  \\
  \hline
QC4b& $\epsilon_q=-1$, $\epsilon_c=1$, $\eta_c=-1$, $\epsilon_{qc}=-1$  & $ R_2 $& \Rtw \\
  \hline
  QC5a& $\epsilon_q=1$, $\epsilon_c=-1$, $\eta_c=1$, $\epsilon_{qc}=1$  & $ R_1^2 $& \Ronon  \\
  \hline
QC5b& $\epsilon_q=-1$, $\epsilon_c=-1$, $\eta_c=1$, $\epsilon_{qc}=1$  & $ R_0 $& \Rze \\
  \hline
  QC6a& $\epsilon_q=1$, $\epsilon_c=-1$, $\eta_c=1$, $\epsilon_{qc}=-1$  & $ C_1 $& \Con  \\
  \hline
QC6b& $\epsilon_q=-1$, $\epsilon_c=-1$, $\eta_c=1$, $\epsilon_{qc}=-1$  & $ R_2 $& \Rtw \\
  \hline
  QC7a& $\epsilon_q=1$, $\epsilon_c=-1$, $\eta_c=-1$, $\epsilon_{qc}=1$  & $ R_5^2 $& \Rfifi \\
  \hline
QC7b& $\epsilon_q=-1$, $\epsilon_c=-1$, $\eta_c=-1$, $\epsilon_{qc}=1$  & $ R_4 $& \Rfo  \\
  \hline
 QC8a& $\epsilon_q=1$, $\epsilon_c=-1$, $\eta_c=-1$, $\epsilon_{qc}=-1$   & $ C_1 $& \Con  \\
  \hline
QC8b& $\epsilon_q=-1$, $\epsilon_c=-1$, $\eta_c=-1$, $\epsilon_{qc}=-1$   & $ R_6 $& \Rsi \\
  \hline
PQC1& \begin{tabular}{c} $\epsilon_c=1$, $\eta_c=1$, $\epsilon_{pq}=1$, $\epsilon_{pc}=1$, $\epsilon_{qc}=1$ \\ \hline $\epsilon_c=-1$, $\eta_c=1$, $\epsilon_{pq}=1$, $\epsilon_{pc}=1$, $\epsilon_{qc}=1$ \end{tabular} & $ R_0^2 $& \Rzeze  \\
  \hline
PQC2& \begin{tabular}{c} $\epsilon_c=1$, $\eta_c=1$, $\epsilon_{pq}=1$, $\epsilon_{pc}=1$, $\epsilon_{qc}=-1$ \\ \hline $\epsilon_c=-1$, $\eta_c=1$, $\epsilon_{pq}=1$, $\epsilon_{pc}=1$, $\epsilon_{qc}=-1$ \end{tabular} & $ C_0 $& \Cze \\
  \hline
PQC3& \begin{tabular}{c} $\epsilon_c=1$, $\eta_c=1$, $\epsilon_{pq}=-1$, $\epsilon_{pc}=-1$, $\epsilon_{qc}=1$ \\ \hline $\epsilon_c=-1$, $\eta_c=-1$, $\epsilon_{pq}=-1$, $\epsilon_{pc}=-1$, $\epsilon_{qc}=-1$ \end{tabular} & $ R_7 $& \Rse \\
  \hline
PQC4& \begin{tabular}{c} $\epsilon_c=1$, $\eta_c=1$, $\epsilon_{pq}=-1$, $\epsilon_{pc}=-1$, $\epsilon_{qc}=-1$ \\ \hline $\epsilon_c=-1$, $\eta_c=-1$, $\epsilon_{pq}=-1$, $\epsilon_{pc}=-1$, $\epsilon_{qc}=1$ \end{tabular} & $  R_5 $& \Rfi  \\
  \hline
 PQC5& \begin{tabular}{c} $\epsilon_c=1$, $\eta_c=-1$, $\epsilon_{pq}=1$, $\epsilon_{pc}=1$, $\epsilon_{qc}=1$ \\ \hline $\epsilon_c=-1$, $\eta_c=-1$, $\epsilon_{pq}=1$, $\epsilon_{pc}=1$, $\epsilon_{qc}=1$ \end{tabular} &$ R_4^2 $& \Rfofo  \\
  \hline
PQC6& \begin{tabular}{c} $\epsilon_c=1$, $\eta_c=-1$, $\epsilon_{pq}=1$, $\epsilon_{pc}=1$, $\epsilon_{qc}=-1$ \\ \hline $\epsilon_c=-1$, $\eta_c=-1$, $\epsilon_{pq}=1$, $\epsilon_{pc}=1$, $\epsilon_{qc}=-1$ \end{tabular} & $ C_0 $& \Cze  \\
  \hline
PQC7& \begin{tabular}{c} $\epsilon_c=1$, $\eta_c=-1$, $\epsilon_{pq}=-1$, $\epsilon_{pc}=-1$, $\epsilon_{qc}=1$ \\ \hline $\epsilon_c=-1$, $\eta_c=1$, $\epsilon_{pq}=-1$, $\epsilon_{pc}=-1$, $\epsilon_{qc}=-1$ \end{tabular} & $ R_3  $& \Rth \\
  \hline
PQC8& \begin{tabular}{c} $\epsilon_c=1$, $\eta_c=-1$, $\epsilon_{pq}=-1$, $\epsilon_{pc}=-1$, $\epsilon_{qc}=-1$ \\ \hline $\epsilon_c=-1$, $\eta_c=1$, $\epsilon_{pq}=-1$, $\epsilon_{pc}=-1$, $\epsilon_{qc}=1$ \end{tabular} & $  R_1 $& \Ron  \\
  \hline
PQC9a& \begin{tabular}{c} $\epsilon_c=1$, $\eta_c=1$, $\epsilon_{pq}=1$, $\epsilon_{pc}=-1$, $\epsilon_{qc}=1$ \\ \hline $\epsilon_c=-1$, $\eta_c=-1$, $\epsilon_{pq}=1$, $\epsilon_{pc}=-1$, $\epsilon_{qc}=1$  \end{tabular} & $ R_6^2 $& \Rsisi  \\
  \hline
  PQC9b& \begin{tabular}{c}  $\epsilon_c=1$, $\eta_c=1$, $\epsilon_{pq}=1$, $\epsilon_{pc}=-1$, $\epsilon_{qc}=-1$ \\ \hline $\epsilon_c=-1$, $\eta_c=-1$, $\epsilon_{pq}=1$, $\epsilon_{pc}=-1$, $\epsilon_{qc}=-1$ \end{tabular} & $ C_0 $& \Cze  \\
  \hline
PQC10a& \begin{tabular}{c} $\epsilon_c=1$, $\eta_c=1$, $\epsilon_{pq}=-1$, $\epsilon_{pc}=1$, $\epsilon_{qc}=1$ \\ \hline $\epsilon_c=-1$, $\eta_c=1$, $\epsilon_{pq}=-1$, $\epsilon_{pc}=1$, $\epsilon_{qc}=-1$  \end{tabular} &$ R_7 $& \Rse  \\
  \hline
  PQC10b& \begin{tabular}{c}  $\epsilon_c=1$, $\eta_c=1$, $\epsilon_{pq}=-1$, $\epsilon_{pc}=1$, $\epsilon_{qc}=-1$ \\ \hline $\epsilon_c=-1$, $\eta_c=1$, $\epsilon_{pq}=-1$, $\epsilon_{pc}=1$, $\epsilon_{qc}=1$ \end{tabular} & $ R_1 $& \Ron \\
  \hline
PQC11a& \begin{tabular}{c} $\epsilon_c=1$, $\eta_c=-1$, $\epsilon_{pq}=1$, $\epsilon_{pc}=-1$, $\epsilon_{qc}=1$ \\ \hline $\epsilon_c=-1$, $\eta_c=1$, $\epsilon_{pq}=1$, $\epsilon_{pc}=-1$, $\epsilon_{qc}=1$  \end{tabular} &$ R_2^2 $& \Rtwtw  \\
  \hline
  PQC11b& \begin{tabular}{c} $\epsilon_c=1$, $\eta_c=-1$, $\epsilon_{pq}=1$, $\epsilon_{pc}=-1$, $\epsilon_{qc}=-1$ \\ \hline $\epsilon_c=-1$, $\eta_c=1$, $\epsilon_{pq}=1$, $\epsilon_{pc}=-1$, $\epsilon_{qc}=-1$ \end{tabular} & $ C_0 $& \Cze  \\
  \hline
PQC12a& \begin{tabular}{c} $\epsilon_c=1$, $\eta_c=-1$, $\epsilon_{pq}=-1$, $\epsilon_{pc}=1$, $\epsilon_{qc}=1$ \\ \hline $\epsilon_c=-1$, $\eta_c=-1$, $\epsilon_{pq}=-1$, $\epsilon_{pc}=1$, $\epsilon_{qc}=-1$  \end{tabular} &$ R_3 $& \Rth  \\
\hline
  PQC12b& \begin{tabular}{c} $\epsilon_c=1$, $\eta_c=-1$, $\epsilon_{pq}=-1$, $\epsilon_{pc}=1$, $\epsilon_{qc}=-1$ \\ \hline $\epsilon_c=-1$, $\eta_c=-1$, $\epsilon_{pq}=-1$, $\epsilon_{pc}=1$, $\epsilon_{qc}=1$ \end{tabular} & $ R_5 $& \Rfi  \\
  \hline
\end{tabular}
\end{center}
\end{table*}
\clearpage \newpage
\twocolumngrid
\section{Examples of Floquet non-Hermitian topological phases \label{SecFermion}}
To get an intuitive understanding of our topological classification and the novel features of FNH 
topological phases, we demonstrate two explicit examples 
in this section. They correspond to the 
FO angle-gapped and FO angle-gapless cases, respectively. And we extract the physical
 meanings of the topological invariants.

\subsection{FO angle-gapped topology}\label{seciva}
\begin{figure}[t]
\centerline{\includegraphics[width=3.2in]{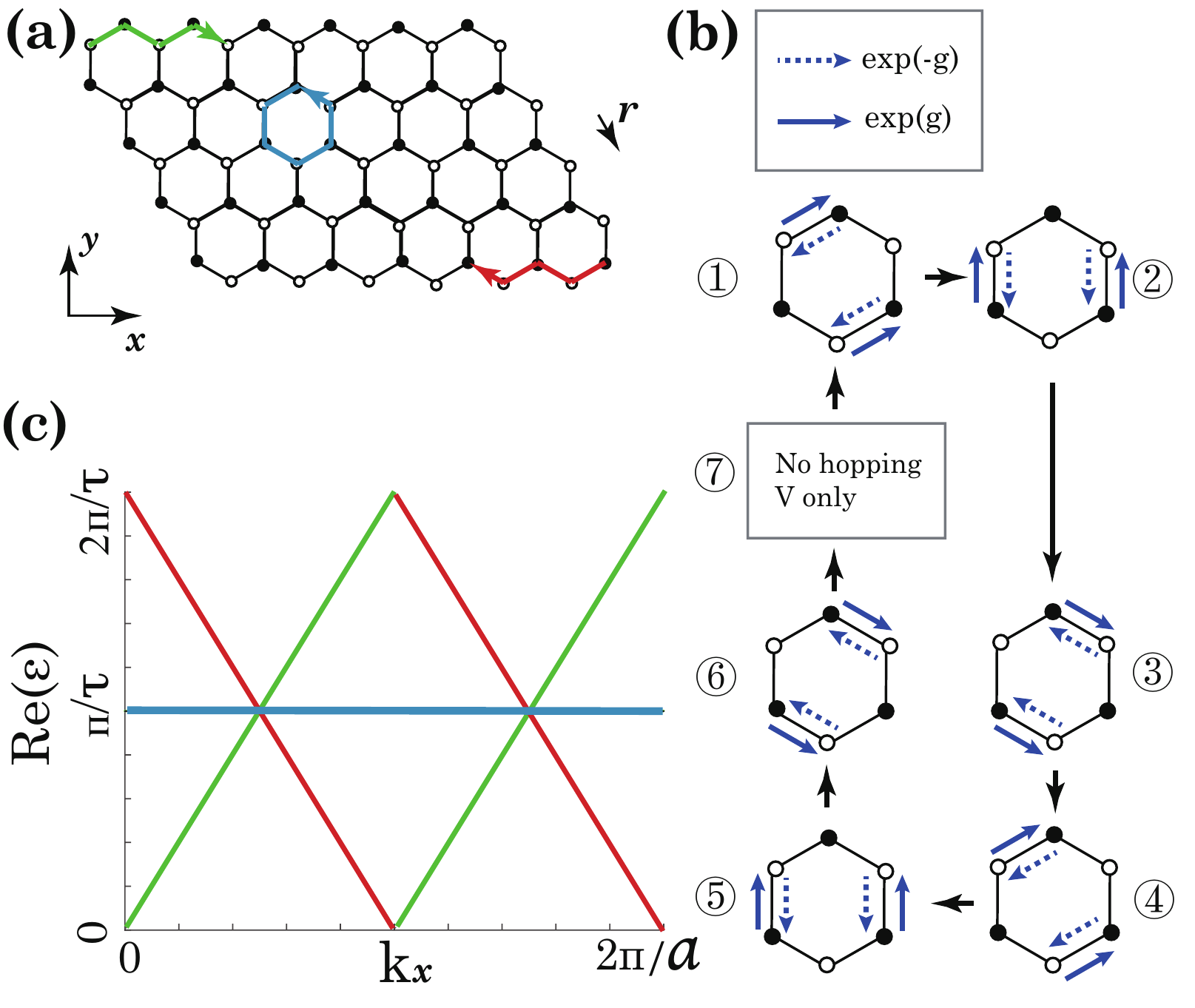}}
\caption{Anomalous non-Hermitian edge modes on a two-dimensional Floquet driven honeycomb lattice. (a) Honeycomb lattice structure. Filled and open circles represent A and B sites, respectively. Red and green arrows depict the trajectories of a particle initially at 
the bottom 
 edge A site and top edge B site in a driving period $t\in[0,\tau]$, respectively. A bulk particle follows the blue trajectory cyclically. (b) Driving protocol. In the first six steps, the spatially homogeneous hopping amplitudes are varied in a chiral way. In each step, only nonreciprocal hoppings along one specific bond are allowed. The dotted blue arrows and solid blue arrows denote 
 hopping amplitudes $e^{-g}$ and $e^g$ , respectively. In the last step, 
 the potential 
 $-V$ and $V$ are applied on the A and B lattice, respectively. (c) Quasienergy spectra (real part). The red and blue 
 bands are for the chiral edge modes at the bottom and top zigzag edges, respectively; the flat bulk bands (blue) are 
 pinned at $\epsilon=\pi/\tau$. The parameters 
 are $V=0$, $\tau=3.5\pi$, and $g=\pi$. \label{F1}}
\end{figure}

Let us consider a periodically driven two-dimensional honeycomb lattice, which contains two sublattices 
(denoted as A and B), as depicted in Fig. \ref{F1}(a). The Floquet dynamics is implemented through a 
seven-step driving sequence, with each step described by a constant Hamiltonian $H_j$ for time-lapse 
$n\tau+\frac{j-1}{7}\tau\leq t<n\tau+\frac{j}{7}\tau$ ($n\in\mathbb{Z}$, $j=1,2,...,7$), $\tau$ is the 
driving period. The driving protocol is schematically shown in Fig. \ref{F1}(b). The first three steps 
involve only nonreciprocal hoppings between the neighbouring A, B sites along three different edges, respectively. Their Hamiltonians are given by
\begin{equation}
  \begin{split}
H_1({\bf k})&=\left[ \begin{array}{cc}
0 & e^{g+i{\bf k}\cdot {\bf a}_1}\\
e^{-g-i{\bf k}\cdot {\bf a}_1} & 0
\end{array}
\right ];  \\
H_2({\bf k})&=\left[ \begin{array}{cc}
  0 & e^{-g+i{\bf k}\cdot {\bf a}_2}\\
  e^{g-i{\bf k}\cdot {\bf a}_2} & 0
  \end{array}
  \right ];\\
H_3({\bf k})&=\left[ \begin{array}{cc}
    0 & e^{-g+i{\bf k}\cdot {\bf a}_3}\\
    e^{g-i{\bf k}\cdot {\bf a}_3} & 0
    \end{array}
  \right ], \label{Hjk4} 
\end{split}
\end{equation}
on the sublattice basis. Here ${\bf a}_1=(-\frac{a}{2},-\frac{a}{2\sqrt{3}})$, ${\bf a}_2=(0
,\frac{a}{\sqrt{3}})$, ${\bf a}_3=(\frac{a}{2},-\frac{a}{2\sqrt{3}})$, and $a=1$ is the lattice constant. $g$ is the
 nonreciprocal coefficient. When $g\neq 0$, the Hamiltonian is non-Hermitian. The next three steps repeat 
 the first three steps, The final step is an onsite chemical 
 potential term, i.e.,
 \begin{equation}
  H_4=H_1, \quad H_5=H_2, \quad H_6=H_3, \quad H_7=-V\sigma_z \label{H4567}
 \end{equation}
 Here $\sigma_z$ is Pauli matrix. This model can be regarded as a 
 non-Hermitian generalization of Kitagawa et al.'s model \cite{kitagawa} and belongs to class Non in the 
 GBL class.

We start from the ideal case with $V=0$, $\tau=3.5\pi$, and consider the motion of a single particle in the geometry depicted in Fig. \ref{F1}(a). The Floquet dynamics are exactly 
 solvable and depicted in Fig. \ref{F1}(a). The bulk Floquet operator is $U({\bf k})=-\mathbb{I}$. An initial particle at the A site in bulk travels along the blue loci cyclically. And the particle returns to its initial position after one whole period. Similar, the particle at the bulk B site also returns to its initial position after one whole period. Thus, the bulk Floquet operator is trivial. The dynamics exhibit different behaviors on the edges. 
First, we consider the top edge. An initial particle located at B site moves along the edge towards the right (green arrow), and after one driving cycle, it passes through two unit cells. 
In comparison, an initial particle located at A site returns to itself. 
Second, we consider the bottom edge. An initial particle at A site moves along the bottom edge towards the left (red arrow) and passes through two unit cells after one driving cycle, while an initial particle at B site returns to itself. The topological edge modes at the top and bottom edge possess different 
imaginary quasienergies, which are also different 
from the imaginary quasienergies of the bulk modes. We can explicitly
 work out the edge dynamics of the ideal case. The effective Floquet Hamiltonians for the four edges are 
 given by (up to $2\pi j/\tau$, $j\in \mathbb{Z}$)
\begin{equation*}
\begin{aligned}
\hat{H}_{T}=\left[ \begin{array}{cc}
  \frac{-\pi}{\tau} & 0\\
0 &   \frac{2{k_x}+4gi}{\tau}
\end{array}
\right];\hat{H}_{B}=\left[\begin{array}{cc}
  -\frac{2{k_x}+4gi}{\tau}  & 0\\
0 &   \frac{-\pi}{\tau}
\end{array}
\right];\label{HUE}
\end{aligned}
\end{equation*}
\begin{equation}
\begin{aligned}
\hat{H}_{L}=\left[ \begin{array}{cc}
  -\frac{2{k_r}}{\tau} & 0\\
0 &   \frac{-\pi}{\tau}
\end{array}\right ];~\hat{H}_{R}=\left[ \begin{array}{cc}
  \frac{-\pi}{\tau}  & 0\\
0 & \frac{2{k_r}}{\tau}
\end{array}
\right ]. \label{HRE}
\end{aligned}
\end{equation}
where $T$, $B$, $L$, and $R$ refer to the top, bottom, left, and right edges, respectively. The $k_x$ and $k_r$ are the momenta along $\bf x$ and $\bf r$ directions labeled in Fig. \ref{F1}(a).
\begin{figure}[t]
\centerline{\includegraphics[width=3.2in]{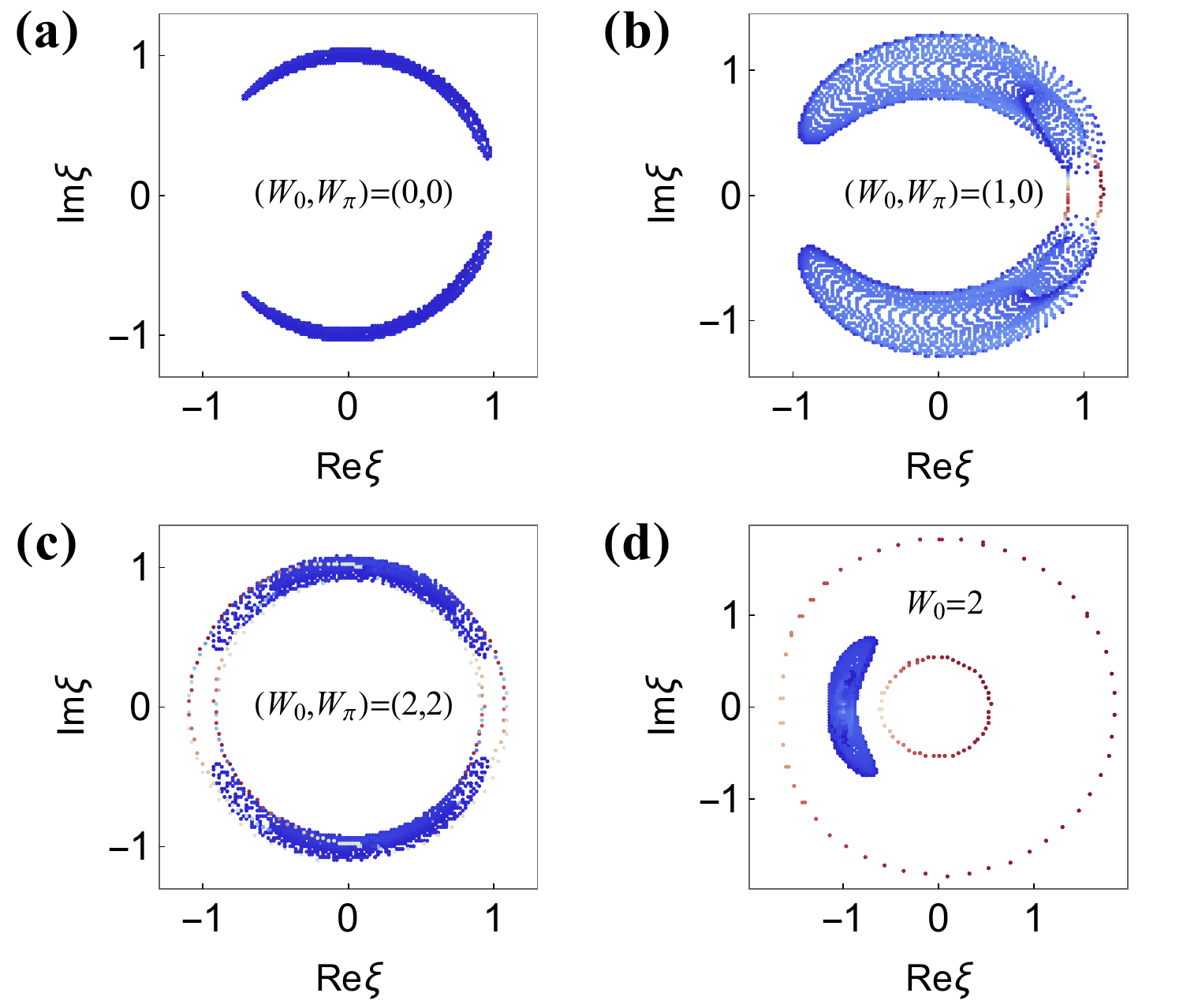}}
\caption{Spectra of the Floquet operator for four topologically distinct FO angle-gapped phases. 
 (a) $g=0.01\pi$, $\tau=\pi$, and $V=0.5\pi$. The $(W_0, W_{\pi})=(0,0)$ phase when both $0$ and $\pi$ FO angle gaps exist. 
  (b)  $g=0.05\pi$, $\tau=\pi$, and $V=0.05\pi$. The $(W_0, W_{\pi})=(1,0)$ phase with both $0$ and $\pi$ FO angle gaps. 
   The colors represent the value of inverse partial ratio (IPR), which is defined as IPR=$(\sum_j|\psi_j|^4)/(\sum_j|\psi_j|^2)$, where $j$ is the site index. The IPR clearly shows the appearance of edge states inside the $0$ FO angle gap.
 (c) $g=0.01\pi$, $\tau=2.5\pi$, and 
 $V=0.5\pi$. The $(W_2, W_{\pi})=(2,2)$ phase with both $0$ and $\pi$ FO angle gaps. This is an anomalous Floquet phase with zero
  Chern number for each Floquet operator band. (d) $g=0.05\pi$, $\tau=3.2\pi$, and $V=0.01\pi$.  The 
 $W_{\pi}=2$ phase is unique to Floquet non-Hermitian system. 
 There is only one FO angle gap at $0$. The edge modes are fully detached from the bulk bands. \label{F2}}
\end{figure}

The quasienergy spectra of the above ideal case are illustrated in Fig. \ref{F1}(c). Besides the flat 
bulk quasienergy bands located at $\pi/\tau$, there exist two sets of chiral 
edge modes shown by the red and green lines. These edge modes describe the chiral motion on 
the top and bottom edges. The appearance of these edge modes in the 
$\pi/\tau$ real gap of the 
quasienergy spectra cannot be understood from the bulk Floquet Hamiltonian itself. These 
chiral edge states are intrinsically dynamical, and the system is in the Floquet anomalous 
topological phase. The dynamical topology can only be revealed from the time-evolution 
operator. According to Table \ref{tab:tableI} for the Non class, the topological 
invariant is $\mathbb{Z}^{\times n}$ with $n$ the number of FO angle gaps. Formally, 
we can define the loop operator from 
$H_{F,\theta}=\frac{i}{\tau}\ln_{-\theta}[U({\bf k})]$ as 
$U_{l,\theta}=U({\bf k},t)*e^{iH_{F,\theta}t}$, here 
$U({\bf k},t)$ is the time evolution operator of the system. The topological invariant defined for the 
loop operator is as follows:
\begin{equation}
\begin{split}
W_{\theta}=&\frac{1}{8\pi^2}\int_0^{2\pi} k_x \int_0^{2\pi} dk_y\int_0^{\tau} dt\times\\
&\textrm{Tr}(U_{l,\theta}^{-1}\partial_{t} U_{l,\theta}
[U_{l,\theta}^{-1}\partial_{k_x} U_{l,\theta},U_{l,\theta}^{-1}\partial_{k_y} U_{l,\theta}]). \label{Wtheta}
\end{split}
\end{equation}
$W_{\theta}$ is the so-called three-winding number in the momentum-time space and directly 
gives the number of chiral edge modes located at the
 FO angle gap $\theta$ (equivalent to FH real gap $\theta/\tau$). For 
 Fig. \ref{F1}(c) case, $W_0=2$, which is consistent with the chiral motion across two unit
   cells in Fig. \ref{F1}(a).
There may exist multiple FO angle gaps when deviating from the ideal case. By tuning the parameters 
$g$, $V$, and $T$ in this model, various FNH topological phases can appear. Let us consider four typical 
examples, with their spectra of the Floquet operator on the complex plane illustrated in Fig. \ref{F2}. 
In Fig. \ref{F2}(a), the spectra exhibit both $0$ and $\pi$ FO angle gaps. There is no topologically nontrivial 
edge state in these FO angle gaps, consistent with their topological invariants $(W_0, W_{\pi})=(0,0)$. By tuning
 parameters, the FO angle gap at $0$ closes and further reopens, accompanied by the emergence of one chiral edge 
 state (at the top edge) at the FO angle gap $0$ as shown in Fig. \ref{F2}(b). For this case, the topological invariants are
  $(W_0,W_{\pi})=(1,0)$. Fig. \ref{F2}(c) depicts an anomalous case when the bulk Floquet Hamiltonian is
   topologically trivial. We can verify the Chern number for each Floquet operator band is zero. However, there 
   exist both edge states inside the $0$ and $\pi$ FO angle gaps, dictated by topological invariants
    $(W_0, W_{\pi})=(2,2)$. In Fig. \ref{F2}(d), we illustrate an unexpected case 
    when there is no FO angle gap at $\pi$, and the Floquet operator possesses a FO angle gap at $0$.
     The topological invariant for 
    the FO angle gap $0$ is $W_{0}=2$. Contrary to the general picture of boundary states connecting bulk bands
     in the two dimensional Hermitian topological phases, the two chiral edge states are fully detached from the bulk. 
     They can be engineered independently without changing the bulk invariants. We emphasize that such a 
     phase is intrinsic to Floquet non-Hermitian system and can be utilized, e.g., in boundary-state 
     engineering in photonic waveguides \cite{fnhano}.

\subsection{FO angle-gapless topology}
\begin{figure}[t]
\centerline{\includegraphics[width=3.3in]{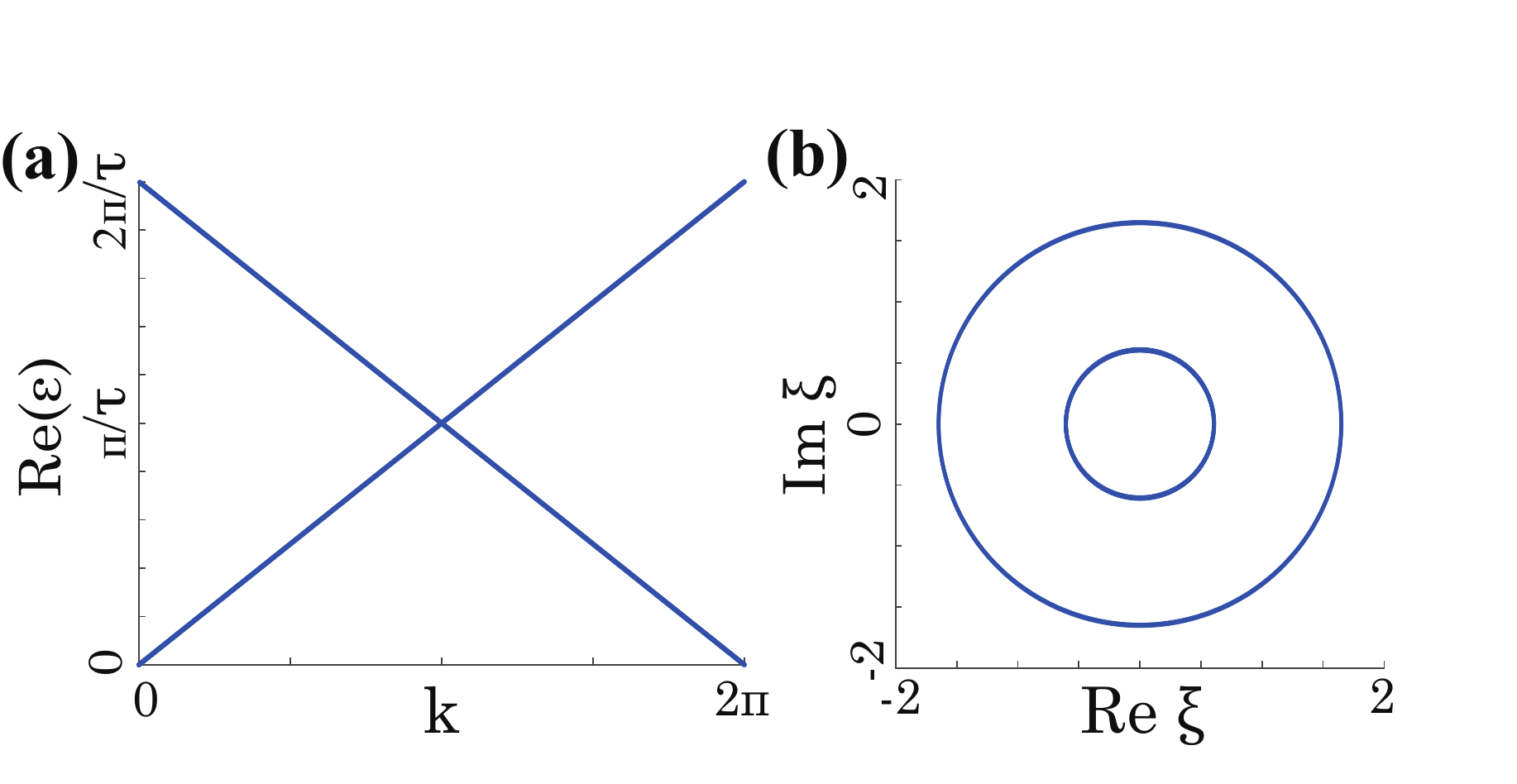}}
\caption{Spectra of the FO angle-gapless system. (a) Quasienergy spectra (real part) with respect to momentum $k$. The two bands with opposite chirality wind around the quasienergy zone. (b) Spectra of the Floquet operator on the complex plane. The inner (outer) circle with winding number $W_{\pm}=\pm1$ encloses the origin counter-clockwise (clockwise) by increasing $k$ from $0$ to $2\pi$. \label{Ffour}}
\end{figure}

When the spectra of the Floquet operator enclose the origin without any FO angle gaps,
 the topology of the system is carried by its Floquet operator. We consider a one-dimensional two-band system with a two-step driving sequence generated by Hamiltonian $\hat{H}_{o1}$ and $\hat{H}_{o2}$. The time-lapse of each step is $\tau/2$. The expressions of $\hat{H}_{o1}$ and $\hat{H}_{o2}$ are
\begin{equation}\begin{aligned}
\hat{H}_{o1}=-\frac{\pi}{\tau} \left[ \begin{array}{cc}
0 & e^{ik}\\
e^{-ik} & 0
\end{array}
\right ];~~ \hat{H}_{o2}=\frac{\pi}{\tau}\left[ \begin{array}{cc}
0 & e^{g}\\
e^{-g} & 0
\end{array}
\right ]. \label{Hhaotj}
\end{aligned}\end{equation}
The Floquet operator $U(k)=e^{-iH_{o2}\tau/2}e^{-iH_{o1}\tau/2}$ is
\begin{equation}
U(k)=\left[ \begin{array}{cc}
e^{-ik+g} & 0\\
0 & e^{ ik- g}
\end{array}
\right ]. \label{Hof}
\end{equation}
This model can be regarded as a non-Hermitian generalization of Budich, Hu, and Zoller's model 
\cite{BudichHuZoller}. The system belongs to the Non class, and Floquet operator $U(k)$ is a reducible
 matrix. The quasienergy spectra of the Floquet Hamiltonian and the spectra of the Floquet operator are 
 depicted in Fig. \ref*{Ffour}(a) and (b), respectively. Due to the absence of the FO angle gap, 
 the Floquet Hamiltonian bands (real part) wind around the quasienergy zone ($[0,2\pi/\tau)$) and possess different imaginary 
 parts (i.e., lifetimes). Such band structures only exist in Floquet driven systems and describe the chiral
  motions of opposite chirality. Each irreducible block ($U_+:=e^{-ik+ g}$ and $U_-:=e^{ ik- g}$) is 
  characterized by a winding number: $W_+=\frac{i}{2\pi}\int_0^{2\pi}dk U_+^{-1}\partial_k U_+=1$ for 
  $U_+$ and $W_-=\frac{i}{2\pi}\int_0^{2\pi}dk U_-^{-1}\partial_k U_-=-1$ for $U_-$. $W_{\pm}$ corresponds to the 
  number of right/left-moving chiral fermions in the Floquet Hamiltonian bands. Similarly, we can consider 
  FO angle-gapless model in other dimensions. The above topological number in the FO angle gapless case gives rise to 
  unidirectional topological charge pumping differs from the physical meaning of angle gapped topological number \cite{fnhpumping}.

\section{Bosonic systems}\label{secv}
Our topological classification can be equally applied to the Floquet driving bosonic systems. Usually, 
the bosonic particle number is not conserved \cite{rmptopophotonics} in realistic experimental settings. Let us formally consider a generic tight-binding bosonic BdG-type Hamiltonian:
\begin{equation}
\hat{H}_b=\sum_{ij;\mu\nu}(h_{i\mu,j\nu}\hat{a}_{i\mu}^\dag \hat{a}_{j\nu}+\frac{1}{2}\Delta_{i\mu,j\nu}\hat{a}_{i\mu}^\dag\hat{a}_{j\nu}^\dag+\frac{1}{2}\Delta_{i\mu,j\nu}^*\hat{a}_{i\mu}\hat{a}_{j\nu}).\label{Hb}
\end{equation}
Here $\hat{a}_{i\mu}^\dag$ ($\hat{a}_{i\mu}$) is the creation (annihilation) operator of bosonic particles. $i$ and $\mu$ label the unit cell and some internal degrees of 
freedom (e.g., spin, sublattice, or orbit), respectively. They satisfy the standard bosonic commutation relation $[\hat{a}_{i\mu},\hat{a}^{\dag}_{j\nu}]=\delta_{ij}\delta_{\mu\nu}$. $\Delta_{i\mu,j\nu}=\Delta_{j\nu,i\mu}$ is the pairing term. We define the two field operators: $\hat{\Psi}= (\hat{a}_{11},...,\hat{a}_{1m},\hat{a}_{21},...,\hat{a}_{2m},...,\hat{a}_{L1},...,\hat{a}_{Lm})^{\bf T}$, and $\bar{\hat{\Psi}}=(\hat{a}_{11}^\dag,...,\hat{a}_{1m}^\dag,\hat{a}_{21}^\dag,...,\hat{a}_{2m}^\dag,...,\hat{a}_{L1}^\dag,...,\hat{a}_{Lm}^\dag)^{\bf T}.$ The Heisenberg equation of motion of the system is
\begin{equation}
\frac{d}{dt}\left(\begin{array}{c}
\hat{\Psi}  \\
\bar{\hat{\Psi}}
\end{array}\right)=-iM(t)\left(\begin{array}{c}
\hat{\Psi}  \\
\bar{\hat{\Psi}}
\end{array}\right), \label{HsbEq}
\end{equation}
where
\begin{equation}
M(t):=
\left(\begin{array}{cc}
h & \Delta  \\
-\Delta^* & -h^{\bf T}
\end{array}\right).\label{mmatrix}
\end{equation}
The single-particle Hamiltonian $h$ can be either Hermitian or non-Hermitian. The dynamics of the bosonic system are fully governed by the $M$ matrices. $M$ plays a similar role as the time-dependent Hamiltonian in the fermionic system. The topological classification of the bosonic system is for the $M(t)$ matrix, and our conclusions of FNH symmetry and topology can be directly applied. Specifically, for a Hermitian $\hat{H}_b$ with BdG pairing, the $M$ matrix fulfills $\tau_x M(t)\tau_x=-M^*(t)$, $\tau_z M(t)\tau_z=M^{\dagger}(t)$ and $\tau_y M(t)\tau_y=-M^{\bf T}(t)$. 
Thus, the $M(t)$ belongs to the GBL class QC8a. According to Table \ref{tab:tableI}, the topological invariant for the bosonic system in 
 even (odd) dimensions is 
$\mathbb{Z}^{n_p}$ ($0$) if the system is FO angle-gapped. 
Here $n_p=1$ if there is only one FO angle gap at $0$ or $\pi$, and $n_p=2$ if there exist both
 $0$ and $\pi$ FO angle gaps. According to Table \ref{tab:tableII}, the topological invariant for the bosonic system in 
 even (odd) dimensions is 
  $0$ ($\mathbb{Z}$) if the system is FO angle-gapless. In the following, we consider two representative examples to illustrate our classifications for the bosonic systems. One is a two-dimensional Floquet bosonic topological superconductor with 
  FO angle gaps, which hosts anomalous Floquet bosonic edge modes without any static counterparts. The other one is a one-dimensional Floquet bosonic system without any FO angle gap, which can exhibit nontrivial spectral windings.
 
\textit{First example: FO angle-gapped bosonic system.} We consider a similar driving protocol as in Fig. \ref{F1}(a) on a bosonic honeycomb lattice, with bosons on neighbouring sites paired together. 
There are seven driven steps, each step determined by driven time $\tau/7$ and 
a static Hamiltonian. Written on the sublattice and BdG basis, the static Hamiltonian for each step is
\begin{equation}
\begin{aligned}
\hat{H}_{bj}=\sum_{\bf k}\left[\left(a_{{\bf k}A}^{\dagger},a_{{\bf k}B}^{\dagger}\right)H_{j}({\bf k})
\left(\begin{array}{c}
a_{{\bf k}A}  \\
a_{{\bf k}B}
\end{array}\right)\right.\\
+\left(a_{{\bf k}A}^{\dagger},a_{{\bf k}B}^{\dagger}\right){\Delta}({\bf k})
\left(\begin{array}{c}
a_{{\bf k}A}^{\dagger}  \\
a_{{\bf k}B}^{\dagger}
\end{array}\right)\\ \left.
+\left(a_{{\bf k}A},a_{{\bf k}B}\right){\Delta}^{\dagger}({\bf k})
\left(\begin{array}{c}
a_{{\bf k}A}  \\
a_{{\bf k}B}
\end{array}\right) \right]. \label{Hhatj}
\end{aligned}
\end{equation}
Here $H_j({\bf k})$ is given by 
 Eqs. (\ref{Hjk4}) and (\ref{H4567}), 
and $g=0$ in these equations. $\Delta({\bf k})=\Delta_0\sigma_0$, and $\Delta_0$ is a complex number. The field operators are denoted by $\hat{\Psi}=(a_{{\bf k} A},a_{{\bf k} B})^{\bf T}$ and $\bar{\hat{\Psi}}=(a_{{\bf k} A}^{\dag},a_{{\bf k} B}^{\dag})^{\bf T}$. When $\Delta_0=0$, 
the model reduces to the bosonic version of the model in Sec. \ref{seciva}. For the generic case with $\Delta_0\ne 0$, the Floquet dynamics is described by the M matrix in Eq. (\ref{mmatrix}).

\begin{figure}[t]
\centerline{\includegraphics[width=3.3in]{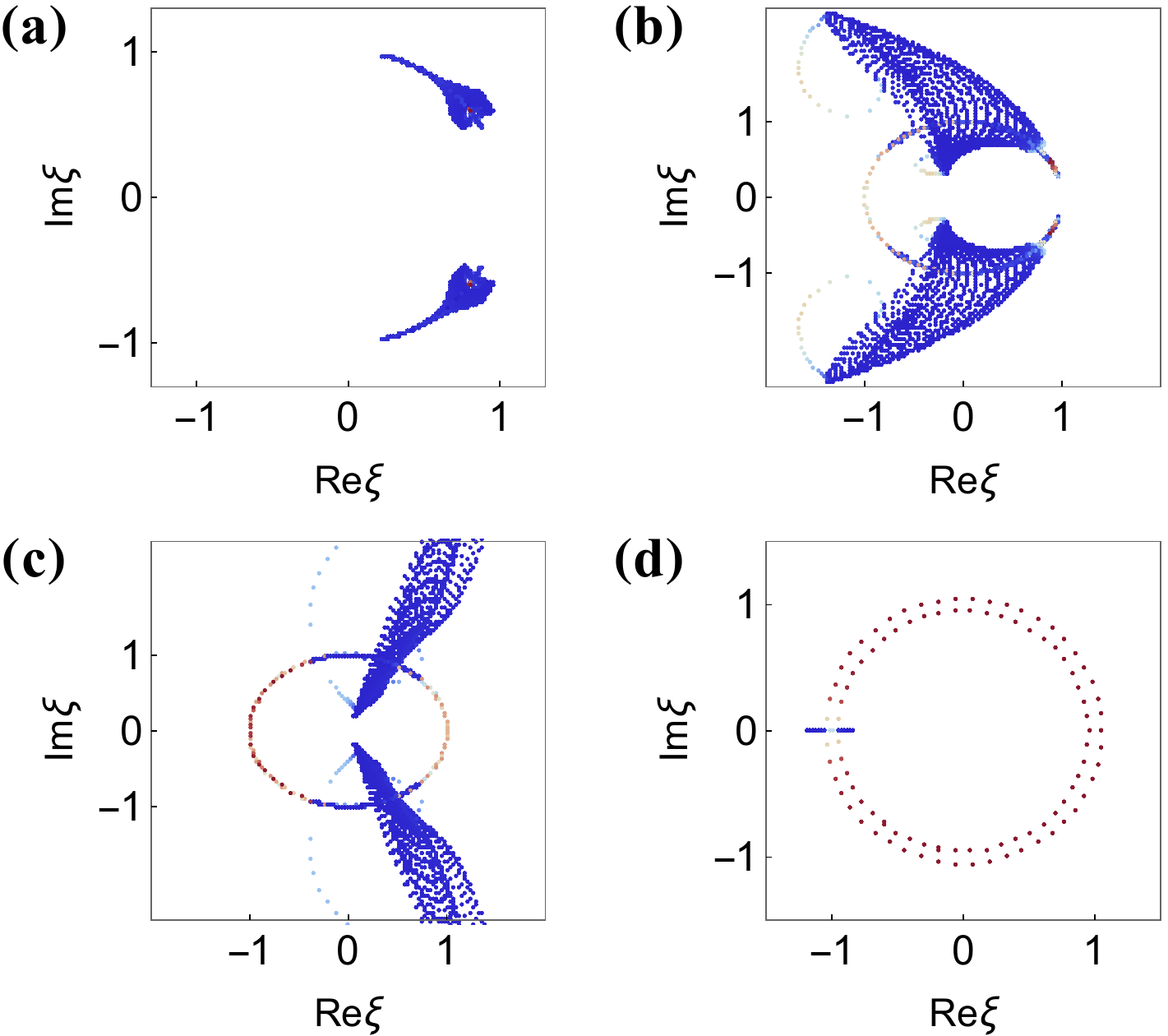}}
\caption{Spectra of the M matrix's Floquet operator 
 for four distinct FO angle-gapped bosonic topological phases. (a) 
 $\Delta_1=0.1\pi$, $\tau=0.5\pi$, and $V=\pi$. 
 The $(W_0,W_{\pi})=(0,0)$ phase when both $0$ and $\pi$ FO angle gaps exist. 
  (b) $\Delta_1=0.1\pi$, $\tau=1.7\pi$, and $V=\pi$. The $(W_0,W_{\pi})=(0,2)$ phase 
 with both $0$ and $\pi$ FO angle gaps.  
 (c) $\Delta_1=0.1\pi$,
 $\tau=3\pi$, and $V=\pi$. The $(W_2,W_{\pi})=(4,4)$ phase with both $0$ and $\pi$ FO angle gaps.  It is an anomalous Floquet phase with zero Chern number 
  for each Floquet operator band. (d) $\Delta_0=0.01\pi$, $\tau=3.5\pi$, and $V=0$. The $W_{0}=4$ phase is unique to Floquet non-Hermitian 
  system. There is only one FO angle gap at
   $0$. The edge modes are detached from the bulk bands. The colors represent the 
   value of IPR, which clearly shows the appearance of 
 edge states. \label{F3}}
\end{figure}
Similar to the fermionic case, we can discuss the spectra and topological invariants of the M matrix. The system exhibits various topological phases by tuning the parameters $\Delta_0$, $V$, and $\tau$. Figs. \ref{F3}(a)-(d) depict 
the spectra of the M matrix's Floquet operator for four typical phases. Their topological invariants $(W_0,W_{\pi})$ are $(0,0)$, $(0,2)$, $(4,4)$, and $(4,\textit{undefined})$, respectively. The third case shown in Fig. \ref{F3}(c) is dynamically anomalous, 
i.e., the bulk Chern number of each Floquet operator band is zero, yet with the appearance of edge states. For the fourth case shown in Fig. \ref{F3}(d), there is only 
one FO angle gap at 0, and the topological invariant $W_{\pi}$ is ill-defined due to the lack of a 
branch cut at the $\pi$ FO angle gap. It is different from the fermionic Non class. Firstly, the 
M matrix's FO angle gaps, if any, must be pinned at 
either $0$ or $\pi$ or pairs at ($\theta_m$, $-\theta_m$), compared to the fermionic Non class where the FO angle gap can be tuned anywhere. 
It is due to the additional symmetry constraints on the M matrix. Secondly, The topological invariants for the bosonic system are doubled due to the existence of BdG pairing.

\textit{Second example: FO angle-gapless bosonic system.} We consider a two equal-step driving sequence described by Hermitian Hamiltonian $H_{bo1}=-\frac{\pi}{\tau}\tau_0 \otimes[\cos(k)\sigma_x- \sin(k)\sigma_y]$, and
\begin{equation} 
H_{bo2}({k})=\frac{\pi}{\tau}(\tau_0\otimes \sigma_x+\Delta_1\tau_x\otimes \sigma_z)
 \label{Hojk4B}
\end{equation}
Here $\tau_0$ is a $2\times 2$ identity matrix, $\tau_{x,y,z}$ are the Pauli matrices in the particle-hole space. $\Delta_1$ is the 
real pairing amplitude. It is easy to check the M matrix of each step for this bosonic system is $M_{boj}=(\tau_z\otimes\sigma_0) H_{Boj}$ ($j=1,2$). 
The Floquet operator of M matrices is $U_{bo}(k,\tau)=e^{-iM_{bo2}\frac{\tau}{2}}e^{-iM_{bo1}\frac{\tau}{2}}=e^{-ik\tau_0\otimes \sigma_z+\frac{\pi\Delta_1}{2}\tau_y\otimes \sigma_z}$. 
The M matrices' Floquet operator is reducible into four irreducible sub-blocks: $U_1=e^{- ik-\pi\Delta_1/2}$, $U_2=e^{- ik+\pi\Delta_1/2}$, $U_3=e^{ik+\pi\Delta_1/2}$, and $U_4=e^{ik-\pi\Delta_1/2}$. The corresponding M matrices' Floquet Hamiltonians for the four blocks are $\frac{1}{\tau}(\pm k\pm i\frac{\pi\Delta_1}{2})$. They describe the chiral bosonic modes, where the pairing term contributes to the lifetime of these modes. For each block, we can define an integer winding number as $W_m=\frac{i}{2\pi }\int_0^{2\pi}dk U_m^{-1}\partial_k U_m$ ($m=1,2,3,4$). It follows that $W_1=W_2=1$ and $W_3=W_4=-1$. The number and chirality of each chiral bosonic mode are given by $|W_m|$ and the sign of $W_m$, respectively.

\section{Conclusions}\label{secsum}
In summary, we have developed a comprehensive classification of FNH topological phases using the $K$-theory 
based on the internal symmetries of the system and the FO angle gaps. We have demonstrated there exist 54 distinct GBL classes
 for time-dependent non-Hermitian Hamiltonians. We have obtained two periodic tables for 
 the FO angle-gapped and FO 
 angle-gapless FNH topological phases, respectively. 
 Our scheme fully covers the previous topological classifications of 
 Floquet Hermitian
  topological insulators and Floquet unitaries. Our classification can also be utilized to 
 characterize the Floquet topological phases of bosonic systems. We have unveiled the physical meanings 
 and consequences of the topological invariants through explicit examples, like the appearance of edge 
 states in the FO angle gaps and charge transport.

Our topological classification of the FNH topological 
 phase is based on the periodic boundary condition or Bloch Hamiltonian. 
According to the previous study of the topological classification of static 
non-Hermitian systems, the conclusions of topological classification 
don't change when the periodic boundary condition is transformed into 
the open boundary condition. The topological phase transition 
points may change due to the non-Hermitian skin effect. 
To recover the topological phase transition 
points, we can use non-Bloch band theory \cite{ne1}, biorthogonal bulk-boundary correspondence \cite{ne2}, 
or real space topological number \cite{RealSpace}. Similarly, 
we think our results of topological classification don't change 
when the periodic boundary condition is transformed into 
the open boundary condition. We may need to use 
non-Bloch band theory, biorthogonal bulk-boundary correspondence, 
or real space topological number to define the topological number in the open boundary condition 
when there is the non-Hermitian skin effect.   

Our framework, together with the previous AZ tenfold way of static Hermitian topological 
phases \cite{AltlandZirnbauer,Chiu,ChiuStone,ChiuYaoRyu,MorimotoFurusaki,Kitaev,Ludwig}, 
Roy and Harper's periodic table of Floquet Hermitian topological insulators, 
\cite{fclass2} and Higashikawa, Nakagawa, and Ueda's periodic table 
of Floquet unitaries
\cite{HigashikawaNakagawaUeda} 
 and the most recent classifications of static non-Hermitian Hamiltonians 
 \cite{GAKTHU,KawabataShiozakiUedaSato,LeeZhou,LiuChen,LiuJiangChen,colladd3}, 
 hence completes the whole classification map based on the internal symmetries. 
 Our scheme may be extended to include other types of symmetries, 
 e.g., crystalline symmetries \cite{LiuJiangChen,MorimotoFurusaki,ChiuYaoRyu,KBWKS}, and to higher-order
  Floquet topological phases \cite{fhotihu,fhoti1,fhoti2,fhoti3,fhoti4,fhoti5}.
  Our scheme should be readily extended to include defects by changing $H({\bf k},t)$ to 
  $H({\bf k},{\bf r},t)$ \cite{LiuChen}, and include Floquet dislocation-induced 
  non-Hermitian skin effect by discussing the wave 
  function of $H_{F}$ \cite{SchindlerPrem,SunZhuHughs}. Beyond its immediate 
  significance for understanding the various topological phases unique to Floquet and non-Hermitian
  systems, our scheme should open a broad avenue to explore the novel dynamical phenomena and 
  topological effects originating from the interplay of non-Hermiticy, topology, and 
  Floquet engineering and further guide the experimental design and application in atomic and photonic systems.

\begin{acknowledgments}
The work is supported by the  NSFC under Grants No.12174436 and No.11974413 and the Strategic Priority Research Program of Chinese Academy of Sciences under Grant No. XDB33000000.
\end{acknowledgments}
\appendix

\section{A full counting of the GBL classes} \label{ap0}
In this section, we give the details of the full count of the GBL classes.
We need to count all possible group structures generated by Eqs. (\ref{eq:symshpr})-(\ref{eq:symshkr}).
From that the Hamiltonian is irreducible, we can get that  
$p^2=\mathbb{I}, q^2=\mathbb{I}, cc^*=\pm\mathbb{I},$ and $kk^*=\pm\mathbb{I}$.
From the P, Q, C, and K commute with each other, we can get Eq. (\ref{cr}).
And we can prove the following three propositions,

{\it Proposition 1:} If there are P and Q symmetry, we can define another $Q$-type symmetry
 labeled as $Q'$ with $\epsilon_q=-\epsilon_{q'}$. 

{\it Proposition 2:} If there are P and C symmetry, we can define another $C$-type symmetry
 labeled as $C'$ with $\epsilon_c=-\epsilon_{c'}$. 

{\it Proposition 3:} If there are P and K symmetry, we can define another $K$-type symmetry
 labeled as $K'$ with $\epsilon_k=-\epsilon_{k'}$. 

 The proof of Proposition 1: 
 The system has Q and P symmetries, then it satisfies Eqs. (\ref{eq:symshqr}) and (\ref{eq:symshkr}).
Then we can get that,
\begin{equation}
  \begin{split}
  H({\bf k},t)=&\epsilon_qqH^\dagger({\bf k},\epsilon_q t) q^{-1}\\
  =&\epsilon_qq[-pH({\bf k},-\epsilon_q t)p^{-1}]^\dagger q^{-1}\\
  =&-\epsilon_qqp H^\dagger ({\bf k},-\epsilon_q t)(qp)^{-1}
  \label{Qtilde}
  \end{split}
  \end{equation}
  According to Eq. (\ref{Qtilde}), we can define another type-Q symmetry,
   and we label it as $Q'$ and
  $\epsilon_{q'}=-\epsilon_q, q'=\sqrt{\epsilon_{pq}}qp$.
  Thus, we get the proof of Propositions 1. 
  Similarly, we can prove Propositions 2 and 3.

  According to Propositions 1, 2, and 3, if the system has type-P symmetry, 
  we can fix $\epsilon_q=\epsilon_c=\epsilon_k=1$. To count all possible group structures,
 we should consider all possible situations:

 (1) When there is no symmetry, the number of group structures is 1.

 (2) When there is only type-P symmetry, the number of group structures is 1.

 (3) When there are only type-P and type-Q symmetries, we can fix 
 $\epsilon_q=1$. $\epsilon_{pq}=\pm 1$, the number of group structures is $2^1=2$.

 (4) When there are only type-P, type-Q, and type C symmetries, we can fix 
 $\epsilon_q=\epsilon_c=1$. $\eta_c,\epsilon_{pq},\epsilon_{pc},\epsilon_{qc}=\pm 1$,
  the number of group structure is $2^4=16$.

  Similarly, when there is only type-Q symmetry, the number of group structures is 2.
   When there is only type-C symmetry, the number of group structures is 4.
   When there is only type-K symmetry, the number of group structures is 4.
   When there are only type-P and type-C symmetries, the number of group structures is $4$.
   When there are only type-P and type-K symmetries, the number of group structures is $4$.
   When there are only type-C and type-Q symmetries, the number of group structures is $16$. 
   We summarize these conclusions in Table \ref{table54class}. 
   Adding all numbers of group structures of different situations, 
   $1+1+2+4+4+2+4+4+16+16=54$. Thus, the total number of group structures is $54$.
   \begin{widetext}
   \begin{center}
    \begin{table}[!htbp]
      \centering
      \caption{\label{table54class} Counting all possible group structures generated by Eqs. 
      (\ref{eq:symshpr})-(\ref{eq:symshkr}).}
      \footnotesize
      \setlength{\tabcolsep}{4pt}
      \renewcommand{\arraystretch}{1.2}
      \begin{tabular}{lccc}
        \hline
       System's symmetries \;\;\;&\;\;\;The variable can be $\pm 1$\;\;\;&Number of group structure\\
      \hline
     Non&&$1$\\
    \hline
    P&&$1$\\
      \hline
    Q&$\epsilon_{q}$&$2^1=2$\\
      \hline
    C&$\epsilon_{c},\eta_{c}$&$2^2=4$\\
      \hline
    K&$\epsilon_{k},\eta_{k}$&$2^2=4$\\
      \hline
    P,Q&$\epsilon_{pq}$&$2^1=2$\\
      \hline
    P,C&$\eta_{c},\epsilon_{pc}$&$2^2=4$\\
      \hline
    P,K&$\eta_{k},\epsilon_{pk}$&$2^2=4$\\
      \hline
    Q,C&$\epsilon_{q},\epsilon_{c},\eta_{c},\epsilon_{qc}$&$2^4=16$\\
    \hline
    P,Q,C&$\eta_{c},\epsilon_{pc},\epsilon_{pk},\epsilon_{qc}$&$2^4=16$\\
      \hline
      \end{tabular}
  \end{table}
 \end{center}
  \newpage
\section{Derivation of Eqs. (\ref{eq:symsup})-(\ref{eq:symsuk}) from Eqs. (\ref{eq:symshp})-(\ref{eq:symshk})} \label{apa}
In the following, we derive the symmetry transformations on the time-evolution operator $U({\bf k}, t)$ [Eqs. (\ref{eq:symsup})-(\ref{eq:symsuk})] from the symmetry transformations on the time-dependent Hamiltonian $H({\bf k}, t)$ [Eqs. (\ref{eq:symshp})-(\ref{eq:symshk})].

1.a. From $H({\bf k},t)=kH^*(-{\bf k}, -t)k^{-1}$ to $U^*(-{\bf k},-t)=k^{-1}U({\bf k}, t)k$.
\begin{equation}
\begin{aligned}
k^{-1}U({\bf k},t)k=&[1-i\Delta t k^{-1}H({\bf k},t)k][1-i\Delta t k^{-1}H({\bf k},t-\Delta t)k]...[1-i\Delta t k^{-1}H({\bf k},\Delta t)k]\\
=&[1-i\Delta t H^*(-{\bf k},-t)][1-i\Delta t H^*(-{\bf k},-t+\Delta t)]...[1-i\Delta t H^*(-{\bf k},-\Delta t)] \\
=&\left\{[1-i\Delta t H(-{\bf k},-\Delta t)]...[1-i\Delta t H(-{\bf k},-t+\Delta t)][1-i\Delta t H(-{\bf k},-t)]\right\}^{-1*}\\
=&[U^*(-{\bf k},0,-t)]^{-1} \\
=&U^*(-{\bf k},-t).
\label{Kp}
\end{aligned}
\end{equation}

1.b. From $H({\bf k},t)=-kH^*(-{\bf k}, t)k^{-1}$ to $U^*(-{\bf k},t)=k^{-1}U({\bf k}, t)k$.
\begin{equation}
\begin{aligned}
k^{-1}U({\bf k},t)k=&[1-i\Delta t k^{-1}H({\bf k},t)k][1-i\Delta t k^{-1}H({\bf k},t-\Delta t)k]...[1-i\Delta t k^{-1}H({\bf k},\Delta t)k]\\
=&[1+i\Delta t H^*(-{\bf k},t)][1+i\Delta t H^*(-{\bf k},t-\Delta t)]...[1+i\Delta t H^*(-{\bf k},\Delta t)] \\
=&\left\{[1-i\Delta t H(-{\bf k},t)][1-i\Delta t H(-{\bf k},t-\Delta t)]...[1-i\Delta t H(-{\bf k},\Delta t)]\right\}^{*}\\
=&U^*(-{\bf k},t).
\label{Km}
\end{aligned}
\end{equation}

2.a. From $H({\bf k},t)=qH^{\dagger}({\bf k}, t)q^{-1}$ to $[U^{\dagger}({\bf k},t)]^{-1}=q^{-1}U({\bf k}, t)q$.
\begin{equation}
\begin{aligned}
q^{-1}U({\bf k},t)q=&[1-i\Delta t q^{-1}H({\bf k},t)q][1-i\Delta t q^{-1}H({\bf k},t-\Delta t)q]...[1-i\Delta t q^{-1}H({\bf k},\Delta t)q]\\
=&[1-i\Delta t H^{\dagger}({\bf k},t)][1-i\Delta t H^{\dagger}({\bf k},t-\Delta t)]...[1-i\Delta t H^{\dagger}({\bf k},\Delta t)] \\
=&\left\{[1-i\Delta t H({\bf k},t)][1-i\Delta t H({\bf k},t-\Delta t)]...[1-i\Delta t H({\bf k},\Delta t)]\right\}^{\dagger-1}\\
=&[U^{\dagger}({\bf k},t)]^{-1}.
\label{Qp}
\end{aligned}
\end{equation}

2.b. From $H({\bf k},t)=- qH^{\dagger}({\bf k}, -t)q^{-1}$ to $[U^{\dagger}({\bf k},-t)]^{-1}=q^{-1}U({\bf k}, t)q$.
\begin{equation}
\begin{aligned}
q^{-1}U({\bf k},t)q=&[1-i\Delta t q^{-1}H({\bf k},t)q][1-i\Delta t q^{-1}H({\bf k},t-\Delta t)q]...[1-i\Delta t q^{-1}H({\bf k},\Delta t)q]\\
=&[1+i\Delta t H^{\dagger}({\bf k},-t)][1+i\Delta t H^{\dagger}({\bf k},-t+\Delta t)]...[1+i\Delta t H^{\dagger}({\bf k},-\Delta t)] \\
=&\left\{[1-i\Delta t H({\bf k},-\Delta t)]...[1-i\Delta t H({\bf k},-t+\Delta t)]  [1-i\Delta t H({\bf k},-t)]\right\}^{\dagger}\\
=&[U({\bf k},0,-t)]^{\dagger} \\
=&[U^{\dagger}({\bf k},-t)]^{-1}.
\label{Qm}
\end{aligned}
\end{equation}

3.a. From $H({\bf k},t)=- cH^{\bf T}(-{\bf k}, t)c^{-1}$ to $[U^{\bf T}(-{\bf k},t)]^{-1}=c^{-1}U({\bf k}, t)c$.
\begin{equation}
\begin{aligned}
c^{-1}U({\bf k},t)c=&[1-i\Delta t c^{-1}H({\bf k},t)c][1-i\Delta t c^{-1}H({\bf k},t-\Delta t)c]...[1-i\Delta t c^{-1}H({\bf k},\Delta t)c]\\
=&[1+i\Delta t H^{\bf T}(-{\bf k},t)][1+i\Delta t H^{\bf T}(-{\bf k},t-\Delta t)]...[1+i\Delta t H^{\bf T}(-{\bf k},\Delta t)] \\
=&\left\{[1-i\Delta t H(-{\bf k},t)][1-i\Delta t H(-{\bf k},t-\Delta t)]...[1-i\Delta t H(-{\bf k},\Delta t)]\right\}^{-1{\bf T}}\\
=&[U^{\bf T}(-{\bf k},t)]^{-1}.
\label{Cm}
\end{aligned}
\end{equation}

3.b. From $H({\bf k},t)=cH^{\bf T}(-{\bf k}, -t)c^{-1}$ to $[U^{\bf T}(-{\bf k},-t)]^{-1}=c^{-1}U({\bf k}, t)c$.
\begin{equation}
\begin{aligned}
c^{-1}U({\bf k},t)c=&[1-i\Delta t c^{-1}H({\bf k},t)c][1-i\Delta t c{-1}H({\bf k},t-\Delta t)c]...[1-i\Delta t c^{-1}H({\bf k},\Delta t)c]\\
=&[1-i\Delta t H^{\bf T}(-{\bf k},-t)][1-i\Delta t H^{\bf T}(-{\bf k},-t+\Delta t)]...[1-i\Delta t H^{\bf T}(-{\bf k},-\Delta t)] \\
=&\left\{[1-i\Delta t H(-{\bf k},-\Delta t)]...[1-i\Delta t H(-{\bf k},-t+\Delta t)][1-i\Delta t H(-{\bf k},-t)]\right\}^{\bf T}\\
=&[U^{\bf T}(-{\bf k},0,-t)]\\
=&[U^{\bf T}(-{\bf k},-t)]^{-1}.
\label{Cp}
\end{aligned}
\end{equation}

4. From $H({\bf k},t)=-pH({\bf k}, -t)p^{-1}$ to $U({\bf k},-t)=p^{-1}U({\bf k}, t)p$.
\begin{equation}
\begin{aligned}
p^{-1}U({\bf k},t)p=&[1-i\Delta t p^{-1}H({\bf k},t)p][1-i\Delta t p^{-1}H({\bf k},t-\Delta t)p]...[1-i\Delta t p^{-1}H({\bf k},\Delta t)p]\\
=&[1+i\Delta t H({\bf k},-t)][1+i\Delta t H({\bf k},-t+\Delta t)]...[1+i\Delta t H({\bf k},-\Delta t)] \\
=&\left\{[1-i\Delta t H({\bf k},-\Delta t)]...[1-i\Delta t H({\bf k},-t+\Delta t)][1-i\Delta t H({\bf k},-t)]\right\}^{-1}\\
=&[U({\bf k},0,-t)]^{-1} \\
=&U({\bf k},-t).
\label{P}
\end{aligned}
\end{equation}
\end{widetext}

\section{Derivation of Eqs. (\ref{eq:symsutp})-(\ref{eq:symsutk}) from Eqs. (\ref{eq:symsup})-(\ref{eq:symsuk})} \label{apb}
First, we need to show:

\begin{lemma}
  $U({\bf k},\epsilon \tau)=[U({\bf k})]^{\epsilon}$, where $\epsilon=\pm 1$.
\end{lemma}

Proof: $\epsilon=1$ is obvious. If $\epsilon=-1$. We have,
 \begin{equation}
\begin{aligned}
U({\bf k},-\tau)=&U({\bf k},-\tau,0) \\
=&U({\bf k},0,\tau) \\
=&[U({\bf k},\tau,0)]^{-1} \\
=&[U({\bf k})]^{-1}.\\
\end{aligned}
\end{equation}
1. From $U^*(-{\bf k},-t)=k^{-1}U({\bf k},\epsilon_k t)k$ to $[U^*(-{\bf k})]^{-\epsilon_k}=k^{-1}U({\bf k})k$. First, we have $U^*(-{\bf k},-\tau)=k^{-1}U({\bf k},\epsilon_k \tau)k$. Thus,
\begin{equation}
\begin{aligned}
&[U^*(-{\bf k})]^{-\epsilon_k} \\
=&[U^*(-{\bf k},-\tau)]^{\epsilon_k}\\
=&[k^{-1}U({\bf k},\epsilon_k \tau)k ]^{\epsilon_k}\\
=&\left\{k^{-1}[U({\bf k})]^{\epsilon_k }k\right\}^{\epsilon_k} \\
=&k^{-1}U({\bf k})k.
\end{aligned}
\end{equation}

2. From $[U^{\dagger}({\bf k},t)]^{-1}=q^{-1}U({\bf k},\epsilon_q t) q$ to $[U^{\dagger}({\bf k})]^{-\epsilon_q}=q^{-1}U({\bf k}) q$. First, we have $[U^{\dagger}({\bf k},\tau)]^{-1}=q^{-1}U({\bf k},\epsilon_q \tau) q$. Thus,
\begin{equation}
\begin{aligned}
&[U^{\dagger}({\bf k})]^{-\epsilon_q} \\
=&[U^{\dagger}({\bf k},\tau)]^{-\epsilon_q} \\
=&[q^{-1}U({\bf k},\epsilon_q \tau) q]^{\epsilon_q}\\
=&\left\{q^{-1}[U({\bf k},\tau)]^{\epsilon_q } q\right\}^{\epsilon_q} \\
=&q^{-1}U({\bf k},\tau) q.
\end{aligned}
\end{equation}

3. From $[U^{\bf T}(-{\bf k},t)]^{-1}=c^{-1}U({\bf k},-\epsilon_c t)c$ to $[U^{\bf T}(-{\bf k})]^{\epsilon_c}=c^{-1}U({\bf k})c$. First, we have $[U^{\bf T}(-{\bf k},\tau)]^{-1}=c^{-1}U({\bf k},-\epsilon_c \tau)c$. Thus,
\begin{equation}
\begin{aligned}
&[U^{\bf T}(-{\bf k})]^{\epsilon_c} \\
=&[U^{\bf T}(-{\bf k},\tau)]^{\epsilon_c} \\
=&[c^{-1}U({\bf k},-\epsilon_c \tau)c]^{-\epsilon_c}\\
=&\left\{c^{-1}[U({\bf k},\tau)]^{-\epsilon_c }c \right\}^{-\epsilon_c} \\
=&c^{-1}U({\bf k},\tau)c.
\end{aligned}
\end{equation}

4. From $U({\bf k},-t)=p^{-1}U({\bf k},t)p$ to $[U({\bf k})]^{-1}=p^{-1}U({\bf k})p$. First, we have $U({\bf k},-\tau)=p^{-1}U({\bf k},\tau)p$. Thus,
\begin{equation}
\begin{aligned}
&[U({\bf k})]^{-1} \\
=&U({\bf k},-\tau) \\
=&p^{-1}U({\bf k},\tau)p\\
=&p^{-1}U({\bf k})p.
\end{aligned}
\end{equation}

\section{Proof of Lemma 1}\label{apc}
Lemma 1 states that the composition of two time-evolution generated by Hamiltonian $H_1({\bf k},t)$ and $H_2({\bf k},t)$ does not alter the underlying symmetry class. Here we consider all the four symmetries $K, Q, C, P$ and denote the composition as $H({\bf k},t)=H_1*H_2$.

1. $K$ symmetry with $\epsilon_k=1$. We have $H_1({\bf k},t)=kH_1^*(-{\bf k},-t)k^{-1}$ and $H_2({\bf k},t)=kH_2^*(-{\bf k},- t)k^{-1}$. We need to derive $H({\bf k},t)=kH^*(-{\bf k},- t)k^{-1}$. The conditions are equivalent to $H_1(-{\bf k},-t)=kH_1^*({\bf k},t)k^{-1}$ and $H_2(-{\bf k},-t)=kH_2^*({\bf k},t)k^{-1}$.

1a. When $0\le t\le \tau/4$,
\begin{equation}
\begin{aligned}
&kH^*({\bf k},t)k^{-1} \\
=&kH_2^*({\bf k},2t)k^{-1} \\
=&H_2(-{\bf k},-2t),
\end{aligned}
\end{equation}
\begin{equation}
\begin{aligned}
&H(-{\bf k},-t) \\
=&H(-{\bf k},\tau-t) \\
=&H_2(-{\bf k},2\tau-2t-\tau)\\
=&H_2(-{\bf k},-2t),\\
\end{aligned}
\end{equation}
Thus, $kH^*({\bf k},t)k^{-1}=H(-{\bf k},-t)$.

1b. When $\tau/4< t< 3\tau/4$,
\begin{equation}
\begin{aligned}
&kH^*({\bf k},t)k^{-1} \\
=&kH_1^*({\bf k},2t-\tau/2)k^{-1} \\
=&H_1(-{\bf k},\tau/2-2t),
\end{aligned}
\end{equation}
\begin{equation}
\begin{aligned}
&H(-{\bf k},-t) \\
=&H(-{\bf k},\tau-t) \\
=&H_1(-{\bf k},2\tau-2t-\tau/2)\\
=&H_1(-{\bf k},\tau/2-2t),\\
\end{aligned}
\end{equation}
Thus, $kH^*({\bf k},t)k^{-1}=H(-{\bf k},-t)$.

1c. When $3\tau/4\le t\le \tau$,
\begin{equation}
\begin{aligned}
&kH^*({\bf k},t)k^{-1} \\
=&kH_2^*({\bf k},2t-\tau)k^{-1} \\
=&H_2(-{\bf k},\tau-2t),
\end{aligned}
\end{equation}
\begin{equation}
\begin{aligned}
&H(-{\bf k},-t) \\
=&H(-{\bf k},\tau-t) \\
=&H_2(-{\bf k},2\tau-2t)\\
=&H_2(-{\bf k},\tau-2t),\\
\end{aligned}
\end{equation}
Thus, $kH^*({\bf k},t)k^{-1}=H(-{\bf k},-t)$

2. $K$ symmetry with $\epsilon_k=-1$. From $H_1({\bf k},t)=-kH_1^*(-{\bf k},t)k^{-1}$ and $H_2({\bf k},t)=-kH_2^*(-{\bf k},t)k^{-1}$ to $H({\bf k},t)=-kH^*(-{\bf k}, t)k^{-1}$.

2a. When $0\le t\le \tau/4$,
\begin{equation}
\begin{aligned}
&-kH^*(-{\bf k},t)k^{-1} \\
=&-kH_2^*(-{\bf k},2t)k^{-1} \\
=&H_2({\bf k},2t) \\
=&H({\bf k},t),\\
\end{aligned}
\end{equation}
Thus, $-kH^*(-{\bf k},t)k^{-1}=H({\bf k},t)$.

2b. When $\tau/4< t< 3\tau/4$,
  \begin{equation}
\begin{aligned}
&-kH^*(-{\bf k},t)k^{-1} \\
=&-kH_1^*(-{\bf k},2t-\tau/2)k^{-1} \\
=&H_1({\bf k},2t-\tau/2) \\
=&H({\bf k},t), \\
\end{aligned}
\end{equation}
Thus, $-kH^*(-{\bf k},t)k^{-1}=H({\bf k},t)$.

2c. When $3\tau/4\le t\le \tau$,
\begin{equation}
\begin{aligned}
&-kH^*(-{\bf k},t)k^{-1} \\
=&-kH_2^*(-{\bf k},2t-\tau)k^{-1} \\
=&H_2({\bf k},2t-\tau) \\
=&H({\bf k},t), \\
\end{aligned}
\end{equation}
Thus, $-kH^*(-{\bf k},t)k^{-1}=H({\bf k},t)$.

3. $Q$ symmetry with $\epsilon_q=-1$. From $H_1({\bf k},t)=-qH_1^\dagger({\bf k},- t) q^{-1}$ and $H_2({\bf k},t)=-qH_2^\dagger({\bf k},- t) q^{-1}$ to $H({\bf k},t)=-qH^\dagger({\bf k},- t) q^{-1}$. It is equivalent to  ``From $H_1({\bf k},-t)=-qH_1^\dagger({\bf k},t) q^{-1}$ and $H_2({\bf k},-t)=-qH_2^\dagger({\bf k},t) q^{-1}$ to $H({\bf k},-t)=-qH^\dagger({\bf k},t) q^{-1}$".

3a. When $0\le t\le \tau/4$,
  \begin{equation}
\begin{aligned}
&-qH^\dagger({\bf k},t) q^{-1} \\
=&-qH_2^\dagger({\bf k},2t) q^{-1} \\
=&H_2({\bf k},-2t),
\end{aligned}
\end{equation}
\begin{equation}
\begin{aligned}
&H({\bf k},-t) \\
=&H({\bf k},\tau-t) \\
=&H_2({\bf k},2\tau-2t-\tau)\\
=&H_2({\bf k},-2t),\\
\end{aligned}
\end{equation}
Thus, $H({\bf k},-t)=-qH^\dagger({\bf k},t) q^{-1}$.

3b. When $\tau/4< t< 3\tau/4$,
\begin{equation}
\begin{aligned}
&-qH^\dagger({\bf k},t) q^{-1} \\
=&-qH_1^\dagger({\bf k},2t-\tau/2) q^{-1} \\
=&H_1({\bf k},\tau/2-2t),
\end{aligned}
\end{equation}
\begin{equation}
\begin{aligned}
&H({\bf k},-t) \\
=&H({\bf k},\tau-t) \\
=&H_1({\bf k},2\tau-2t-\tau/2)\\
=&H_1({\bf k},\tau/2-2t),\\
\end{aligned}
\end{equation}
Thus, $H({\bf k},-t)=-qH^\dagger({\bf k},t) q^{-1}$.

3c. When $3\tau/4\le t\le \tau$,
\begin{equation}
\begin{aligned}
&-qH^\dagger({\bf k},t) q^{-1} \\
=&-qH_2^\dagger({\bf k},2t-\tau) q^{-1} \\
=&H_2({\bf k},\tau-2t),
\end{aligned}
\end{equation}
\begin{equation}
\begin{aligned}
&H({\bf k},-t) \\
=&H({\bf k},\tau-t) \\
=&H_2({\bf k},2\tau-2t-\tau)\\
=&H_2({\bf k},\tau-2t),\\
\end{aligned}
\end{equation}
Thus, $H({\bf k},-t)=-qH^\dagger({\bf k},t) q^{-1}$.

4. $Q$ symmetry with $\epsilon_q=1$. From $H_1({\bf k},t)=qH_1^\dagger({\bf k},t) q^{-1}$ and $H_2({\bf k},t)=qH_2^\dagger({\bf k},t) q^{-1}$ to $H({\bf k},t)=qH^\dagger({\bf k},t) q^{-1}$.

4a. When $0\le t\le \tau/4$,
\begin{equation}
\begin{aligned}
&qH^\dagger({\bf k},t) q^{-1} \\
=&qH_2^\dagger({\bf k},2t) q^{-1} \\
=&H_2({\bf k},2t) \\
=&H({\bf k},t), \\
\end{aligned}
\end{equation}
Thus, $H({\bf k},t)=qH^\dagger({\bf k},t) q^{-1}$.

4b. When $\tau/4< t< 3\tau/4$,
\begin{equation}
\begin{aligned}
&qH^\dagger({\bf k},t) q^{-1} \\
=&qH_1^\dagger({\bf k},2t-\tau/2) q^{-1} \\
=&H_1({\bf k},2t-\tau/2) \\
=&H({\bf k},t),\\
\end{aligned}
\end{equation}
Thus, $H({\bf k},t)=qH^\dagger({\bf k},t) q^{-1}$.

4c. When $3\tau/4\le t\le \tau$,
\begin{equation}
\begin{aligned}
&qH^\dagger({\bf k},t) q^{-1} \\
=&qH_2^\dagger({\bf k},2t-\tau) q^{-1} \\
=&H_2({\bf k},2t-\tau) \\
=&H({\bf k},t), \\
\end{aligned}
\end{equation}
Thus, $H({\bf k},t)=qH^\dagger({\bf k},t) q^{-1}$.

5. $C$ symmetry with $\epsilon_c=1$. From $H_1({\bf k},t)=cH_1^{\bf T}(-{\bf k},- t)c^{-1}$ and $H_2({\bf k},t)=cH_2^{\bf T}(-{\bf k},- t)c^{-1}$ to $H({\bf k},t)=cH^{\bf T}(-{\bf k},- t)c^{-1}$. It is equivalent to  ``From $H_1(-{\bf k},-t)=cH_1^{\bf T}({\bf k},t)c^{-1}$ and $H_2(-{\bf k},-t)=cH_2^{\bf T}({\bf k},t)c^{-1}$ to $H(-{\bf k},-t)=cH^{\bf T}({\bf k},t)c^{-1}$".

5a. When $0\le t\le \tau/4$,
\begin{equation}
\begin{aligned}
&cH^{\bf T}({\bf k},t)c^{-1} \\
=&cH_2^{\bf T}({\bf k},2t)c^{-1} \\
=&H_2(-{\bf k},-2t),\\
\end{aligned}
\end{equation}
\begin{equation}
\begin{aligned}
&H(-{\bf k},-t) \\
=&H(-{\bf k},\tau-t) \\
=&H_2(-{\bf k},2\tau-2t-\tau)\\
=&H_2(-{\bf k},-2t),\\
\end{aligned}
\end{equation}
Thus, $H(-{\bf k},-t)=cH^{\bf T}({\bf k},t)c^{-1}$.

5b. When $\tau/4 <t <3\tau/4$,
\begin{equation}
\begin{aligned}
&cH^{\bf T}({\bf k},t)c^{-1} \\
=&cH_1^{\bf T}({\bf k},2t-\tau/2)c^{-1} \\
=&H_1(-{\bf k},\tau/2-2t),\\
\end{aligned}
\end{equation}
\begin{equation}
\begin{aligned}
&H(-{\bf k},-t) \\
=&H(-{\bf k},\tau-t) \\
=&H_1(-{\bf k},2\tau-2t-\tau/2)\\
=&H_1(-{\bf k},\tau/2-2t),\\
\end{aligned}
\end{equation}
Thus, $H(-{\bf k},-t)=cH^{\bf T}({\bf k},t)c^{-1}$.

5c. When $3\tau/4\le t\le \tau$,
\begin{equation}
\begin{aligned}
&cH^{\bf T}({\bf k},t)c^{-1} \\
=&cH_2^{\bf T}({\bf k},2t-\tau)c^{-1} \\
=&H_2(-{\bf k},\tau-2t),\\
\end{aligned}
\end{equation}
\begin{equation}
\begin{aligned}
&H(-{\bf k},-t) \\
=&H(-{\bf k},\tau-t) \\
=&H_2(-{\bf k},2\tau-2t)\\
=&H_2(-{\bf k},\tau-2t),\\
\end{aligned}
\end{equation}
Thus, $H(-{\bf k},-t)=cH^{\bf T}({\bf k},t)c^{-1}$.

6. $C$ symmetry with $\epsilon_c=-1$. From $H_1({\bf k},t)=-cH_1^{\bf T}(-{\bf k},t)c^{-1}$ and $H_2({\bf k},t)=-cH_2^{\bf T}(-{\bf k},t)c^{-1}$ to $H({\bf k},t)=-cH^{\bf T}(-{\bf k},t)c^{-1}$.

6a. When $0\le t\le \tau/4$,
\begin{equation}
\begin{aligned}
&-cH^{\bf T}(-{\bf k},t)c^{-1} \\
=&-cH_2^{\bf T}(-{\bf k},2t)c^{-1} \\
=&H_2({\bf k},2t)\\
=&H({\bf k},t),\\
\end{aligned}
\end{equation}

Thus, $H({\bf k},t)=-cH^{\bf T}(-{\bf k},t)c^{-1}$.

6b. When $\tau/4 < t< 3\tau/4$,
\begin{equation}
\begin{aligned}
&-cH^{\bf T}(-{\bf k},t)c^{-1} \\
=&-cH_1^{\bf T}(-{\bf k},2t-\tau/2)c^{-1} \\
=&H_1({\bf k},2t-\tau/2)\\
=&H({\bf k},t),\\
\end{aligned}
\end{equation}

Thus, $H({\bf k},t)=-cH^{\bf T}(-{\bf k},t)c^{-1}$.

6c. When $3\tau/4\le t\le \tau$,
\begin{equation}
\begin{aligned}
&-cH^{\bf T}(-{\bf k},t)c^{-1} \\
=&-cH_2^{\bf T}(-{\bf k},2t-\tau)c^{-1} \\
=&H_2({\bf k},2t-\tau)\\
=&H({\bf k},t),\\
\end{aligned}
\end{equation}

Thus, $H({\bf k},t)=-cH^{\bf T}(-{\bf k},t)c^{-1}$.

7. $P$ symmetry. From $-H_1({\bf k},t)=pH_1({\bf k},-t)p^{-1}$ and $-H_2({\bf k},t)=pH_2({\bf k},-t)p^{-1}$ to $-H({\bf k},t)=pH({\bf k},-t)p^{-1}$. It is equivalent to ``From $-H_1({\bf k},-t)=pH_1({\bf k},t)p^{-1}$ and $-H_2({\bf k},-t)=pH_2({\bf k},t)p^{-1}$ to $-H({\bf k},-t)=pH({\bf k},t)p^{-1}$".

7a. When $0\le t\le \tau/4$,
\begin{equation}
\begin{aligned}
&pH({\bf k},t)p^{-1} \\
=&pH_2({\bf k},2t)p^{-1} \\
=&-H_2({\bf k},-2t),\\
\end{aligned}
\end{equation}
\begin{equation}
\begin{aligned}
&-H({\bf k},-t) \\
=&-H({\bf k},\tau-t) \\
=&-H_2({\bf k},2\tau-2t-\tau)\\
=&-H_2({\bf k},-2t),\\
\end{aligned}
\end{equation}
Thus, $-H({\bf k},-t)=pH({\bf k},t)p^{-1}$.

7b. When $\tau/4< t< 3\tau/4$,
\begin{equation}
\begin{aligned}
&pH({\bf k},t)p^{-1} \\
=&pH_1({\bf k},2t-\tau/2)p^{-1} \\
=&-H_1({\bf k},\tau/2-2t),\\
\end{aligned}
\end{equation}

\begin{equation}
\begin{aligned}
&-H({\bf k},-t) \\
=&-H({\bf k},\tau-t) \\
=&-H_1({\bf k},2\tau-2t-\tau/2)\\
=&-H_1({\bf k},\tau/2-2t),\\
\end{aligned}
\end{equation}
Thus, $-H({\bf k},-t)=pH({\bf k},t)p^{-1}$.

7c. When $3\tau/4\le t\le \tau$,
\begin{equation}
\begin{aligned}
&pH({\bf k},t)p^{-1} \\
=&pH_2({\bf k},2t-\tau)p^{-1} \\
=&-H_2({\bf k},\tau-2t),\\
\end{aligned}
\end{equation}
\begin{equation}
\begin{aligned}
&-H({\bf k},-t) \\
=&-H({\bf k},\tau-t) \\
=&-H_2({\bf k},2\tau-2t)\\
=&-H_2({\bf k},\tau-2t),\\
\end{aligned}
\end{equation}
Thus, $-H({\bf k},-t)=pH({\bf k},t)p^{-1}$.

\section{Proof of Lemma 2} \label{apd}

To prove Lemma 2, we first need to show Lemma 4-9 as listed below.

\begin{lemma}
  For a complex number $\lambda$ and $\ln_{\theta}(\lambda)$ is well defined, $\ln_{\theta}(\lambda)+\ln_{-\theta}(\lambda^{-1})=2\pi i$.
\end{lemma}

Proof:
There always exist $\phi_1\in(\theta,\theta+2\pi)$ and $\phi_2$, satisfying $e^{i\phi_1+\phi_2}=\lambda$. $\ln_{\theta}(\lambda) =i\phi_1+\phi_2$ and $-\phi_1\in(-\theta-2\pi,-\theta)$. Thus, $-\phi_1+2\pi\in(-\theta,-\theta+2\pi)$. $\ln_{-\theta}(\lambda^{-1})=\ln_{-\theta}(e^{-i\phi_1-\phi_2})=-i\phi_1+2\pi i-\phi_2$. Hence $\ln_{\theta}(\lambda)+\ln_{-\theta}(\lambda^{-1})=2\pi i$.
\begin{lemma}
  For a complex number $\lambda$ and $\ln_{-\theta}(\lambda)$ is well defined, $\ln_{-\theta}(\lambda^*)-[\ln_{\theta}(\lambda)]^*=2\pi i$.
\end{lemma}

Proof:
There always exist $\phi_1\in(\theta,\theta+2\pi)$ and $\phi_2$, satisfying $e^{i\phi_1+\phi_2}=\lambda$. $\ln_{\theta}(\lambda) =i\phi_1+\phi_2$ and $-\phi_1\in(-\theta-2\pi,-\theta)$. Thus, $-\phi_1+2\pi\in(-\theta,-\theta+2\pi)$. $\ln_{-\theta}(\lambda^*)=\ln_{-\theta}(e^{-i\phi_1+\phi_2})=-i\phi_1+2\pi i+\phi_2$. Hence $\ln_{-\theta}(\lambda^*)-[\ln_{\theta}(\lambda)]^*=2\pi i$.
\begin{lemma}
  For a complex number $\lambda$ and $\ln_{\theta}(\lambda)$ is well defined, $\ln_{\theta}(\lambda)+2\pi i=\ln_{\theta+2\pi}(\lambda)$.
\end{lemma}

Proof:
There always exist $\phi_1\in(\theta,\theta+2\pi)$ and $\phi_2$, satisfying $e^{i\phi_1+\phi_2}=\lambda$. $\ln_{\theta}(\lambda) =i\phi_1+\phi_2$. And $\phi_1+2\pi\in(\theta+2\pi,\theta+4\pi)$. Thus, $\ln_{\theta+2\pi}(\lambda)=\ln_{\theta+2\pi}(e^{i\phi_1+\phi_2})=i\phi_1+2\pi i+\phi_2$. Hence $\ln_{\theta}(\lambda)+2\pi i=\ln_{\theta+2\pi}(\lambda)$.

\begin{widetext}
  \begin{lemma}
    $H_{F,\theta}+\frac{2\pi}{\tau}=H_{F,\theta+2\pi}$.
  \end{lemma}

Proof:
\begin{equation}
\begin{aligned}
&H_{F,\theta}+\frac{2\pi}{\tau}=\frac{i}{\tau}\ln_{-\theta}[U({\bf k})]+\frac{2\pi}{\tau} \\
=& \sum_n \frac{i}{\tau}\ln_{-\theta}(\lambda_n)|\psi_{n,R}\rangle \langle \psi_{n,L}|+\frac{2\pi}{\tau} \\
=& \sum_n \frac{i}{\tau}(\ln_{-\theta}(\lambda_n)-2\pi i)|\psi_{n,R}\rangle \langle \psi_{n,L}|\\
=& \sum_n \frac{i}{\tau}(\ln_{-\theta-2\pi}(\lambda_n))|\psi_{n,R}\rangle \langle \psi_{n,L}|=H_{F,\theta+2\pi}.
\end{aligned}
\end{equation}
\begin{lemma}
  For any unitary matrix $p_1$ and any evolution operator $U({\bf k})$, $\ln_{\theta}[p_1^{-1}U({\bf k})p_1]=p_1^{-1}\ln_{\theta}[U({\bf k})]p_1$.
\end{lemma}

Proof:
\begin{equation}
\begin{aligned}
&\ln_{\theta}[p_1^{-1}U({\bf k})p_1] \notag\\
=& \sum_n \frac{i}{\tau}\ln_{\theta}(\lambda_n)p_1^{-1}|\psi_{n,R}\rangle \langle \psi_{n,L}|p_1 \\
=& p_1^{-1}[\sum_n \frac{i}{\tau}\ln_{\theta}(\lambda_n)|\psi_{n,R}\rangle \langle \psi_{n,L}|]p_1 \\
=& p_1^{-1}\ln_{\theta}[U({\bf k})]p_1\\
\end{aligned}
\end{equation}
\begin{lemma}
  If $H({\bf k},t)$ satisfies Eqs. (\ref{eq:symshp})-(\ref{eq:symshk}), then $H_{F,\theta}$ respectively satisfies Eqs. (\ref{eq:symshfp})-(\ref{eq:symshfk}) below:
\begin{align}
H_{F,\theta}({\bf k})=-\frac{(1-\epsilon_k)\pi}{\tau}+\epsilon_kkH_{F,\epsilon_k\theta}^*(-{\bf k})k^{-1}&, ~~kk^*=\eta_k \mathbb{I}, &K \textrm{ sym.} \label{eq:symshfp} \\
H_{F,\theta}({\bf k})=-\frac{(1-\epsilon_q)\pi}{\tau}+\epsilon_qqH_{F,\epsilon_q\theta}^\dagger({\bf k}) q^{-1}&, ~~q^2=\mathbb{I}, &Q \textrm{ sym.}\label{eq:symshfq}\\
H_{F,\theta}({\bf k})=-\frac{(1-\epsilon_c)\pi}{\tau}+\epsilon_c cH_{F,\epsilon_c\theta}^{\bf T}(-{\bf k})c^{-1}&, ~~cc^*=\eta_c \mathbb{I},&C \textrm{ sym.}\label{eq:symshfc}\\
H_{F,\theta}({\bf k})=-\frac{2\pi}{\tau}-pH_{F,-\theta}({\bf k})p^{-1}&, ~~p^2=\mathbb{I}. &P \textrm{ sym.}\label{eq:symshfk}
\end{align}
\end{lemma}

Proof: We go through the four symmetries one by one.

1. From $U^*(-{\bf k})=k^{-1}U({\bf k})k$ to $H_{F,\theta}({\bf k})=-\frac{2\pi}{\tau}-kH_{F,-\theta}^*(-{\bf k})k^{-1}$. It is equivalent to ``From $\frac{i}{\tau}\ln_{-\theta}[U^*(-{\bf k})]=\frac{i}{\tau}\ln_{-\theta}[k^{-1}U({\bf k})k]$ to $k^{-1}H_{F,\theta}({\bf k})k=-\frac{2\pi}{\tau}-H_{F,-\theta}^*(-{\bf k})$".

\begin{equation}
\begin{aligned}
&\frac{i}{\tau}\ln_{-\theta}[U^*(-{\bf k})] \\
=& \sum_n \frac{i}{\tau}\ln_{-\theta}[\lambda_n^*(-{\bf k})]|\psi_{n,R}^*(-{\bf k})\rangle \langle \psi_{n,L}^*(-{\bf k})| \\
=& \sum_n \frac{i}{\tau}[[\ln_{\theta}[\lambda_n(-{\bf k})]]^*+2\pi i]|\psi_{n,R}^*(-{\bf k})\rangle \langle \psi_{n,L}^*(-{\bf k})| \\
=& -\frac{2\pi}{\tau}+\sum_n \frac{i}{\tau}[\ln_{\theta}[\lambda_n(-{\bf k})]]^*|\psi_{n,R}^*(-{\bf k})\rangle \langle \psi_{n,L}^*(-{\bf k})| \\
=& -\frac{2\pi}{\tau}-[\sum_n \frac{i}{\tau}\ln_{\theta}[\lambda_n(-{\bf k})]|\psi_{n,R}(-{\bf k})\rangle \langle \psi_{n,L}(-{\bf k})|]^* \\
=& -\frac{2\pi}{\tau}-H_{F,-\theta}^*(-{\bf k}).
\end{aligned}
\end{equation}
And we have
\begin{equation}
\frac{i}{\tau}\ln_{-\theta}[k^{-1}U({\bf k})k]=k^{-1}\frac{i}{\tau}\ln_{-\theta}[U({\bf k})]k= k^{-1}H_{F,\theta}({\bf k})k.
\end{equation}
Thus, $k^{-1}H_{F,\theta}({\bf k})k=-\frac{2\pi}{\tau}-H_{F,-\theta}^*(-{\bf k})$.

2. From $U^{*-1}(-{\bf k})=k^{-1}U({\bf k})k$ to $H_{F,\theta}({\bf k})=kH_{F,\theta}^*(-{\bf k})k^{-1}$. It is equivalent to ``From $\frac{i}{\tau}\ln_{-\theta}[U^{*-1}(-{\bf k})]=\frac{i}{\tau}\ln_{-\theta}[k^{-1}U({\bf k})k]$ to $k^{-1}H_{F,\theta}({\bf k})k=H_{F,\theta}^*(-{\bf k})$".
\begin{equation}
\begin{aligned}
&\frac{i}{\tau}\ln_{-\theta}[U^{*-1}(-{\bf k})] \\
=& \sum_n \frac{i}{\tau}\ln_{-\theta}[\lambda_n^{*-1}(-{\bf k})]|\psi_{n,R}^*(-{\bf k})\rangle \langle \psi_{n,L}^*(-{\bf k})| \\
=& \sum_n \frac{i}{\tau}[[\ln_{\theta}[\lambda_n^{-1}(-{\bf k})]]^*+2\pi i]|\psi_{n,R}^*(-{\bf k})\rangle \langle \psi_{n,L}^*(-{\bf k})| \\
=& \sum_n \frac{i}{\tau}[[-\ln_{-\theta}[\lambda_n(-{\bf k})]+2\pi i]^*+2\pi i]|\psi_{n,R}^*(-{\bf k})\rangle \langle \psi_{n,L}^*(-{\bf k})| \\
=& \sum_n \frac{i}{\tau}[-\ln_{-\theta}[\lambda_n(-{\bf k})]]^*|\psi_{n,R}^*(-{\bf k})\rangle \langle \psi_{n,L}^*(-{\bf k})| \\
=& [\sum_n \frac{i}{\tau}\ln_{-\theta}[\lambda_n(-{\bf k})]|\psi_{n,R}(-{\bf k})\rangle \langle \psi_{n,L}(-{\bf k})|]^* \\
=& H_{F,\theta}^*(-{\bf k})
\end{aligned}
\end{equation}
And we have
\begin{equation}
\begin{aligned}
\frac{i}{\tau}\ln_{-\theta}[k^{-1}U({\bf k})k]= k^{-1}\frac{i}{\tau}\ln_{-\theta}[U({\bf k})]k= k^{-1}H_{F,\theta}({\bf k})k.
\end{aligned}
\end{equation}
Thus, $k^{-1}H_{F,\theta}({\bf k})k=H_{F,\theta}^*(-{\bf k})$.

3. From $[U^{\dagger}({\bf k})]=q^{-1}U({\bf k}) q$ to $H_{F,\theta}({\bf k})=-\frac{2\pi}{\tau}-qH_{F,-\theta}^\dagger({\bf k}) q^{-1}$. It is equivalent to ``From $\frac{i}{\tau}\ln_{-\theta}[U^{\dagger}({\bf k})]=\frac{i}{\tau}\ln_{-\theta}[q^{-1}U({\bf k}) q]$ to $q^{-1}H_{F,\theta}({\bf k})q=-\frac{2\pi}{\tau}-H_{F,-\theta}^\dagger({\bf k}) $".
\begin{equation}
\begin{aligned}
&\frac{i}{\tau}\ln_{-\theta}[U^{\dagger}({\bf k})] \\
=& \sum_n \frac{i}{\tau}\ln_{-\theta}[\lambda_n^*({\bf k})]|\psi_{n,L}({\bf k})\rangle \langle \psi_{n,R}({\bf k})| \\
=& \sum_n \frac{i}{\tau}[[\ln_{\theta}[\lambda_n({\bf k})]]^*+2\pi i]|\psi_{n,L}({\bf k})\rangle \langle \psi_{n,R}({\bf k})| \\
=& -\frac{2\pi}{\tau}+\sum_n \frac{i}{\tau}[\ln_{\theta}[\lambda_n({\bf k})]]^*|\psi_{n,L}({\bf k})\rangle \langle \psi_{n,R}({\bf k})| \\
=& -\frac{2\pi}{\tau}-[\sum_n \frac{i}{\tau}\ln_{\theta}[\lambda_n({\bf k})]|\psi_{n,R}({\bf k})\rangle \langle \psi_{n,L}({\bf k})|]^{\dagger} \\
=& -\frac{2\pi}{\tau}-H_{F,-\theta}^{\dagger}({\bf k}).
\end{aligned}
\end{equation}
And we have
\begin{equation}
\begin{aligned}
\frac{i}{\tau}\ln_{-\theta}[q^{-1}U({\bf k}) q]= q^{-1}\frac{i}{\tau}\ln_{-\theta}[U({\bf k}) ]q = q^{-1}H_{F,\theta}({\bf k})q.
\end{aligned}
\end{equation}
Thus, $q^{-1}H_{F,\theta}({\bf k})q=-\frac{2\pi}{\tau}-H_{F,-\theta}^\dagger({\bf k}) $.

4. From $[U^{\dagger}({\bf k})]^{-1}=q^{-1}U({\bf k}) q$ to $H_{F,\theta}({\bf k})=qH_{F,\theta}^\dagger({\bf k}) q^{-1}$. It is equivalent to ``From $\frac{i}{\tau}\ln_{-\theta}[U^{\dagger}({\bf k})]^{-1}=\frac{i}{\tau}\ln_{-\theta}[q^{-1}U({\bf k}) q]$ to $q^{-1}H_{F,\theta}({\bf k})q=H_{F,\theta}^\dagger({\bf k}) $".
\begin{equation}
\begin{aligned}
&\frac{i}{\tau}\ln_{-\theta}[U^{\dagger}({\bf k})]^{-1} \\
=& \sum_n \frac{i}{\tau}\ln_{-\theta}[\lambda_n^{*-1}({\bf k})]|\psi_{n,L}({\bf k})\rangle \langle \psi_{n,R}({\bf k})| \\
=& \sum_n \frac{i}{\tau}[[\ln_{\theta}[\lambda_n^{-1}({\bf k})]]^*+2\pi i]|\psi_{n,L}({\bf k})\rangle \langle \psi_{n,R}({\bf k})| \\
=& \sum_n \frac{i}{\tau}[[-\ln_{-\theta}[\lambda_n({\bf k})]+2\pi i]^*+2\pi i]|\psi_{n,L}({\bf k})\rangle \langle \psi_{n,R}({\bf k})| \\
=& \sum_n \frac{i}{\tau}[-\ln_{-\theta}[\lambda_n({\bf k})]]^*|\psi_{n,L}({\bf k})\rangle \langle \psi_{n,R}({\bf k})| \\
=& [\sum_n \frac{i}{\tau}\ln_{-\theta}[\lambda_n({\bf k})]|\psi_{n,R}({\bf k})\rangle \langle \psi_{n,L}({\bf k})|]^{\dagger} \\
=& H_{F,\theta}^\dagger({\bf k}).
\end{aligned}
\end{equation}
And we have
\begin{equation}
\begin{aligned}
\frac{i}{\tau}\ln_{-\theta}[q^{-1}U({\bf k}) q] = q^{-1}\frac{i}{\tau}\ln_{-\theta}[U({\bf k})] q = q^{-1}H_{F,\theta}({\bf k})q.
\end{aligned}
\end{equation}
Thus, $q^{-1}H_{F,\theta}({\bf k})q=H_{F,\theta}^\dagger({\bf k}) $.

5. From $[U^{\bf T}(-{\bf k})]=c^{-1}U({\bf k})c$ to $H_{F,\theta}({\bf k})= cH_{F,\theta}^{\bf T}(-{\bf k})c^{-1}$. It is equivalent to ``From $\frac{i}{\tau}\ln_{-\theta}[U^{\bf T}(-{\bf k})]=\frac{i}{\tau}\ln_{-\theta}[c^{-1}U({\bf k})c]$ to $c^{-1}H_{F,\theta}({\bf k})c= H_{F,\theta}^{\bf T}(-{\bf k})$".
\begin{equation}
\begin{aligned}
&\frac{i}{\tau}\ln_{-\theta}[U^{\bf T}(-{\bf k})] \\
=& \sum_n \frac{i}{\tau}\ln_{-\theta}[\lambda_n(-{\bf k})]|\psi_{n,L}^*(-{\bf k})\rangle \langle \psi_{n,R}^*(-{\bf k})| \\
=& [\sum_n \frac{i}{\tau}\ln_{-\theta}[\lambda_n(-{\bf k})]|\psi_{n,R}(-{\bf k})\rangle \langle \psi_{n,L}(-{\bf k})|]^{\bf T} \\
=& H_{F,\theta}^{\bf T}(-{\bf k}).
\end{aligned}
\end{equation}
And we have
\begin{equation}
\begin{aligned}
\frac{i}{\tau}\ln_{-\theta}[c^{-1}U({\bf k})c] =c^{-1} \frac{i}{\tau}\ln_{-\theta}[U({\bf k})] c = c^{-1}H_{F,\theta}({\bf k})c.
\end{aligned}
\end{equation}
Thus, $c^{-1}H_{F,\theta}({\bf k})c= H_{F,\theta}^{\bf T}(-{\bf k})$.

6. From $[U^{\bf T}(-{\bf k})]^{-1}=c^{-1}U({\bf k})c$ to $H_{F,\theta}({\bf k})=-\frac{2\pi}{\tau}- cH_{F,-\theta}^{\bf T}(-{\bf k})c^{-1}$. It is equivalent to ``From $\frac{i}{\tau}\ln_{-\theta}[U^{\bf T}(-{\bf k})]^{-1}=\frac{i}{\tau}\ln_{-\theta}[c^{-1}U({\bf k})c]$ to $c^{-1}H_{F,\theta}({\bf k})c=-\frac{2\pi}{\tau}- H_{F,-\theta}^{\bf T}(-{\bf k})$".
\begin{equation}
\begin{aligned}
&\frac{i}{\tau}\ln_{-\theta}[U^{\bf T}(-{\bf k})]^{-1} \\
=& \sum_n \frac{i}{\tau}\ln_{-\theta}[\lambda_n^{-1}(-{\bf k})]|\psi_{n,L}^*(-{\bf k})\rangle \langle \psi_{n,R}^*(-{\bf k})| \\
=& \sum_n \frac{i}{\tau}[-\ln_{\theta}[\lambda_n(-{\bf k})]+2\pi i]|\psi_{n,L}^*(-{\bf k})\rangle \langle \psi_{n,R}^*(-{\bf k})| \\
=&-\frac{2\pi}{\tau}+\sum_n \frac{i}{\tau}[-\ln_{\theta}[\lambda_n(-{\bf k})]]|\psi_{n,L}^*(-{\bf k})\rangle \langle \psi_{n,R}^*(-{\bf k})| \\
=&-\frac{2\pi}{\tau}-[\sum_n \frac{i}{\tau}\ln_{\theta}[\lambda_n(-{\bf k})]|\psi_{n,R}(-{\bf k})\rangle \langle \psi_{n,L}(-{\bf k})|]^{\bf T} \\
=&-\frac{2\pi}{\tau}- H_{F,-\theta}^{\bf T}(-{\bf k}).\\
\end{aligned}
\end{equation}
And we have
\begin{equation}
\begin{aligned}
\frac{i}{\tau}\ln_{-\theta}[c^{-1}U({\bf k})c] =c^{-1}\frac{i}{\tau}\ln_{-\theta}[U({\bf k})]c = c^{-1}H_{F,\theta}({\bf k})c.
\end{aligned}
\end{equation}
Thus, $c^{-1}H_{F,\theta}({\bf k})c=-\frac{2\pi}{\tau}- H_{F,-\theta}^{\bf T}(-{\bf k})$.

7. From $[U({\bf k})]^{-1}=p^{-1}U({\bf k})p$ to $H_{F,\theta}({\bf k})=-\frac{2\pi}{\tau}-pH_{F,-\theta}({\bf k})p^{-1}$. It is equivalent to ``From $\frac{i}{\tau}\ln_{-\theta}[U({\bf k})]^{-1}=\frac{i}{\tau}\ln_{-\theta}[p^{-1}U({\bf k})p]$ to $p^{-1}H_{F,\theta}({\bf k})p=-\frac{2\pi}{\tau}-H_{F,-\theta}({\bf k})$".
\begin{equation}
\begin{aligned}
&\frac{i}{\tau}\ln_{-\theta}[U({\bf k})]^{-1} \\
=& \sum_n \frac{i}{\tau}\ln_{-\theta}[\lambda_n^{-1}({\bf k})]|\psi_{n,R}({\bf k})\rangle \langle \psi_{n,L}({\bf k})| \\
=& \sum_n \frac{i}{\tau}[-\ln_{\theta}[\lambda_n({\bf k})]+2\pi i]|\psi_{n,R}({\bf k})\rangle \langle \psi_{n,L}({\bf k})| \\
=&-\frac{2\pi}{\tau}-\sum_n \frac{i}{\tau}\ln_{\theta}[\lambda_n({\bf k})]|\psi_{n,R}({\bf k})\rangle \langle \psi_{n,L}({\bf k})| \\
=&-\frac{2\pi}{\tau}-H_{F,-\theta}({\bf k}).\\
\end{aligned}
\end{equation}
And we have
\begin{equation}
\begin{aligned}
\frac{i}{\tau}\ln_{-\theta}[p^{-1}U({\bf k})p] = p^{-1}\frac{i}{\tau}\ln_{-\theta}[U({\bf k})]p = p^{-1}H_{F,\theta}({\bf k})p.
\end{aligned}
\end{equation}
Thus, $p^{-1}H_{F,\theta}({\bf k})p=-\frac{2\pi}{\tau}-H_{F,-\theta}({\bf k})$. Combining 1-7, we get the proof of Lemma 9. Now we are ready to prove Lemma 2 in the main text. According to Lemma 9, if $H({\bf k},t)$ satisfies Eqs. (\ref{eq:symshp})-(\ref{eq:symshk}), then $H_{F,\pi}$ satisfies Eqs. (\ref{eq:symshfp1})-(\ref{eq:symshfk1}) below:
\begin{align}
H_{F,\pi}({\bf k})=-\frac{(1-\epsilon_k)\pi}{\tau}+\epsilon_kkH_{F,\epsilon_k\pi}^*(-{\bf k})k^{-1}&, ~~kk^*=\eta_k \mathbb{I}, &K \textrm{ sym.} \label{eq:symshfp1} \\
H_{F,\pi}({\bf k})=-\frac{(1-\epsilon_q)\pi}{\tau}+\epsilon_qqH_{F,\epsilon_q\pi}^\dagger({\bf k}) q^{-1}&, ~~q^2=\mathbb{I}, &Q \textrm{ sym.}\label{eq:symshfq1}\\
H_{F,\pi}({\bf k})=-\frac{(1-\epsilon_c)\pi}{\tau}+\epsilon_c cH_{F,\epsilon_c\pi}^{\bf T}(-{\bf k})c^{-1}&, ~~cc^*=\eta_c \mathbb{I},&C \textrm{ sym.}\label{eq:symshfc1}\\
H_{F,\pi}({\bf k})=-\frac{2\pi}{\tau}-pH_{F,-\pi}({\bf k})p^{-1}&, ~~p^2=\mathbb{I}. &P \textrm{ sym.}\label{eq:symshfk1}
\end{align}
According to Lemma 7, we have:
\begin{align}
-\frac{(1-\epsilon_k)\pi}{\tau}+\epsilon_kkH_{F,\epsilon_k\pi}^*(-{\bf k})k^{-1}= \epsilon_kkH_{F,\epsilon_k\pi+(1-\epsilon_k)\pi}^*(-{\bf k})k^{-1}=\epsilon_kkH_{F,\pi}^*(-{\bf k})k^{-1} , \label{eq:symshfp2} \\
-\frac{(1-\epsilon_q)\pi}{\tau}+\epsilon_qqH_{F,\epsilon_q\pi}^\dagger({\bf k}) q^{-1}=\epsilon_qqH_{F,\epsilon_q\pi+(1-\epsilon_q)\pi}^\dagger({\bf k}) q^{-1}=\epsilon_qqH_{F,\pi}^\dagger({\bf k}) q^{-1}, \label{eq:symshfq2}\\
-\frac{(1-\epsilon_c)\pi}{\tau}+\epsilon_c cH_{F,\epsilon_c\pi}^{\bf T}(-{\bf k})c^{-1}=\epsilon_c cH_{F,\epsilon_c\pi+(1-\epsilon_c)\pi}^{\bf T}(-{\bf k})c^{-1}=\epsilon_c cH_{F,\pi}^{\bf T}(-{\bf k})c^{-1}, \label{eq:symshfc2}\\
-\frac{2\pi}{\tau}-pH_{F,-\pi}({\bf k})p^{-1}=-pH_{F,-\pi+2\pi}({\bf k})p^{-1}=-pH_{F,\pi}({\bf k})p^{-1}. \label{eq:symshfk2}
\end{align}
Combining Eqs. (\ref{eq:symshfp1})-(\ref{eq:symshfk2}), we can conclude that if $H({\bf k},t)$ satisfies Eqs. (\ref{eq:symshp})-(\ref{eq:symshk}), then $H_{F,\pi}$ satisfies:
\begin{align}
H_{F,\pi}({\bf k})=\epsilon_kkH_{F,\pi}^*(-{\bf k})k^{-1}&, ~~kk^*=\eta_k \mathbb{I}, &K \textrm{ sym.} \label{eq:symshfp3} \\
H_{F,\pi}({\bf k})=\epsilon_qqH_{F,\pi}^\dagger({\bf k}) q^{-1}&, ~~q^2=\mathbb{I}, &Q \textrm{ sym.}\label{eq:symshfq3}\\
H_{F,\pi}({\bf k})=\epsilon_c cH_{F,\pi}^{\bf T}(-{\bf k})c^{-1}&, ~~cc^*=\eta_c \mathbb{I},&C \textrm{ sym.}\label{eq:symshfc3}\\
H_{F,\pi}({\bf k})=-pH_{F,\pi}({\bf k})p^{-1}&, ~~p^2=\mathbb{I}. &P \textrm{ sym.}\label{eq:symshfk3}
\end{align}
Thus, $H_{F,\pi}({\bf k})$ and $H({\bf k},t)$ belong to same GBL class. The loop operator is $U_{l,\pi}({\bf k},t)=U({\bf k},t)*e^{iH_{F,\pi}({\bf k})t}$. According to Lemma 1 in the main text, $H_{F,\pi}({\bf k})$, $U_{l,\pi}$, and $H({\bf k},t)$ belong to same GBL class. We remark that $H_{F,\pi}({\bf k})$ has (continuous) time-translation symmetry but $H({\bf k})$ only has discrete time-translation symmetry. In general $H_{F,\pi}({\bf k})$ has more symmetries than $H({\bf k})$.
\end{widetext}

\section{Proof of Theorem 1}\label{ape}
The proof can be done in a similar vein with that of the Hermitian case (see Appendix C of Ref. \cite{fclass2}). Here we decompose the proof into 
Lemma 10 and Lemma 11.
\begin{lemma}
  The time evolution operator $U({\bf k},t)$ with FH real gap at $\pi/\tau$ can continuously transform to $U_f({\bf k},t)=L * C$, with $L$ a loop operator and $C$ a constant (time-independent) Hamiltonian evolution. One of the $L$ is $L=U_{l,\pi}({\bf k},t)$. And the corresponding $C$ is the constant evolution with Hamiltonian $H_{F,\pi}$. The continuous transformation preserves the GBL symmetries.
\end{lemma}

To see this, we define the constant evolution $C_{\pm}(s):=e^{\mp isH_{F,\pi}t}$. According to Lemma 9, $C_{\pm}(s)$ and $U({\bf k},t)$ belong to same GBL class. We further define $U_0(s):=U({\bf k},t)*C_{-}(s)$ and $h(s):=U_0(s)*C_{+}(s)$. According to Lemma 1, $C_{\pm}(s)$, $U({\bf k},t)$, $U_0(s)$, and $h(s)$ belong to same GBL class. $h(s)$ is a continuous function and satisfies:
\begin{equation}
\begin{aligned}
h(0)=&U({\bf k},t);\\
h(1)=&U_0(1)*C_{+}(1) =U_{l,\pi}*e^{-iH_{F,\pi}t}.
\end{aligned}
\end{equation}
\begin{lemma}
  In Lemma 10, if we have two $U_f({\bf k},t)$ labeled as $U_{f1}$ and $U_{f2}$. And $U_{f1}=L_1 * C_1$ and $U_{f2}=L_2 * C_2$. We then have that $L_1\approx L_2$ and $C_1\approx C_2$. Here $\approx$ is the homotopy equivalence.
\end{lemma}

As $U({\bf k},t)\approx U_{f1}$ and $U({\bf k},t)\approx U_{f2}$, we have $U_{f1}\approx U_{f2}$. Thus, 
there exists 
continuous function $g_{s}({\bf k},t)$ such that 
$g_{0}=U_{f1}=L_1*C_1$, $g_{1}=U_{f2}=L_2*C_2$, and 
time evolution operator $g_{s}({\bf k},t)$ 
has FH real gap at $\pi/\tau$ for any $s$. 
And $U({\bf k},t)$ and $g_s$ belong to same GBL class. We can define the continuous function:
\begin{align}
h^{F}_s({\bf k}):=\frac{i}{\tau}\ln_{-\pi}[g_{s}({\bf k},\tau)],  \\
g^{\pm}_s({\bf k},t):=e^{\mp ih_s^Ft},\\
g^1_s({\bf k},t):=g_s*g_s^-  .
\end{align}
According to Lemma 9, $U({\bf k},t))$, and $g^{\pm}_s({\bf k},t)$ belong to same GBL class. According to Lemma 1, $U({\bf k},t)$, $g^{\pm}_s({\bf k},t)$, and $g^1_s({\bf k},t)$ belong to same GBL class, and
\begin{align}
g^{+}_0({\bf k},t)=e^{- ih_0^Ft}=C_1;\\
g^{+}_1({\bf k},t)=e^{- ih_1^Ft}=C_2;\\
g^1_0({\bf k},t):=g_0*g_0^-\approx L_1; \\
g^1_1({\bf k},t):=g_1*g_1^-\approx L_2 .
\end{align}

\section{Derivation of Eqs. (\ref{eq:symshtp})-(\ref{eq:symshtk}) from Eqs. (\ref{eq:symsup})-(\ref{eq:symsuk})}\label{apf}
1. From $U_{l,\pi}^*(-{\bf k},-t)=k^{-1}U_{l,\pi}({\bf k},\epsilon_k t)k$ to $\tilde {H}(-{\bf k}, -\epsilon_kt)=\tilde{K}\tilde {H}({\bf k}, t)\tilde{K}^{-1}$.
\begin{equation}
\begin{aligned}
&\tilde{K}\tilde {H}({\bf k}, t)\tilde{K}^{-1}\\
=&\left[ \begin{array}{cc}
                k & 0\\
                0 & k
                \end{array}
                \right ]
\left[ \begin{array}{cc}
                0 & U_{l,\pi}^*({\bf k}, t)\\
                U_{l,\pi}^{\bf T}({\bf k}, t) & 0
                \end{array}
                \right ]
          \left[ \begin{array}{cc}
                k^{-1} & 0\\
                0 & k^{-1}
                \end{array}
                \right ] \\
=&\left[ \begin{array}{cc}
                0 & kU_{l,\pi}^*({\bf k}, t)k^{-1}\\
                kU_{l,\pi}^{\bf T}({\bf k}, t)k^{-1} & 0
                \end{array}
                \right ] \\
=&\left[ \begin{array}{cc}
                0 & U_{l,\pi}(-{\bf k},-\epsilon_k t)\\
               U_{l,\pi}^{\dagger}(-{\bf k},-\epsilon_k t) & 0
                \end{array}
                \right ]\\
=&\tilde {H}(-{\bf k}, -\epsilon_kt).
  \end{aligned}
\end{equation}

2. From $[U_{l,\pi}^{\dagger}({\bf k},t)]^{-1}=q^{-1}U_{l,\pi}({\bf k},\epsilon_q t) q$ to $\tilde {H}({\bf k}, \epsilon_q t)=\tilde{Q}\tilde {H}({\bf k}, t) \tilde{Q}^{-1}$.
  \begin{equation}
\begin{aligned}
&\tilde{Q}\tilde {H}({\bf k}, t) \tilde{Q}^{-1}\\
=&\left[ \begin{array}{cc}
                q & 0\\
                0 & q
                \end{array}
                \right ]
\left[ \begin{array}{cc}
                0 & U_{l,\pi}({\bf k}, t)\\
                U_{l,\pi}^{\dagger}({\bf k}, t) & 0
                \end{array}
                \right ]
          \left[ \begin{array}{cc}
                q^{-1} & 0\\
                0 & q^{-1}
                \end{array}
                \right ] \\
=&\left[ \begin{array}{cc}
                0 & qU_{l,\pi}({\bf k}, t)q^{-1}\\
                qU_{l,\pi}^{\dagger}({\bf k}, t)q^{-1} & 0
                \end{array}
                \right ] \\
=&\left[ \begin{array}{cc}
                0 & (U_{l,\pi}^{\dagger}({\bf k}, \epsilon_q t))^{-1} \\
               (U_{l,\pi}({\bf k}, \epsilon_q t))^{-1} & 0
                \end{array}
                \right ]\\
=&\left[ \begin{array}{cc}
                0 &   U_{l,\pi}({\bf k}, \epsilon_q t)\\
                U_{l,\pi}^{\dagger}({\bf k}, \epsilon_q t) & 0
                \end{array}
                \right ]^{-1}\\
=&[\tilde {H}({\bf k}, \epsilon_q t)]^{-1}\\
=&\tilde {H}({\bf k}, \epsilon_q t).\\
  \end{aligned}
\end{equation}

3. From $[U_{l,\pi}^{\bf T}(-{\bf k},t)]^{-1}=c^{-1}U_{l,\pi}({\bf k},-\epsilon_c t)c$ to $ \tilde {H}(-{\bf k}, -\epsilon_c t)=\tilde{C}\tilde {H}({\bf k}, t)\tilde{C}^{-1}$.
  \begin{equation}
\begin{aligned}
&\tilde{C}\tilde {H}({\bf k}, t)\tilde{C}^{-1}\\
=&\left[ \begin{array}{cc}
                c & 0\\
                0 & c
                \end{array}
                \right ]
\left[ \begin{array}{cc}
                0 & U_{l,\pi}^*({\bf k}, t)\\
                U_{l,\pi}^{\bf T}({\bf k}, t) & 0
                \end{array}
                \right ]
          \left[ \begin{array}{cc}
                c^{-1} & 0\\
                0 & c^{-1}
                \end{array}
                \right ] \\
=&\left[ \begin{array}{cc}
                0 & cU_{l,\pi}^*({\bf k}, t)c^{-1}\\
                cU_{l,\pi}^{\bf T}({\bf k}, t)c^{-1} & 0
                \end{array}
                \right ] \\
=&\left[ \begin{array}{cc}
                0 & (U_{l,\pi}^{\dagger}(-{\bf k},-\epsilon_c t))^{-1} \\
               (U_{l,\pi}(-{\bf k},-\epsilon_c t))^{-1} & 0
                \end{array}
                \right ]\\
=&\left[ \begin{array}{cc}
                0 &   U_{l,\pi}(-{\bf k},-\epsilon_c t) \\
                U_{l,\pi}^{\dagger}(-{\bf k},-\epsilon_c t) & 0
                \end{array}
                \right ]^{-1}\\
=&[\tilde {H}(-{\bf k}, -\epsilon_c t)]^{-1}\\
=&\tilde {H}(-{\bf k}, -\epsilon_c t).\\
  \end{aligned}
\end{equation}

4. From $U_{l,\pi}({\bf k},-t)=p^{-1}U_{l,\pi}({\bf k},t)p$ to $ \tilde {H}({\bf k}, - t)=\tilde{P}\tilde {H}({\bf k}, t)\tilde{P}^{-1} $.
  \begin{equation}
\begin{aligned}
&\tilde{P}\tilde {H}({\bf k}, t)\tilde{P}^{-1} \\
=&\left[ \begin{array}{cc}
                p & 0\\
                0 & p
                \end{array}
                \right ]
\left[ \begin{array}{cc}
                0 & U_{l,\pi}({\bf k}, t)\\
                U_{l,\pi}^{\dagger}({\bf k}, t) & 0
                \end{array}
                \right ]
          \left[ \begin{array}{cc}
                p^{-1} & 0\\
                0 & p^{-1}
                \end{array}
                \right ] \\
=&\left[ \begin{array}{cc}
                0 & pU_{l,\pi}({\bf k}, t) p^{-1}\\
                pU_{l,\pi}^{\dagger}({\bf k}, t) p^{-1}& 0
                \end{array}
                \right ]  \\
=&\left[ \begin{array}{cc}
                0 & U_{l,\pi}({\bf k},-t)\\
                U_{l,\pi}^{\dagger}({\bf k},-t)  & 0
                \end{array}
                \right ] \\
=&\tilde {H}({\bf k}, - t). \\
  \end{aligned}
\end{equation}

\section{Derivation of Eqs. (\ref{eq:symshttp})-(\ref{eq:symshttk}) from Eqs. (\ref{eq:symsutp})-(\ref{eq:symsutk})}\label{apg}
1. From $[U^*(-{\bf k})]^{-1}=k^{-1}U({\bf k})k $ to $\tilde {H}(-{\bf k})=\bar{K}\tilde {H}({\bf k})\bar{K}^{-1}$ and $\bar{K}=\sigma_x \otimes k\mathcal{K}$. First, we have that $kU^*(-{\bf k})k^{-1}=[U({\bf k})]^{-1}$. Thus,
 \begin{equation}
\begin{aligned}
&\bar{K}\tilde {H}({\bf k})\bar{K}^{-1} \\
=&\left[ \begin{array}{cc}
                0 & k\\
                k & 0
                \end{array}
                \right ]
\left[ \begin{array}{cc}
                0 & U^*({\bf k})\\
                U^{\bf T}({\bf k}) & 0
                \end{array}
                \right ]
         \left[ \begin{array}{cc}
                0 & k^{-1}\\
                k^{-1} & 0
                \end{array}
                \right ] \\
=&\left[ \begin{array}{cc}
                0 & kU^{\bf T}({\bf k})k^{-1}\\
                kU^*({\bf k})k^{-1} & 0
                \end{array}
                \right ]  \\
=&\left[ \begin{array}{cc}
                0 & (U^{\dagger}(-{\bf k}))^{-1}\\
                (U(-{\bf k}))^{-1}  & 0
                \end{array}
                \right ] \\
 =&\left[ \begin{array}{cc}
                0 & U(-{\bf k})\\
                U^{\dagger}(-{\bf k}) & 0
                \end{array}
                \right ]^{-1} \\
=&[\tilde {H}(-{\bf k})]^{-1} =\tilde {H}(-{\bf k}). \\
  \end{aligned}
\end{equation}

2. From $U^*(-{\bf k})=k^{-1}U({\bf k})k $ to $\tilde {H}(-{\bf k})=\bar{K}\tilde {H}({\bf k})\bar{K}^{-1}$ and $\bar{K}=\sigma_0 \otimes k\mathcal{K}$. First, we have that $kU^*(-{\bf k})k^{-1}=U({\bf k})$. Thus,
 \begin{equation}
\begin{aligned}
&\bar{K}\tilde {H}({\bf k})\bar{K}^{-1} \\
=&\left[ \begin{array}{cc}
                k & 0\\
                0 & k
                \end{array}
                \right ]
\left[ \begin{array}{cc}
                0 & U^*({\bf k})\\
                U^{\bf T}({\bf k}) & 0
                \end{array}
                \right ]
         \left[ \begin{array}{cc}
                k^{-1}& 0\\
                0 & k^{-1}
                \end{array}
                \right ] \\
=&\left[ \begin{array}{cc}
                0 & kU^*({\bf k})k^{-1}\\
                kU^{\bf T}({\bf k}) k^{-1} & 0
                \end{array}
                \right ]  \\
=&\left[ \begin{array}{cc}
                0 & U(-{\bf k})\\
                U^{\dagger}(-{\bf k}) & 0
                \end{array}
                \right ] \\
=&\tilde {H}(-{\bf k}). \\
  \end{aligned}
\end{equation}

3. From $[U^{\dagger}({\bf k})]^{-1}=q^{-1}U({\bf k}) q $ to $\tilde {H}({\bf k})=\bar{Q}\tilde {H}({\bf k}) \bar{Q}^{-1}$ and $\bar{Q}=\sigma_0\otimes q $.
 \begin{equation}
\begin{aligned}
&\bar{Q}\tilde {H}({\bf k}) \bar{Q}^{-1} \\
=&\left[ \begin{array}{cc}
                q & 0\\
                0 & q
                \end{array}
                \right ]
\left[ \begin{array}{cc}
                0 & U({\bf k})\\
                U^{\dagger}({\bf k}) & 0
                \end{array}
                \right ]
\left[ \begin{array}{cc}
                q^{-1} & 0\\
                0 & q^{-1}
                \end{array}
                \right ] \\
=&\left[ \begin{array}{cc}
                0 & qU({\bf k})q^{-1}\\
                qU^{\dagger}({\bf k})q^{-1} & 0
                \end{array}
                \right ]  \\
=&\left[ \begin{array}{cc}
                0 & (U^{\dagger}({\bf k}))^{-1} \\
                (U({\bf k}))^{-1}  & 0
                \end{array}
                \right ] \\
 =&\left[ \begin{array}{cc}
                0 &  U({\bf k})\\
                U^{\dagger}({\bf k}) & 0
                \end{array}
                \right ]^{-1} \\
=&[\tilde {H}({\bf k})]^{-1} =\tilde {H}({\bf k}). \\
  \end{aligned}
\end{equation}

4. From $U^{\dagger}({\bf k})=q^{-1}U({\bf k}) q $ to $\tilde {H}({\bf k})=\bar{Q}\tilde {H}({\bf k}) \bar{Q}^{-1}$ and $\bar{Q}=\sigma_x\otimes q $.
 \begin{equation}
\begin{aligned}
&\bar{Q}\tilde {H}({\bf k}) \bar{Q}^{-1} \\
=&\left[ \begin{array}{cc}
                0 & q\\
                q & 0
                \end{array}
                \right ]
\left[ \begin{array}{cc}
                0 & U({\bf k})\\
                U^{\dagger}({\bf k}) & 0
                \end{array}
                \right ]
\left[ \begin{array}{cc}
                0 &  q^{-1}\\
                q^{-1}& 0
                \end{array}
                \right ] \\
=&\left[ \begin{array}{cc}
                0 &   qU^{\dagger}({\bf k})q^{-1}\\
                qU({\bf k})q^{-1} & 0
                \end{array}
                \right ]  \\
=&\left[ \begin{array}{cc}
                0 &   U({\bf k})\\
                U^{\dagger}({\bf k}) & 0
                \end{array}
                \right ]  \\
=&\tilde {H}({\bf k}). \\
  \end{aligned}
\end{equation}

5. From $U^T(-{\bf k})=c^{-1}U({\bf k})c $ to $\tilde {H}(-{\bf k})=\bar{C}\tilde {H}^*({\bf k})\bar{C}^{-1}$ and $\bar{C}=\sigma_x\otimes c\mathcal{K}$.
 \begin{equation}
\begin{aligned}
&\bar{C}\tilde {H}^*({\bf k})\bar{C}^{-1} \\
=&\left[ \begin{array}{cc}
                0 & c\\
                c & 0
                \end{array}
                \right ]
\left[ \begin{array}{cc}
                0 & U^*({\bf k})\\
                U^{\bf T}({\bf k}) & 0
                \end{array}
                \right ]
\left[ \begin{array}{cc}
                0 & c^{-1}\\
                c^{-1} & 0
                \end{array}
                \right ] \\
 =&\left[ \begin{array}{cc}
                0 &   cU^{\bf T}({\bf k})c^{-1}\\
                cU^*({\bf k})c^{-1} & 0
                \end{array}
                \right ]  \\
=&\left[ \begin{array}{cc}
                0 &  U(-{\bf k})\\
                U^{\dagger}(-{\bf k}) & 0
                \end{array}
                \right ] \\
=&\tilde {H}(-{\bf k}). \\
  \end{aligned}
\end{equation}

6. From $[U^T(-{\bf k})]^{-1}=c^{-1}U({\bf k})c $ to $\tilde {H}(-{\bf k})=\bar{C}\tilde {H}^*({\bf k})\bar{C}^{-1}$ and $\bar{C}=\sigma_0\otimes c\mathcal{K}$.
 \begin{equation}
\begin{aligned}
&\bar{C}\tilde {H}^*({\bf k})\bar{C}^{-1} \\
=&\left[ \begin{array}{cc}
                c & 0\\
                0 & c
                \end{array}
                \right ]
\left[ \begin{array}{cc}
                0 & U^*({\bf k})\\
                U^{\bf T}({\bf k}) & 0
                \end{array}
                \right ]
\left[ \begin{array}{cc}
                c^{-1} & 0\\
                0 & c^{-1}
                \end{array}
                \right ] \\
 =&\left[ \begin{array}{cc}
                0 &   cU^*({\bf k})c^{-1}\\
                 cU^{\bf T}({\bf k})c^{-1}& 0
                \end{array}
                \right ]  \\
=&\left[ \begin{array}{cc}
                0 &   (U^{\dagger}(-{\bf k}))^{-1}\\
                (U(-{\bf k}))^{-1} & 0
                \end{array}
                \right ] \\
=&\left[ \begin{array}{cc}
                0 & U(-{\bf k}) \\
                U^{\dagger}(-{\bf k})& 0
                \end{array}
                \right ]^{-1} \\
=&[\tilde {H}(-{\bf k})]^{-1} =\tilde {H}(-{\bf k}) .\\
  \end{aligned}
\end{equation}

7. From $[U({\bf k})]^{-1}=p^{-1}U({\bf k})p $ to $\tilde {H}({\bf k})=\bar{P}\tilde {H}({\bf k})\bar{P}^{-1}$ and $\bar{P}=\sigma_x \otimes p$.
 \begin{equation}
\begin{aligned}
&\bar{P}\tilde {H}({\bf k})\bar{P}^{-1} \\
=&\left[ \begin{array}{cc}
                0 & p\\
                p & 0
                \end{array}
                \right ]
\left[ \begin{array}{cc}
                0 & U({\bf k})\\
                U^{\dagger}({\bf k}) & 0
                \end{array}
                \right ]
\left[ \begin{array}{cc}
                0 & p^{-1}\\
                p^{-1} & 0
                \end{array}
                \right ] \\
 =&\left[ \begin{array}{cc}
                0 &  pU^{\dagger}({\bf k})p^{-1}\\
                pU({\bf k})p^{-1} & 0
                \end{array}
                \right ]  \\
=&\left[ \begin{array}{cc}
                0 & (U^{\dagger}({\bf k}))^{-1}\\
                (U({\bf k}))^{-1}& 0
                \end{array}
                \right ] \\
=&\left[ \begin{array}{cc}
                0 &  U({\bf k})\\
                U^{\dagger}({\bf k})& 0
                \end{array}
                \right ]^{-1} \\
=&[\tilde {H}({\bf k})]^{-1} =\tilde {H}({\bf k}) .\\
  \end{aligned}
\end{equation}

\section{Clifford algebra's extension for $\tilde{H}({\bf k},t)$ and $\tilde{H}({\bf k})$}\label{aph}
Table \ref{HktCliAlg} and Table \ref{HkCliAlg} list the Clifford algebra's extensions for $\tilde{H}({\bf k},t)$ and $\tilde{H}({\bf k})$, respectively.

\begin{lemma}
  For any GBL class in Table \ref{HktCliAlg} and Table \ref{HkCliAlg}, the classification space is $C_n$, $C_n^2$, $R_m$, or $R_m^2$($n=0,1$, $m=0,1,...,7$). The space of mass term of the corresponding GBL class in dimension $d$ is $C_{n-d}$, $C_{n-d}^2$, $R_{m-d}$, or $R_{m-d}^2$, respectively.
\end{lemma}

 Proof: First for classes with $\tilde{K}$ and $\tilde{C}$. We note that the anti-unitary symmetry operators $\tilde{K}$ and $\tilde{C}$ always reverse the momentum (${\bf k}\rightarrow -{\bf k}$) while the unitary symmetry operators $\tilde{P}$ and $\tilde{Q}$ don't. When adding $\gamma_1,\gamma_2$,...,$\gamma_d$ into the Clifford algebra's extension in Table \ref{HktCliAlg} and Table \ref{HkCliAlg}, the difference between the coefficient of $\gamma_1,\gamma_2$,...,$\gamma_d$ and the coefficient of $\gamma_0$ can always be chosen as $J$ or $-J$. For class without $\tilde{K}$ and $\tilde{C}$, when adding $\gamma_1,\gamma_2$,...,$\gamma_d$ into the Clifford algebra's extension in Table \ref{HktCliAlg} and Table \ref{HkCliAlg}, the coefficients of $\gamma_1,\gamma_2$,...,$\gamma_d$ and the coefficient of $\gamma_0$ are can be chosen as the same. Hence Lemma 12 is proved. Let us consider for example:

{\it Class P in Table \ref{HktCliAlg}.}  The Clifford algebra's extension in $d$ dimensions is $\left\{\tilde{\gamma}_t,\gamma_1,\gamma_2,...,\gamma_d,\Sigma,\tilde{\gamma}_t\tilde{P}\right\} \rightarrow
\left\{\tilde{\gamma}_0,\gamma_1,\gamma_2,...,\gamma_d,\tilde{\gamma}_t,\Sigma,\tilde{\gamma}_t\tilde{P}\right\}=Cl_{3+d}\rightarrow
Cl_{4+d}$. The space of mass term is $C_{3+d}\simeq C_{1-d}$.

{\it Class Qa in Table \ref{HktCliAlg}.}  The Clifford algebra's extension in $d$ dimensions is $\left\{\tilde{\gamma}_t,\gamma_1,\gamma_2,...,\gamma_d,\Sigma\right\}\otimes\left\{\tilde{Q}\right\} \rightarrow
\left\{\tilde{\gamma}_0,\gamma_1,\gamma_2,...,\gamma_d,\tilde{\gamma}_t,\Sigma\right\}\otimes\left\{\tilde{Q}\right\} =Cl_{2+d}\otimes Cl_1\rightarrow  Cl_{3+d}\otimes Cl_1$. The space of mass term is $C_{2+d}^2\simeq C_{-d}^2$.

{\it Class K1a in Table \ref{HktCliAlg}.}  The Clifford algebra's extension in $d$ dimensions is $\left\{\tilde{\gamma}_t,\gamma_1,\gamma_2,...,\gamma_d,J\Sigma,\tilde{K},J\tilde{K}\right\} \rightarrow
\left\{J\tilde{\gamma}_0,\gamma_1,\gamma_2,...,\gamma_d,\tilde{\gamma}_t,J\Sigma,\tilde{K},J\tilde{K}\right\}=Cl_{1,3+d}\rightarrow
Cl_{2,3+d}$. The space of mass term is $R_{-d}$.

{\it Class QC1a in Table \ref{HktCliAlg}.}  The Clifford algebra's extension in $d$ dimensions is $\left\{\tilde{\gamma}_t,\gamma_1,\gamma_2,...,\gamma_d,J\Sigma,\tilde{C},J\tilde{C}\right\}
\otimes\left\{\tilde{Q}\right\} \rightarrow
\left\{J\tilde{\gamma}_0,\gamma_1,\gamma_2,...,\gamma_d,\tilde{\gamma}_t,J\Sigma,\tilde{C},J\tilde{C},\right\}\otimes\left\{\tilde{Q}\right\}
 =Cl_{1,3+d}\otimes Cl_{0,1}\rightarrow  Cl_{2,3+d}\otimes Cl_{0,1}$. The space of mass term is $R_{-d}^2$.

 \onecolumngrid
 \begin{table*}[htbp]\footnotesize
  \begin{center}
  \caption{\label{tab:tableIII}  The construction of Clifford algebra's extension and its corresponding classifying space (Cl) for $\tilde{H}({\bf k},t)$ in all GBL classes.  $J=i$ is the imaginary unit.\label{HktCliAlg}}
\begin{tabular}{|c|c|c|}
    \hline
    \;\;\;GBL\;\;\;&$\qquad \qquad \qquad \qquad \qquad \qquad \qquad \qquad \;\;\;$Clifford algebra's extension$\;\;\; \qquad \qquad \qquad \qquad \qquad \qquad \qquad \qquad$&Cl\\
  \hline
Non &$\left\{\tilde{\gamma}_1,...,\tilde{\gamma}_d,\tilde{\gamma}_t,\Sigma\right\} \rightarrow
\left\{\tilde{\gamma}_0,\tilde{\gamma}_1,...,\tilde{\gamma}_d,\tilde{\gamma}_t,\Sigma\right\}=Cl_{d+2}\rightarrow
Cl_{d+3}$&$C_0$\\
  \hline
P &  $\left\{\tilde{\gamma}_t,\Sigma,\tilde{\gamma}_t\tilde{P}\right\} \rightarrow
\left\{\tilde{\gamma}_0,\tilde{\gamma}_t,\Sigma,\tilde{\gamma}_t\tilde{P}\right\}=Cl_{3}\rightarrow
Cl_{4}$&$C_1 $\\
  \hline
Qa &  $\left\{\tilde{\gamma}_t,\Sigma\right\}\otimes\left\{\tilde{Q}\right\} \rightarrow
\left\{\tilde{\gamma}_0,\tilde{\gamma}_t,\Sigma\right\}\otimes\left\{\tilde{Q}\right\} =Cl_{2}\otimes Cl_1\rightarrow  Cl_{3}\otimes Cl_1$ & $ C_0^{\times 2} $ \\
  \hline
  Qb&  $\left\{\tilde{\gamma}_t,\Sigma,\tilde{\gamma}_t\tilde{Q}\right\} \rightarrow
  \left\{\tilde{\gamma}_0,\tilde{\gamma}_t,\Sigma,\tilde{\gamma}_t \tilde{Q}\right\}=Cl_{3}\rightarrow
  Cl_{4}$  & $ C_1$\\
  \hline
K1a &  $\left\{\tilde{\gamma}_t,J\Sigma,\tilde{K},J\tilde{K}\right\} \rightarrow
\left\{J\tilde{\gamma}_0,\tilde{\gamma}_t,J\Sigma,\tilde{K},J\tilde{K}\right\}=Cl_{1,3}\rightarrow
Cl_{2,3}$  & $ R_0 $  \\
  \hline
  K1b &  $\left\{J\tilde{\gamma}_t,J\Sigma,\tilde{K},J\tilde{K}\right\} \rightarrow
  \left\{J\tilde{\gamma}_0,J\tilde{\gamma}_t,J\Sigma,\tilde{K},J\tilde{K}\right\}=Cl_{2,2}\rightarrow
  Cl_{3,2}$   & $ R_2 $ \\
  \hline

K2a & $\left\{\tilde{\gamma}_t,J\Sigma,\tilde{K},J\tilde{K}\right\} \rightarrow
\left\{J\tilde{\gamma}_0,\tilde{\gamma}_t,J\Sigma,\tilde{K},J\tilde{K}\right\}=Cl_{3,1}\rightarrow
Cl_{4,1}$  & $ R_4 $\\
  \hline
  K2b & $\left\{J\tilde{\gamma}_t,J\Sigma,\tilde{K},J\tilde{K}\right\} \rightarrow
  \left\{J\tilde{\gamma}_0,J\tilde{\gamma}_t,J\Sigma,\tilde{K},J\tilde{K}\right\}=Cl_{4,0}\rightarrow
Cl_{5,0}$ &$ R_6 $\\
  \hline

C1 & $\left\{\tilde{\gamma}_t,J\Sigma,\tilde{C},J\tilde{C}\right\} \rightarrow
\left\{J\tilde{\gamma}_0,\tilde{\gamma}_t,J\Sigma,\tilde{C},J\tilde{C}\right\}=Cl_{1,3}\rightarrow
Cl_{2,3}$ & $ R_0 $\\
  \hline
C2 & $\left\{\tilde{\gamma}_t,J\Sigma,\tilde{C},J\tilde{C}\right\} \rightarrow
\left\{J\tilde{\gamma}_0,\tilde{\gamma}_t,J\Sigma,\tilde{C},J\tilde{C}\right\}=Cl_{3,1}\rightarrow
Cl_{4,1}$ & $ R_4 $\\
  \hline
C3 &   $\left\{J\tilde{\gamma}_t,J\Sigma,\tilde{C},J\tilde{C}\right\} \rightarrow
\left\{J\tilde{\gamma}_0,J\tilde{\gamma}_t,J\Sigma,\tilde{C},J\tilde{C}\right\}=Cl_{2,2}\rightarrow
Cl_{3,2}$ & $ R_2 $\\
  \hline
C4 &  $\left\{J\tilde{\gamma}_t,J\Sigma,\tilde{C},J\tilde{C}\right\} \rightarrow
\left\{J\tilde{\gamma}_0,J\tilde{\gamma}_t,J\Sigma,\tilde{C},J\tilde{C}\right\}=Cl_{4,0}\rightarrow
Cl_{5,0}$ & $ R_6 $\\
  \hline

PQ1 & $\left\{\tilde{\gamma}_t,\Sigma,\tilde{\gamma}_tP\right\}\otimes\left\{\tilde{Q}\right\} \rightarrow
\left\{\tilde{\gamma}_0,\tilde{\gamma}_t,\Sigma,\tilde{\gamma}_tP\right\}\otimes\left\{\tilde{Q}\right\} =Cl_{3}\otimes Cl_1\rightarrow  Cl_{4}\otimes Cl_1$  & $ C_1^{\times 2} $\\
  \hline
PQ2 &  $\left\{\tilde{\gamma}_t,\Sigma,\tilde{\gamma}_t\tilde{P},\tilde{\gamma}_t\tilde{P}\tilde{Q}\right\} \rightarrow
\left\{\tilde{\gamma}_0,\tilde{\gamma}_t,\Sigma,\tilde{\gamma}_t\tilde{P},\tilde{\gamma}_t\tilde{P}\tilde{Q}\right\}=Cl_{4}\rightarrow
Cl_{5}$ & $ C_0  $\\
  \hline
PK1 &  $\left\{\tilde{\gamma}_t,J\Sigma,\gamma_t\tilde{P},\tilde{K},J\tilde{K}\right\} \rightarrow
\left\{J\tilde{\gamma}_0,\tilde{\gamma}_t,J\Sigma,\gamma_t\tilde{P},\tilde{K},J\tilde{K}\right\}=Cl_{2,3}\rightarrow
Cl_{3,3}$ & $ R_1 $\\
  \hline
PK2  & $\left\{\tilde{\gamma}_t,J\Sigma,\gamma_t\tilde{P},\tilde{K},J\tilde{K}\right\} \rightarrow
\left\{J\tilde{\gamma}_0,\tilde{\gamma}_t,J\Sigma,\gamma_t\tilde{P},\tilde{K},J\tilde{K}\right\}=Cl_{4,1}\rightarrow
Cl_{5,1}$ & $ R_5  $\\
  \hline
PK3a & $\left\{\tilde{\gamma}_t,J\Sigma,J\gamma_t\tilde{P},\tilde{K},J\tilde{K}\right\}
 \rightarrow \left\{J\tilde{\gamma}_0,\tilde{\gamma}_t,J\Sigma,J\gamma_t\tilde{P},\tilde{K},J\tilde{K}\right\}=Cl_{1,4}\rightarrow
Cl_{2,4}$  & $ R_7 $\\
  \hline
PK3b & $\left\{\tilde{\gamma}_t,J\Sigma,J\gamma_t\tilde{P},\tilde{K},J\tilde{K}\right\}
\rightarrow \left\{J\tilde{\gamma}_0,\tilde{\gamma}_t,J\Sigma,J\gamma_t\tilde{P},\tilde{K},J\tilde{K}\right\}=Cl_{3,2}\rightarrow
Cl_{4,2}$    & $ R_3 $ \\
  \hline
PC1  &
$\left\{\tilde{\gamma}_t,J\Sigma,J\tilde{P},\tilde{C},J\tilde{C}\right\}
\rightarrow \left\{J\tilde{\gamma}_0,\tilde{\gamma}_t,J\Sigma,J\tilde{P},\tilde{C},J\tilde{C}\right\}=Cl_{2,3}\rightarrow
Cl_{3,3}$  & $ R_1 $\\
  \hline
PC2  &
$\left\{\tilde{\gamma}_t,J\Sigma,J\tilde{\gamma}_t\tilde{P},\tilde{C},J\tilde{C}\right\}
\rightarrow \left\{J\tilde{\gamma}_0,\tilde{\gamma}_t,J\Sigma,J\tilde{\gamma}_t\tilde{P},\tilde{C},J\tilde{C}\right\}=Cl_{1,4}\rightarrow
Cl_{2,4}$& $ R_7 $\\
  \hline
PC3  &
$\left\{\tilde{\gamma}_t,J\Sigma,\gamma_t\tilde{P},\tilde{C},J\tilde{C}\right\} \rightarrow
\left\{J\tilde{\gamma}_0,\tilde{\gamma}_t,J\Sigma,\gamma_t\tilde{P},\tilde{C},J\tilde{C}\right\}=Cl_{4,1}\rightarrow
Cl_{5,1}$ & $ R_5 $\\
  \hline
PC4  &
$\left\{\tilde{\gamma}_t,J\Sigma,J\gamma_t\tilde{P},\tilde{C},J\tilde{C}\right\}
\rightarrow \left\{J\tilde{\gamma}_0,\tilde{\gamma}_t,J\Sigma,J\gamma_t\tilde{P},\tilde{C},J\tilde{C}\right\}=Cl_{3,2}\rightarrow
Cl_{4,2}$ & $ R_3$\\
  \hline
QC1a & $\left\{\tilde{\gamma}_t,J\Sigma,\tilde{C},J\tilde{C}\right\}
\otimes\left\{\tilde{Q}\right\} \rightarrow
\left\{J\tilde{\gamma}_0,\tilde{\gamma}_t,J\Sigma,\tilde{C},J\tilde{C},\right\}\otimes\left\{\tilde{Q}\right\}
 =Cl_{1,3}\otimes Cl_{0,1}\rightarrow  Cl_{2,3}\otimes Cl_{0,1}$ & $ R_0^{\times 2}$\\
  \hline
  QC1b &  $\left\{\tilde{\gamma}_t,J\Sigma,\tilde{C},J\tilde{C},\gamma_t\tilde{Q}\right\}
  \rightarrow \left\{J\tilde{\gamma}_0,\tilde{\gamma}_t,J\Sigma,\tilde{C},J\tilde{C},\gamma_t\tilde{Q}\right\}=Cl_{2,3}\rightarrow
  Cl_{3,3}$  & $ R_1 $\\
  \hline

QC2a &  $\left\{\tilde{\gamma}_t,J\Sigma,\tilde{C},J\tilde{C}\right\}
\otimes\left\{J\tilde{Q}\right\} \rightarrow
\left\{J\tilde{\gamma}_0,\tilde{\gamma}_t,J\Sigma,\tilde{C},J\tilde{C},\right\}\otimes\left\{\tilde{JQ}\right\}
 =Cl_{1,3}\otimes Cl_{1,0}\rightarrow  Cl_{2,3}\otimes Cl_{1,0}$ & $ C_0 $\\
  \hline
  QC2b &  $\left\{\tilde{\gamma}_t,J\Sigma,\tilde{C},J\tilde{C},J\gamma_t\tilde{Q}\right\}
  \rightarrow \left\{J\tilde{\gamma}_0,\tilde{\gamma}_t,J\Sigma,\tilde{C},J\tilde{C},J\gamma_t\tilde{Q}\right\}=Cl_{1,4}\rightarrow
  Cl_{2,4}$  & $ R_7 $\\
  \hline

QC3a & $\left\{\tilde{\gamma}_t,J\Sigma,\tilde{C},J\tilde{C}\right\}
\otimes\left\{\tilde{Q}\right\} \rightarrow
\left\{J\tilde{\gamma}_0,\tilde{\gamma}_t,J\Sigma,\tilde{C},J\tilde{C},\right\}\otimes\left\{\tilde{Q}\right\}
 =Cl_{3,1}\otimes Cl_{0,1}\rightarrow  Cl_{4,1}\otimes Cl_{0,1}$  & $ R_4^{\times 2}$\\
  \hline
  QC3b &  $\left\{\tilde{\gamma}_t,J\Sigma,\tilde{C},J\tilde{C},\gamma_t\tilde{Q}\right\}
  \rightarrow \left\{J\tilde{\gamma}_0,\tilde{\gamma}_t,J\Sigma,\tilde{C},J\tilde{C},\gamma_t\tilde{Q}\right\}=Cl_{4,1}\rightarrow
  Cl_{5,1}$  & $ R_5$\\
  \hline

QC4a &  $\left\{\tilde{\gamma}_t,J\Sigma,\tilde{C},J\tilde{C}\right\}
\otimes\left\{J\tilde{Q}\right\} \rightarrow
\left\{J\tilde{\gamma}_0,\tilde{\gamma}_t,J\Sigma,\tilde{C},J\tilde{C},\right\}\otimes\left\{\tilde{JQ}\right\}
 =Cl_{3,1}\otimes Cl_{1,0}\rightarrow  Cl_{4,1}\otimes Cl_{1,0}$ & $ C_0 $\\
  \hline
  QC4b &  $\left\{\tilde{\gamma}_t,J\Sigma,\tilde{C},J\tilde{C},J\gamma_t\tilde{Q}\right\}
  \rightarrow \left\{J\tilde{\gamma}_0,\tilde{\gamma}_t,J\Sigma,\tilde{C},J\tilde{C},J\gamma_t\tilde{Q}\right\}=Cl_{3,2}\rightarrow
  Cl_{4,2}$  & $ R_3 $\\
  \hline

  \multicolumn{3}{c}{continued on next page}
\end{tabular}
\end{center}
\end{table*}

\begin{table*}[htbp]\footnotesize
\begin{center}
\begin{tabular}{|c|c|c|}
\multicolumn{3}{c}{TABLE III  --- continued} \\ \hline

QC5a & $\left\{J\tilde{\gamma}_t,J\Sigma,\tilde{C},J\tilde{C}\right\}
\otimes\left\{\tilde{Q}\right\} \rightarrow
\left\{J\tilde{\gamma}_0,J\tilde{\gamma}_t,J\Sigma,\tilde{C},J\tilde{C},\right\}\otimes\left\{\tilde{Q}\right\}
 =Cl_{2,2}\otimes Cl_{0,1}\rightarrow  Cl_{3,2}\otimes Cl_{0,1}$   & $ R_2^{\times 2} $\\
  \hline
  QC5b &  $\left\{J\tilde{\gamma}_t,J\Sigma,\tilde{C},J\tilde{C},J\gamma_t\tilde{Q}\right\}
  \rightarrow \left\{J\tilde{\gamma}_0,J\tilde{\gamma}_t,J\Sigma,\tilde{C},J\tilde{C},J\gamma_t\tilde{Q}\right\}=Cl_{2,3}\rightarrow
  Cl_{3,3}$  & $ R_1 $\\
  \hline

QC6a &  $\left\{J\tilde{\gamma}_t,J\Sigma,\tilde{C},J\tilde{C}\right\}
\otimes\left\{\tilde{JQ}\right\} \rightarrow
\left\{J\tilde{\gamma}_0,J\tilde{\gamma}_t,J\Sigma,\tilde{C},J\tilde{C},\right\}\otimes\left\{\tilde{JQ}\right\}
 =Cl_{2,2}\otimes Cl_{1,0}\rightarrow  Cl_{3,2}\otimes Cl_{1,0}$   & $ C_0 $\\
  \hline
  QC6b &  $\left\{J\tilde{\gamma}_t,J\Sigma,\tilde{C},J\tilde{C},\gamma_t\tilde{Q}\right\}
  \rightarrow \left\{J\tilde{\gamma}_0,J\tilde{\gamma}_t,J\Sigma,\tilde{C},J\tilde{C},\gamma_t\tilde{Q}\right\}=Cl_{3,2}\rightarrow
  Cl_{4,2}$ & $ R_3 $\\
  \hline

QC7a &  $\left\{J\tilde{\gamma}_t,J\Sigma,\tilde{C},J\tilde{C}\right\}
\otimes\left\{\tilde{Q}\right\} \rightarrow
\left\{J\tilde{\gamma}_0,J\tilde{\gamma}_t,J\Sigma,\tilde{C},J\tilde{C},\right\}\otimes\left\{\tilde{Q}\right\}
 =Cl_{4,0}\otimes Cl_{0,1}\rightarrow  Cl_{5,0}\otimes Cl_{0,1}$  & $ R_6^{\times 2}$\\
  \hline
  QC7b &  $\left\{J\tilde{\gamma}_t,J\Sigma,\tilde{C},J\tilde{C},J\gamma_t\tilde{Q}\right\}
  \rightarrow \left\{J\tilde{\gamma}_0,J\tilde{\gamma}_t,J\Sigma,\tilde{C},J\tilde{C},J\gamma_t\tilde{Q}\right\}=Cl_{4,1}\rightarrow
  Cl_{5,1}$   & $ R_5 $\\
  \hline

QC8a  &  $\left\{J\tilde{\gamma}_t,J\Sigma,\tilde{C},J\tilde{C}\right\}
\otimes\left\{\tilde{JQ}\right\} \rightarrow
\left\{J\tilde{\gamma}_0,J\tilde{\gamma}_t,J\Sigma,\tilde{C},J\tilde{C},\right\}\otimes\left\{\tilde{JQ}\right\}
 =Cl_{4,0}\otimes Cl_{1,0}\rightarrow  Cl_{5,0}\otimes Cl_{1,0}$ & $ C_0 $\\
  \hline
  QC8b &   $\left\{J\tilde{\gamma}_t,J\Sigma,\tilde{C},J\tilde{C},\gamma_t\tilde{Q}\right\}
  \rightarrow \left\{J\tilde{\gamma}_0,J\tilde{\gamma}_t,J\Sigma,\tilde{C},J\tilde{C},\gamma_t\tilde{Q}\right\}=Cl_{5,0}\rightarrow
  Cl_{6,0}$ & $ R_7 $\\
  \hline

PQC1 & $\left\{\tilde{\gamma}_t,J\Sigma,\tilde{\gamma}_t\tilde{P},\tilde{C},J\tilde{C}\right\}
\otimes\left\{\tilde{Q}\right\} \rightarrow
\left\{J\tilde{\gamma}_0,\tilde{\gamma}_t,J\Sigma,\tilde{\gamma}_t\tilde{P},\tilde{C},J\tilde{C}\right\}\otimes\left\{\tilde{Q}\right\}
 =Cl_{2,3}\otimes Cl_{0,1}\rightarrow  Cl_{3,3}\otimes Cl_{0,1}$  & $ R_1^{\times 2} $\\
  \hline
PQC2& $\left\{\tilde{\gamma}_t,J\Sigma,\tilde{\gamma}_t\tilde{P},\tilde{C},J\tilde{C}\right\}
\otimes\left\{J\tilde{Q}\right\} \rightarrow
\left\{J\tilde{\gamma}_0,\tilde{\gamma}_t,J\Sigma,\tilde{\gamma}_t\tilde{P},\tilde{C},J\tilde{C}\right\}\otimes\left\{J\tilde{Q}\right\}
 =Cl_{2,3}\otimes Cl_{1,0}\rightarrow  Cl_{3,3}\otimes Cl_{1,0}$ & $ C_1 $\\
  \hline

PQC3&  $\left\{\tilde{\gamma}_t,J\Sigma,J\tilde{\gamma}_t\tilde{P},\tilde{C},J\tilde{C},J\tilde{\gamma}_t\tilde{P}\tilde{Q}\right\}
 \rightarrow
\left\{J\tilde{\gamma}_0,\tilde{\gamma}_t,J\Sigma,J\tilde{\gamma}_t\tilde{P},\tilde{C},J\tilde{C},J\tilde{\gamma}_t\tilde{P}\tilde{Q}\right\}
 =Cl_{2,4}\rightarrow  Cl_{3,4}$  & $ R_0 $\\
  \hline
PQC4&  $\left\{\tilde{\gamma}_t,J\Sigma,J\tilde{\gamma}_t\tilde{P},\tilde{C},J\tilde{C},\tilde{\gamma}_t\tilde{P}\tilde{Q}\right\}
\rightarrow
\left\{J\tilde{\gamma}_0,\tilde{\gamma}_t,J\Sigma,J\tilde{\gamma}_t\tilde{P},\tilde{C},J\tilde{C},\tilde{\gamma}_t\tilde{P}\tilde{Q}\right\}
=Cl_{1,5}\rightarrow  Cl_{2,5}$  & $ R_6 $\\
  \hline
PQC5 & $\left\{\tilde{\gamma}_t,J\Sigma,\tilde{\gamma}_t\tilde{P},\tilde{C},J\tilde{C}\right\}
\otimes\left\{\tilde{Q}\right\} \rightarrow
\left\{J\tilde{\gamma}_0,\tilde{\gamma}_t,J\Sigma,\tilde{\gamma}_t\tilde{P},\tilde{C},J\tilde{C}\right\}\otimes\left\{\tilde{Q}\right\}
 =Cl_{4,1}\otimes Cl_{0,1}\rightarrow  Cl_{5,1}\otimes Cl_{0,1}$ &$ R_5^{\times 2} $\\
  \hline
PQC6 & $\left\{\tilde{\gamma}_t,J\Sigma,\tilde{\gamma}_t\tilde{P},\tilde{C},J\tilde{C}\right\}
\otimes\left\{J\tilde{Q}\right\} \rightarrow
\left\{J\tilde{\gamma}_0,\tilde{\gamma}_t,J\Sigma,\tilde{\gamma}_t\tilde{P},\tilde{C},J\tilde{C}\right\}\otimes\left\{J\tilde{Q}\right\}
 =Cl_{4,1}\otimes Cl_{1,0}\rightarrow  Cl_{5,1}\otimes Cl_{1,0}$ & $ C_1$\\
  \hline
PQC7 &  $\left\{\tilde{\gamma}_t,J\Sigma,J\tilde{\gamma}_t\tilde{P},\tilde{C},J\tilde{C},J\tilde{\gamma}_t\tilde{P}\tilde{Q}\right\}
 \rightarrow
\left\{J\tilde{\gamma}_0,\tilde{\gamma}_t,J\Sigma,J\tilde{\gamma}_t\tilde{P},\tilde{C},J\tilde{C},J\tilde{\gamma}_t\tilde{P}\tilde{Q}\right\}
 =Cl_{4,2}\rightarrow  Cl_{5,2}$ & $ R_4 $\\
  \hline
PQC8 &  $\left\{\tilde{\gamma}_t,J\Sigma,J\tilde{\gamma}_t\tilde{P},\tilde{C},J\tilde{C},\tilde{\gamma}_t\tilde{P}\tilde{Q}\right\}
\rightarrow
\left\{J\tilde{\gamma}_0,\tilde{\gamma}_t,J\Sigma,J\tilde{\gamma}_t\tilde{P},\tilde{C},J\tilde{C},\tilde{\gamma}_t\tilde{P}\tilde{Q}\right\}
=Cl_{3,3}\rightarrow  Cl_{4,3}$  & $ R_2  $\\
  \hline

PQC9a &  $\left\{\tilde{\gamma}_t,J\Sigma,J\tilde{\gamma}_t\tilde{P},\tilde{C},J\tilde{C}\right\}
\otimes\left\{\tilde{Q}\right\} \rightarrow
\left\{J\tilde{\gamma}_0,\tilde{\gamma}_t,J\Sigma,J\tilde{\gamma}_t\tilde{P},\tilde{C},J\tilde{C}\right\}\otimes\left\{\tilde{Q}\right\}
 =Cl_{1,4}\otimes Cl_{0,1}\rightarrow  Cl_{2,4}\otimes Cl_{0,1}$  & $ R_7^{\times 2} $\\
  \hline
  PQC9b & $\left\{\tilde{\gamma}_t,J\Sigma,J\tilde{\gamma}_t\tilde{P},\tilde{C},J\tilde{C}\right\}
  \otimes\left\{J\tilde{Q}\right\} \rightarrow
  \left\{J\tilde{\gamma}_0,\tilde{\gamma}_t,J\Sigma,J\tilde{\gamma}_t\tilde{P},\tilde{C},J\tilde{C}\right\}\otimes\left\{J\tilde{Q}\right\}
   =Cl_{1,4}\otimes Cl_{1,0}\rightarrow  Cl_{2,4}\otimes Cl_{1,0}$  & $ C_1 $\\
  \hline

PQC10a & $\left\{\tilde{\gamma}_t,J\Sigma,\tilde{\gamma}_t\tilde{P},\tilde{C},J\tilde{C},\tilde{\gamma}_t\tilde{P}\tilde{Q}\right\}
\rightarrow
\left\{J\tilde{\gamma}_0,\tilde{\gamma}_t,J\Sigma,\tilde{\gamma}_t\tilde{P},\tilde{C},J\tilde{C},\tilde{\gamma}_t\tilde{P}\tilde{Q}\right\}
=Cl_{2,4}\rightarrow  Cl_{3,4}$  & $ R_0 $\\
  \hline
  PQC10b & $\left\{\tilde{\gamma}_t,J\Sigma,\tilde{\gamma}_t\tilde{P},\tilde{C},J\tilde{C},J\tilde{\gamma}_t\tilde{P}\tilde{Q}\right\}
  \rightarrow
  \left\{J\tilde{\gamma}_0,\tilde{\gamma}_t,J\Sigma,\tilde{\gamma}_t\tilde{P},\tilde{C},J\tilde{C},J\tilde{\gamma}_t\tilde{P}\tilde{Q}\right\}
  =Cl_{3,3}\rightarrow  Cl_{4,3}$  & $ R_2 $\\
  \hline

PQC11a & $\left\{\tilde{\gamma}_t,J\Sigma,J\tilde{\gamma}_t\tilde{P},\tilde{C},J\tilde{C}\right\}
\otimes\left\{\tilde{Q}\right\} \rightarrow
\left\{J\tilde{\gamma}_0,\tilde{\gamma}_t,J\Sigma,J\tilde{\gamma}_t\tilde{P},\tilde{C},J\tilde{C}\right\}\otimes\left\{\tilde{Q}\right\}
 =Cl_{3,2}\otimes Cl_{0,1}\rightarrow  Cl_{4,2}\otimes Cl_{0,1}$ & $ R_3^{\times 2} $\\
  \hline
  PQC11b & $\left\{\tilde{\gamma}_t,J\Sigma,J\tilde{\gamma}_t\tilde{P},\tilde{C},J\tilde{C}\right\}
  \otimes\left\{J\tilde{Q}\right\} \rightarrow
  \left\{J\tilde{\gamma}_0,\tilde{\gamma}_t,J\Sigma,J\tilde{\gamma}_t\tilde{P},\tilde{C},J\tilde{C}\right\}\otimes\left\{J\tilde{Q}\right\}
   =Cl_{3,2}\otimes Cl_{1,0}\rightarrow  Cl_{4,2}\otimes Cl_{1,0}$ & $ C_1 $\\
  \hline

PQC12a & $\left\{\tilde{\gamma}_t,J\Sigma,\tilde{\gamma}_t\tilde{P},\tilde{C},J\tilde{C},\tilde{\gamma}_t\tilde{P}\tilde{Q}\right\}
\rightarrow
\left\{J\tilde{\gamma}_0,\tilde{\gamma}_t,J\Sigma,\tilde{\gamma}_t\tilde{P},\tilde{C},J\tilde{C},\tilde{\gamma}_t\tilde{P}\tilde{Q}\right\}
=Cl_{4,2}\rightarrow  Cl_{5,2}$ & $ R_4 $\\
  \hline
  PQC12b & $\left\{\tilde{\gamma}_t,J\Sigma,\tilde{\gamma}_t\tilde{P},\tilde{C},J\tilde{C},J\tilde{\gamma}_t\tilde{P}\tilde{Q}\right\}
  \rightarrow
  \left\{J\tilde{\gamma}_0,\tilde{\gamma}_t,J\Sigma,\tilde{\gamma}_t\tilde{P},\tilde{C},J\tilde{C},J\tilde{\gamma}_t\tilde{P}\tilde{Q}\right\}
  =Cl_{5,1}\rightarrow  Cl_{6,1}$ & $ R_6 $\\
  \hline
\end{tabular}
\end{center}
\end{table*}

\begin{table*}[h]\footnotesize
  \begin{center}
  \caption{\label{tab:tableIV}  The construction of Clifford algebra's extension and its corresponding classifying space (Cl) for $\tilde{H}({\bf k})$ in all GBL classes. $J=i$ is imaginary unit. For convenience, we use $P,Q,C,K$ to represent $\bar{P},\bar{Q},\bar{C},\bar{K}$ when we construct the Clifford algebra's extension (in the second column of the table).\label{HkCliAlg}}

  \begin{tabular}{|c|c|c|}
    \hline
    GBL&  Clifford algebra's extension &Cl\\
  \hline
Non&  $\left\{\gamma_1,...,\gamma_d,\Sigma\right\} \rightarrow
\left\{\gamma_0,\gamma_1,...,\gamma_d,\Sigma\right\}=Cl_{d+1}\rightarrow
Cl_{d+2}$  &$C_1$\\
  \hline
P&  $\left\{\Sigma,P\Sigma\right\} \rightarrow
\left\{\gamma_0,\Sigma,P\Sigma\right\}=Cl_{2}\rightarrow
Cl_{3}$ &$ C_0 $ \\
  \hline
Qa& $\left\{\Sigma\right\}\otimes\left\{Q\right\} \rightarrow
\left\{\gamma_0,\Sigma\right\}\otimes\left\{Q\right\} =Cl_{1}\otimes Cl_1\rightarrow  Cl_{2}\otimes Cl_1$  & $ C_1^2 $\\
  \hline
  Qb&  $\left\{\Sigma,Q\Sigma\right\} \rightarrow
  \left\{\gamma_0,\Sigma,Q\Sigma\right\}=Cl_{2}\rightarrow
  Cl_{3}$  & $  C_0$ \\
  \hline

K1a& $\left\{\Sigma,\Sigma K,J\Sigma K\right\} \rightarrow
\left\{\gamma_0,\Sigma,\Sigma K,J\Sigma K\right\}=Cl_{2,1}\rightarrow
Cl_{2,2}$ & $ R_7 $ \\
  \hline
  K1b&  $\left\{J\Sigma, K,J K\right\} \rightarrow
  \left\{J\gamma_0,J\Sigma, K,J K\right\}=Cl_{1,2}\rightarrow
  Cl_{2,2}$  & $ R_1 $  \\
  \hline

K2a&  $\left\{\Sigma,\Sigma K,J\Sigma K\right\} \rightarrow
\left\{\gamma_0,\Sigma,\Sigma K,J\Sigma K\right\}=Cl_{0,3}\rightarrow
Cl_{0,4}$ & $ R_3 $ \\
  \hline
  K2b& $\left\{J\Sigma, K,J K\right\} \rightarrow
  \left\{J\gamma_0,J\Sigma, K,J K\right\}=Cl_{3,0}\rightarrow
  Cl_{4,0}$  & $ R_5 $  \\
  \hline

C1& $\left\{\Sigma,C,JC \right\} \rightarrow
\left\{J\gamma_0,\Sigma,C,JC\right\}=Cl_{0,3}\rightarrow
Cl_{1,3}$   & $  R_7$  \\
  \hline
C2&  $\left\{\Sigma,C,JC \right\} \rightarrow
\left\{J\gamma_0,\Sigma,C,JC\right\}=Cl_{2,1}\rightarrow
Cl_{3,1}$  & $ R_3 $   \\
  \hline
C3&  $\left\{J\Sigma,C,JC \right\} \rightarrow
\left\{J\gamma_0,J\Sigma,C,JC\right\}=Cl_{1,2}\rightarrow
Cl_{2,2}$  & $ R_1 $  \\
  \hline
C4& $\left\{J\Sigma,C,JC \right\} \rightarrow
\left\{J\gamma_0,J\Sigma,C,JC\right\}=Cl_{3,0}\rightarrow
Cl_{4,0}$  & $ R_5  $  \\
  \hline
PQ1&  $\left\{\Sigma,\Sigma P\right\}\otimes\left\{Q\right\} \rightarrow
\left\{\gamma_0,\Sigma,\Sigma P\right\}\otimes\left\{Q\right\} =Cl_{2}\otimes Cl_1\rightarrow  Cl_{3}\otimes Cl_1$  & $ C_0^2 $    \\
  \hline
PQ2& $\left\{\Sigma,\Sigma P,\Sigma P Q\right\} \rightarrow
\left\{\gamma_0,\Sigma,\Sigma P,\Sigma P Q\right\}=Cl_{3}\rightarrow
Cl_{4}$   & $ C_1 $   \\
  \hline
PK1&  $\left\{\Sigma,\Sigma P,K, J K\right\} \rightarrow
\left\{J\gamma_0,\Sigma,\Sigma P,K, J K\right\}=Cl_{1,3}\rightarrow
Cl_{2,3}$  & $ R_0 $   \\
  \hline
PK2&  $\left\{\Sigma,\Sigma P,K, J K\right\} \rightarrow
\left\{J\gamma_0,\Sigma,\Sigma P,K, J K\right\}=Cl_{3,1}\rightarrow
Cl_{4,1}$ & $ R_4 $   \\
  \hline

PK3a&  $\left\{\Sigma,J\Sigma P,K, J K\right\} \rightarrow
\left\{J\gamma_0,\Sigma,J\Sigma P,K, J K\right\}=Cl_{0,4}\rightarrow
Cl_{1,4}$ & $ R_6 $   \\
  \hline
  PK3b& $\left\{\Sigma,J\Sigma P,K, J K\right\} \rightarrow
  \left\{J\gamma_0,\Sigma,J\Sigma P,K, J K\right\}=Cl_{2,2}\rightarrow
  Cl_{3,2}$   & $  R_2 $   \\
  \hline
PC1&  $\left\{\Sigma,\Sigma P,C, J C\right\} \rightarrow
\left\{J\gamma_0,\Sigma,\Sigma P,C, J C\right\}=Cl_{1,3}\rightarrow
Cl_{2,3}$  & $ R_0 $  \\
  \hline
PC2&  $\left\{\Sigma,J\Sigma P,C, J C\right\} \rightarrow
\left\{J\gamma_0,\Sigma,J\Sigma P,C, J C\right\}=Cl_{0,4}\rightarrow
Cl_{1,4}$ & $ R_6 $   \\
  \hline
PC3&  $\left\{\Sigma,\Sigma P,C, J C\right\} \rightarrow
\left\{J\gamma_0,\Sigma,\Sigma P,C, J C\right\}=Cl_{3,1}\rightarrow
Cl_{4,1}$   & $ R_4 $  \\
  \hline
PC4&  $\left\{\Sigma,J\Sigma P,C, J C\right\} \rightarrow
\left\{J\gamma_0,\Sigma,J\Sigma P,C, J C\right\}=Cl_{2,2}\rightarrow
Cl_{3,2}$  & $ R_2 $   \\
  \hline

QC1a&  $\left\{\Sigma,C,J C\right\}\otimes\left\{Q\right\} \rightarrow
\left\{J\gamma_0,\Sigma,C,J C\right\}\otimes\left\{Q\right\} =Cl_{0,3}\otimes Cl_{0,1}\rightarrow  Cl_{1,3}\otimes Cl_{0,1}$    & $ R_7^2 $    \\
  \hline
  QC1b&  $\left\{\Sigma,C,J C, \Sigma Q\right\} \rightarrow
  \left\{J\gamma_0,\Sigma,C,J C,\Sigma Q\right\} =Cl_{1,3}\rightarrow  Cl_{2,3}$    & $ R_0 $    \\
  \hline

 QC2a&  $\left\{\Sigma,C,J C\right\}\otimes\left\{JQ\right\} \rightarrow
 \left\{J\gamma_0,\Sigma,C,J C\right\}\otimes\left\{JQ\right\} =Cl_{0,3}\otimes Cl_{1,0}\rightarrow  Cl_{1,3}\otimes Cl_{1,0}$ & $ C_1 $   \\
  \hline
QC2b&  $\left\{\Sigma,C,J C, J\Sigma Q\right\} \rightarrow
\left\{J\gamma_0,\Sigma,C,J C,J\Sigma Q\right\} =Cl_{0,4}\rightarrow  Cl_{1,4}$  & $ R_6 $   \\
  \hline

  QC3a&   $\left\{\Sigma,C,J C\right\}\otimes\left\{Q\right\} \rightarrow
  \left\{J\gamma_0,\Sigma,C,J C\right\}\otimes\left\{Q\right\} =Cl_{2,1}\otimes Cl_{0,1}\rightarrow  Cl_{3,1}\otimes Cl_{0,1}$  & $ R_3^2 $   \\
  \hline
QC3b&  $\left\{\Sigma,C,J C, \Sigma Q\right\} \rightarrow
\left\{J\gamma_0,\Sigma,C,J C,\Sigma Q\right\} =Cl_{3,1}\rightarrow  Cl_{4,1}$    & $ R_4 $   \\
  \hline

  QC4a&  $\left\{\Sigma,C,J C\right\}\otimes\left\{JQ\right\} \rightarrow
  \left\{J\gamma_0,\Sigma,C,J C\right\}\otimes\left\{JQ\right\} =Cl_{2,1}\otimes Cl_{1,0}\rightarrow  Cl_{3,1}\otimes Cl_{1,0}$  & $ C_1 $   \\
  \hline
QC4b&  $\left\{\Sigma,C,J C, J\Sigma Q\right\} \rightarrow
\left\{J\gamma_0,\Sigma,C,J C,J\Sigma Q\right\} =Cl_{2,2}\rightarrow  Cl_{3,2}$ & $ R_2 $   \\
  \hline

  QC5a&  $\left\{J\Sigma,C,J C\right\}\otimes\left\{Q\right\} \rightarrow
  \left\{J\gamma_0,J\Sigma,C,J C\right\}\otimes\left\{Q\right\} =Cl_{1,2}\otimes Cl_{0,1}\rightarrow  Cl_{2,2}\otimes Cl_{0,1}$  & $ R_1^2 $    \\
  \hline
QC5b& $\left\{J\Sigma,C,J C, J\Sigma Q\right\} \rightarrow
\left\{J\gamma_0,J\Sigma,C,J C,J\Sigma Q\right\} =Cl_{1,3}\rightarrow  Cl_{2,3}$  & $ R_0 $   \\
  \hline

  QC6a&  $\left\{J\Sigma,C,J C\right\}\otimes\left\{JQ\right\} \rightarrow
  \left\{J\gamma_0,J\Sigma,C,J C\right\}\otimes\left\{JQ\right\} =Cl_{1,2}\otimes Cl_{1,0}\rightarrow  Cl_{2,2}\otimes Cl_{1,0}$  & $ C_1 $    \\
  \hline
QC6b&  $\left\{J\Sigma,C,J C, \Sigma Q\right\} \rightarrow
\left\{J\gamma_0,J\Sigma,C,J C,\Sigma Q\right\} =Cl_{2,2}\rightarrow  Cl_{3,2}$   & $ R_2 $   \\
  \hline

  QC7a&  $\left\{J\Sigma,C,J C\right\}\otimes\left\{Q\right\} \rightarrow
  \left\{J\gamma_0,J\Sigma,C,J C\right\}\otimes\left\{Q\right\} =Cl_{3,0}\otimes Cl_{0,1}\rightarrow  Cl_{4,0}\otimes Cl_{0,1}$   & $ R_5^2 $   \\
  \hline
QC7b&  $\left\{J\Sigma,C,J C, J\Sigma Q\right\} \rightarrow
\left\{J\gamma_0,J\Sigma,C,J C,J\Sigma Q\right\} =Cl_{3,1}\rightarrow  Cl_{4,1}$  & $ R_4 $    \\
  \hline

 QC8a&  $\left\{J\Sigma,C,J C\right\}\otimes\left\{JQ\right\} \rightarrow
 \left\{J\gamma_0,J\Sigma,C,J C\right\}\otimes\left\{JQ\right\} =Cl_{3,0}\otimes Cl_{1,0}\rightarrow  Cl_{4,0}\otimes Cl_{1,0}$  & $ C_1 $    \\
  \hline
QC8b&  $\left\{J\Sigma,C,J C, \Sigma Q\right\} \rightarrow
\left\{J\gamma_0,J\Sigma,C,J C,\Sigma Q\right\} =Cl_{4,0}\rightarrow  Cl_{5,0}$ & $ R_6 $   \\
  \hline

PQC1&  $\left\{\Sigma,\Sigma P,C,J C\right\}\otimes\left\{Q\right\} \rightarrow
\left\{J\gamma_0,\Sigma,\Sigma P, C,J C\right\}\otimes\left\{Q\right\} =Cl_{1,3}\otimes Cl_{0,1}\rightarrow  Cl_{2,3}\otimes Cl_{0,1}$   & $ R_0^2 $    \\
  \hline
PQC2&  $\left\{\Sigma,\Sigma P,C,J C\right\}\otimes\left\{JQ\right\} \rightarrow
\left\{J\gamma_0,\Sigma,\Sigma P, C,J C\right\}\otimes\left\{JQ\right\} =Cl_{1,3}\otimes Cl_{1,0}\rightarrow  Cl_{2,3}\otimes Cl_{1,0}$   & $ C_0 $   \\
  \hline
PQC3&  $\left\{\Sigma,J\Sigma P,C,J C,J\Sigma P Q\right\} \rightarrow
\left\{J\gamma_0,\Sigma,J\Sigma P,C,J C,J\Sigma P Q\right\} =Cl_{1,4}\rightarrow  Cl_{2,4}$    & $ R_7 $   \\
  \hline
PQC4&  $\left\{\Sigma,J\Sigma P,C,J C,\Sigma P Q\right\} \rightarrow
\left\{J\gamma_0,\Sigma,J\Sigma P,C,J C,\Sigma P Q\right\} =Cl_{0,5}\rightarrow  Cl_{1,5}$   & $  R_5 $   \\
  \hline

 PQC5&  $\left\{\Sigma,\Sigma P,C,J C\right\}\otimes\left\{Q\right\} \rightarrow
 \left\{J\gamma_0,\Sigma,\Sigma P, C,J C\right\}\otimes\left\{Q\right\} =Cl_{3,1}\otimes Cl_{0,1}\rightarrow  Cl_{4,1}\otimes Cl_{0,1}$  &$ R_4^2 $ \\
  \hline
PQC6&  $\left\{\Sigma,\Sigma P,C,J C\right\}\otimes\left\{JQ\right\} \rightarrow
\left\{J\gamma_0,\Sigma,\Sigma P, C,J C\right\}\otimes\left\{JQ\right\} =Cl_{3,1}\otimes Cl_{1,0}\rightarrow  Cl_{4,1}\otimes Cl_{1,0}$  & $ C_0 $  \\
  \hline
PQC7&  $\left\{\Sigma,J\Sigma P,C,J C,J\Sigma P Q\right\} \rightarrow
\left\{J\gamma_0,\Sigma,J\Sigma P,C,J C,J\Sigma P Q\right\} =Cl_{3,2}\rightarrow  Cl_{4,2}$  & $ R_3  $ \\
  \hline
PQC8&  $\left\{\Sigma,J\Sigma P,C,J C,\Sigma P Q\right\} \rightarrow
\left\{J\gamma_0,\Sigma,J\Sigma P,C,J C,\Sigma P Q\right\} =Cl_{2,3}\rightarrow  Cl_{3,3}$  & $  R_1 $ \\
  \hline

PQC9a&  $\left\{\Sigma,J\Sigma P,C,J C\right\}\otimes\left\{Q\right\} \rightarrow
\left\{J\gamma_0,\Sigma,J\Sigma P, C,J C\right\}\otimes\left\{Q\right\} =Cl_{0,,4}\otimes Cl_{0,1}\rightarrow  Cl_{1,4}\otimes Cl_{0,1}$  & $ R_6^2 $ \\
  \hline
  PQC9b&  $\left\{\Sigma,J\Sigma P,C,J C\right\}\otimes\left\{JQ\right\} \rightarrow
  \left\{J\gamma_0,\Sigma,J\Sigma P, C,J C\right\}\otimes\left\{JQ\right\} =Cl_{0,,4}\otimes Cl_{1,0}\rightarrow  Cl_{1,4}\otimes Cl_{1,0}$  & $ C_0 $  \\
  \hline

PQC10a&  $\left\{\Sigma,\Sigma P,C,J C,\Sigma P Q\right\} \rightarrow
\left\{J\gamma_0,\Sigma,\Sigma P,C,J C,\Sigma P Q\right\} =Cl_{1,4}\rightarrow  Cl_{2,4}$   &$ R_7 $ \\
  \hline
  PQC10b&  $\left\{\Sigma,\Sigma P,C,J C,J\Sigma P Q\right\} \rightarrow
  \left\{J\gamma_0,\Sigma,\Sigma P,C,J C,J\Sigma P Q\right\} =Cl_{2,3}\rightarrow  Cl_{3,3}$  & $ R_1 $  \\
  \hline

PQC11a& $\left\{\Sigma,J\Sigma P,C,J C\right\}\otimes\left\{Q\right\} \rightarrow
\left\{J\gamma_0,\Sigma,J\Sigma P, C,J C\right\}\otimes\left\{Q\right\} =Cl_{2,2}\otimes Cl_{0,1}\rightarrow  Cl_{3,2}\otimes Cl_{0,1}$   &$ R_2^2 $  \\
  \hline
  PQC11b&  $\left\{\Sigma,J\Sigma P,C,J C\right\}\otimes\left\{JQ\right\} \rightarrow
  \left\{J\gamma_0,\Sigma,J\Sigma P, C,J C\right\}\otimes\left\{JQ\right\} =Cl_{2,2}\otimes Cl_{1,0}\rightarrow  Cl_{3,2}\otimes Cl_{1,0}$  & $ C_0 $  \\
  \hline

PQC12a&   $\left\{\Sigma,\Sigma P,C,J C,\Sigma P Q\right\} \rightarrow
\left\{J\gamma_0,\Sigma,\Sigma P,C,J C,\Sigma P Q\right\} =Cl_{3,2}\rightarrow  Cl_{4,2}$   &$ R_3 $  \\
\hline
  PQC12b& $\left\{\Sigma,\Sigma P,C,J C,J\Sigma P Q\right\} \rightarrow
  \left\{J\gamma_0,\Sigma,\Sigma P,C,J C,J\Sigma P Q\right\} =Cl_{4,1}\rightarrow  Cl_{5,1}$   & $ R_5 $ \\
  \hline
\end{tabular}
\end{center}
\end{table*}

\clearpage
\newpage
\twocolumngrid

\end{document}